\documentclass[longauth]{aa}

\pdfoutput=1

\usepackage{natbib}
\usepackage{txfonts}
\usepackage{gensymb}
\usepackage{inputenc}
\usepackage{float}

\usepackage{soul}

\bibpunct{(}{)}{;}{a}{}{,}

\usepackage{color}
\usepackage{url}

\usepackage{subcaption}

\usepackage[breaklinks=true,colorlinks,citecolor=blue]{hyperref}

\usepackage{graphicx}

\begin{document}
        
\title{The VIMOS Public Extragalactic Redshift Survey (VIPERS)}
        
\subtitle{The complexity of galaxy populations at $0.4< z<1.3$ revealed with unsupervised machine-learning algorithms\thanks{based on
                        observations collected at the European Southern Observatory, Cerro Paranal, Chile, using the Very Large Telescope under programs 182.A-0886 and partly 070.A-9007.
                        Also based on observations obtained with MegaPrime/MegaCam, a joint project of CFHT and CEA/DAPNIA, at the Canada-France-Hawaii Telescope (CFHT), which is operated by the
                        National Research Council (NRC) of Canada, the Institut National des Sciences de l’Univers of the Centre National de la Recherche Scientifique (CNRS) of France, and the University of Hawaii. This work is based in part on data products produced at TERAPIX and the Canadian Astronomy Data Centre as part of the Canada-France-Hawaii Telescope Legacy Survey, a collaborative project of NRC and CNRS. The VIPERS web site is  \url{http://www.vipers.inaf.it/}.}}

   \titlerunning{The complexity of galaxy populations at $z\sim0.7$}
\author{
        %
        M.~Siudek\inst{\ref{warsaw-theory},\ref{warsaw-nucl}}
        \and K.~Ma{\l}ek\inst{\ref{warsaw-nucl}}
        \and A.~Pollo\inst{\ref{warsaw-nucl},\ref{krakow}}
        \and T.~Krakowski\inst{\ref{warsaw-nucl}}
        \and A.~Iovino\inst{\ref{brera}}
        \and M.~Scodeggio\inst{\ref{iasf-mi}}  
        \and T.~Moutard\inst{\ref{halifax},\ref{lam}}  
        \and G.~Zamorani\inst{\ref{oabo}}
        %
        %
        \and L.~Guzzo\inst{\ref{unimi},\ref{brera}}      
        \and B.~Garilli\inst{\ref{iasf-mi}}          
        \and B.~R.~Granett\inst{\ref{brera},\ref{unimi}}
        \and M.~Bolzonella\inst{\ref{oabo}}      
        \and S.~de la Torre\inst{\ref{lam}}       
        %
        %
        \and U.~Abbas\inst{\ref{oa-to}}
        \and C.~Adami\inst{\ref{lam}}
        \and D.~Bottini\inst{\ref{iasf-mi}}
        \and A.~Cappi\inst{\ref{oabo},\ref{nice}}
        \and O.~Cucciati\inst{\ref{oabo}}           
        \and I.~Davidzon\inst{\ref{lam},\ref{oabo}}   
        \and P.~Franzetti\inst{\ref{iasf-mi}}   
        \and A.~Fritz\inst{\ref{iasf-mi}}       
        \and J.~Krywult\inst{\ref{kielce}}
        \and V.~Le Brun\inst{\ref{lam}}
        \and O.~Le F\`evre\inst{\ref{lam}}
        \and D.~Maccagni\inst{\ref{iasf-mi}}
        \and F.~Marulli\inst{\ref{unibo},\ref{infn-bo},\ref{oabo}} 
        \and M.~Polletta\inst{\ref{iasf-mi},\ref{marseille-uni},\ref{toulouse}}
        \and L.A.M.~Tasca\inst{\ref{lam}}
        \and R.~Tojeiro\inst{\ref{st-andrews}}  
        \and D.~Vergani\inst{\ref{oabo}}
        \and A.~Zanichelli\inst{\ref{ira-bo}}
        %
        %
        \and S.~Arnouts\inst{\ref{lam},\ref{cfht}} 
        \and J.~Bel\inst{\ref{cpt}}
        \and E.~Branchini\inst{\ref{roma3},\ref{infn-roma3},\ref{oa-roma}}
        \and J.~Coupon\inst{\ref{geneva}}
        \and G.~De Lucia\inst{\ref{oats}}
        \and O.~Ilbert\inst{\ref{lam}}
        %
        %
        %
        %
        %
        \and C.~P.~Haines\inst{\ref{brera}}  
        \and L.~Moscardini\inst{\ref{unibo},\ref{infn-bo},\ref{oabo}}   
        \and T.~T.~Takeuchi\inst{\ref{nagoya}}
        %
        %
        %
        %
        %
        %
}
   \offprints{M. Siudek \\ \email{gsiudek@cft.edu.pl}}
\institute{
        Center for Theoretical Physics, Al. Lotnikow 32/46, 02-668 Warsaw, Poland \label{warsaw-theory} 
        \and National Centre for Nuclear Research, ul. Hoza 69, 00-681 Warszawa, Poland \label{warsaw-nucl}
        \and Astronomical Observatory of the Jagiellonian University, Orla 171, 30-001 Cracow, Poland \label{krakow} 
        \and  INAF - Osservatorio Astronomico di Brera, Via Brera 28, 20122 Milano
        --  via E. Bianchi 46, 23807 Merate, Italy \label{brera}
        \and INAF - Istituto di Astrofisica Spaziale e Fisica Cosmica Milano, via Bassini 15, 20133 Milano, Italy \label{iasf-mi}
        \and Department of Astronomy \& Physics, Saint Mary's University, 923 Robie Street, Halifax, Nova Scotia, B3H 3C3, Canada \label{halifax}
        \and Aix Marseille Univ, CNRS, LAM, Laboratoire d'Astrophysique de
        Marseille, Marseille, France  \label{lam}
        \and INAF - Osservatorio di Astrofisica e Scienza dello Spazio di Bologna,via Gobetti 93/3, 40129 Bologna, Italy \label{oabo} 
        \and Universit\`{a} degli Studi di Milano, via G. Celoria 16, 20133 Milano, Italy \label{unimi}
        \and INAF - Osservatorio Astrofisico di Torino, 10025 Pino Torinese, Italy \label{oa-to}
        \and Laboratoire Lagrange, UMR7293, Universit\'e de Nice Sophia Antipolis, CNRS, Observatoire de la C\^ote d’Azur, 06300 Nice, France \label{nice}
        \and Dipartimento di Fisica e Astronomia - Alma Mater Studiorum Universit\`{a} di Bologna, via Gobetti 93/2, I-40129, Bologna, Italy \label{unibo}
        \and Institute of Physics, Jan Kochanowski University, ul. Swietokrzyska 15, 25-406 Kielce, Poland \label{kielce}
        \and INFN, Sezione di Bologna, viale Berti Pichat 6/2, I-40127, Bologna, Italy \label{infn-bo}
        \and IRAP, Universit\`{e} de Toulouse, CNRS, UPS, Toulouse, France \label{marseille-uni}
        \and IRAP,  9 av. du colonel Roche, BP 44346, F-31028 Toulouse cedex 4, France \label{toulouse} 
        \and School of Physics and Astronomy, University of St Andrews, St Andrews KY16 9SS, UK \label{st-andrews}
        \and INAF - Istituto di Radioastronomia, via Gobetti 101, I-40129,
        Bologna, Italy \label{ira-bo}
        \and Canada-France-Hawaii Telescope, 65--1238 Mamalahoa Highway, Kamuela, HI 96743, USA \label{cfht}
        \and Aix Marseille Univ, Univ Toulon, CNRS, CPT, Marseille, France \label{cpt}
        \and Dipartimento di Matematica e Fisica, Universit\`{a} degli Studi Roma Tre, via della Vasca Navale 84, 00146 Roma, Italy\label{roma3} 
        \and INFN, Sezione di Roma Tre, via della Vasca Navale 84, I-00146 Roma, Italy \label{infn-roma3}
        \and INAF - Osservatorio Astronomico di Roma, via Frascati 33, I-00040 Monte Porzio Catone (RM), Italy \label{oa-roma}
        \and Department of Astronomy, University of Geneva, ch. d’Ecogia 16, 1290 Versoix, Switzerland \label{geneva}
        \and INAF - Osservatorio Astronomico di Trieste, via G. B. Tiepolo 11, 34143 Trieste, Italy \label{oats}
        \and Institute of Cosmology and Gravitation, Dennis Sciama Building, University of Portsmouth, Burnaby Road, Portsmouth, PO13FX \label{icg}
        \and Institut Universitaire de France \label{inst-france}
        \and Institute d'Astrophysique de Paris, UMR7095 CNRS, Universit\'{e} Pierre et Marie Curie, 98 bis Boulevard Arago, 75014 Paris, France \label{iap} 
        \and Institute for Astronomy, University of Edinburgh, Royal
        Observatory, Blackford Hill, Edinburgh EH9 3HJ, UK \label{roe}
        \and Division of Particle and Astrophysical Science, Nagoya University, Furo-cho, Chikusa-ku, 464-8602 Nagoya, Japan \label{nagoya}
        %
}

        %
  \abstract
{}
{Various galaxy classification schemes have been developed so far to constrain the main physical processes regulating  evolution of different galaxy types. In the era of a deluge of astrophysical information and recent progress in machine learning, a new approach to galaxy classification has become imperative.  }
{In this paper, we employ a Fisher Expectation-Maximization
  (FEM) unsupervised algorithm working in a parameter space
  of 12 rest-frame magnitudes and spectroscopic redshift.  The model
  (DBk) and the number of classes (12) were established based on the
  joint analysis of standard statistical criteria and confirmed by the
  analysis of the galaxy distribution with respect to a number of
  classes and their properties. This new approach allows us
  to classify galaxies based on only their redshifts and ultraviolet to near-infrared ($UV-NIR)$
  spectral energy distributions. 
   }
{The FEM unsupervised algorithm has automatically distinguished
  12 classes: 11 classes of VIPERS galaxies and an
    additional class of broad-line active galactic nuclei (AGNs). After a first broad
  division into blue, green, and red categories, we obtained a further sub-division into: three red, three green, and five blue
  galaxy classes. The FEM classes follow the galaxy sequence from the earliest to the latest types, which is reflected in their colours (which are constructed from rest-frame magnitudes used in the classification       procedure) but also their morphological, physical, and spectroscopic properties (not included in the classification scheme). We demonstrate that the members of each  class share similar physical and spectral properties. In particular, we are able to find three different classes of red passive galaxy populations. Thus, we demonstrate the potential of an unsupervised approach to galaxy classification and we retrieve the complexity of galaxy populations at $z\sim0.7$, a task that usual, simpler, colour-based approaches cannot fulfil.   }
{}


\maketitle


        \section{Introduction}\label{sec:intr}
        
The problem of classification of galaxies and dividing them into different types is as old as the notion of "extragalactic nebulae" \citep{hubble}. 
As \citet{Sandage1975}, and more recently \citet{buta2011} and \citet{Buta2011b} point out, classification of objects is the first step in the development of most sciences, and applies to galaxy studies no less than to any field of research. 
Only once we find common features of studied objects and use them to
sort them into categories, do we obtain a starting point for the further analysis. 
Identifying similarities and differences between the selected groups allows us to then build theoretical models, which can ultimately lead us to the global picture of physical mechanisms at the origin of their properties. 

Galaxies in the local Universe display a variety of shapes and structural properties. 
The main classification system still in use is the Hubble tuning fork diagram \citep{hubble,Hubble1936}, with all the refinements introduced by \cite{sandage} and \cite{vauc}, based on the morphological properties of galaxies \citep[see][for a detailed discussion]{vandenBergh1998,buta2011}. 
In the modern context, we alternatively refer to continuity of types in the morphological parameter space, where numerous morphological features are taken into account \citep{Lintott2008,Lintott2011,Buta2010,Kartaltepe2015}. 
The basic Hubble classification of galaxies into "early" and "late" types (and their subtypes) has survived because, among other reasons, these types correlate well with other properties of galaxies, such as colours, stellar content, neutral hydrogen content and so on. \citep{kennicutt92,Roberts1994,Buta1994,Strateva2001,Deng2010,moutard16a}. 

Indeed, many types of galaxy properties display bimodal distributions: photometric parameters, such as colours
\citep[e.g.][]{Bell2004,balogh2004,baldry2006,Franzetti07,taylor2015},
morphological parameters like the S\'ersic index \citep[e.g.][]{sersic63,Strateva2001,Driver2006,krywult}, the strength of spectral features \citep[e.g.][]{balogh,kauffmann03,siudek17} and so on. 
Therefore, these properties are often used as the basis for galaxy classification, especially at higher redshifts, $z$, where detailed galaxy morphologies are difficult to observe. 
In particular, colour-colour diagrams \citep[e.g. the $(NUV-r) - (r-K)$ diagram (hereafter: $NUVrK$), $NURrJ$, $BzK$, $NUViB$, introduced/used by][ respectively]{arnouts,Bundy10,Daddi2004,Cibinel13} are often used for the purpose of galaxy classification. 
More refined selection processes can be based on the multi-modality criterion, which selects red passive galaxies, intermediate "green valley" objects, and blue star-forming galaxies based on their rest-frame colours, spectral parameters, or colour and colour-S\'{e}rsic index distributions simultaneously~\citep[e.g.][]{Bell2004,baldry2006,Franzetti07,Bruce14,lange2015,krywult,haines16}. 
The bimodality criterion can be enriched by a variable cut in galaxy colours that evolves with redshift \citep[][]{peng,fritz14,moutard16b,siudek17}, as a non-evolving cut applied for high-redshift galaxies can result in the selection of the reddest and most luminous red-type galaxies in  one group and  a mixture of star-forming and less massive red galaxies in the second group.
 
The methods presented above are powerful tools, but they are sensitive only to a few specific properties. 
A disadvantage of the methods presented above is the small number of groups which can typically be obtained: selection based on bimodality of the distribution of a certain property or a set of correlated properties usually allows for selection of only two or three groups (blue star-forming galaxies -- intermediate types -- red passive galaxies). Some two-dimensional (2D)  colour-colour diagrams, like the $NUVrK$, are used for a more detailed classification \cite[e.g. ][]{arnouts,moutard16a,moutard16b,davidzon16} but are still limited to a relatively small number of groups.

Moreover, classifications based on the standard 2D cuts suffer from multi-fold selection-effect problems. 
For example, the properties of red passive galaxies selected using
different criteria (photometry, morphology, and spectroscopy) differ
from one selection to another~\citep[e.g.][]{renzini, morescocolor}. 
Red passive galaxy samples are mostly affected by some level of contamination from dust-reddened galaxies with relatively low levels of star formation activity that may strongly affect their mean properties. 
\cite{morescocolor} showed that the selection of the purest sample of red passive galaxies demands the combination of different criteria (in this case, morphological, spectroscopic and photometric information) confirming the necessity of multidimensional approaches in order to avoid obtaining a biased sample of different galaxy types.       

Two-dimensional diagrams based on the flux ratios (or equivalent
widths) of spectral lines can also be a powerful tool, for example for AGN diagnostic and classification \citep[e.g. Baldwin, Phillips \& Terlevich "BPT"  diagrams based on the ratios of "blue" and "red" lines:][]{Baldwin81,2010Lamareille}. 
The BPT diagram allows for separation of: (1) star-forming galaxies, (2) Seyferts, (3) low-ionisation nuclear emission-line regions (LINERS), and the two composite groups, which consist of: (4) star-forming galaxies and Seyferts, and (5) star-forming galaxies and LINERS. 
        
However, it becomes clear that any classification based on a small number of parameters, even carefully chosen, is far too simple to reflect the huge range of different cosmic objects. 
        
While classical methods  of classification are still common and very
useful, recent advancements in automatic machine learning have opened
up new possibilities for the classification of distant sources. 
In principle, they allow us to operate in a multi-parameter space, combining all the available pieces of information: photometric measurements, redshifts, spectral lines, and morphologies. 
In principle, such an approach can immensely improve the galaxy classification across a wide redshift range. However, there is also a risk of including too much redundant or indiscriminative information which would blur the final result or lead to the unjustified subdivision of types.

\cite{Ball10} and \cite{FBD15} gave a comprehensive review of different methods for clustering objects into synthetic groups in astrophysics,   showing that classification in multi-dimensional parameter space, backed by sophisticated multivariate statistical tools, leads to a selection of sources that is more accurate than, for example, the  colour-colour method. 
In general, we can distinguish two main groups of algorithms: supervised and unsupervised learning algorithms.

Briefly, supervised algorithms classify data into classes that
  have previously been defined and anticipated. The disadvantage of this method is the requirement to create  a training sample a priori and, at the same time, no possibility to define new classes of objects. 
Unsupervised learning algorithms (such as those used in our analysis)  search for clusters of objects characterised by some pattern in the data and try to discover new classification schemes without any prior assumptions.
The unsupervised algorithm fits the input vector data to a statistical model. 
The algorithm then tries to optimise the parameters of the model in iterative cycles to find the best fit to the data with an optimised number of classes.
Once the defined satisfactory criteria are fulfilled, the iterations are stopped. 
The best known unsupervised learning algorithms include: (a) expectation-maximisation \citep[hereafter: EM]{Bilmes98} algorithms - used to deal with complex data structures, for example, clusters; (b) k-means~\citep{Salman2011} - whose aim is to assign observations to clusters in which each observation belongs to the nearest mean; and
(c) hierarchical clustering \citep{Balcan2014} - treating each point
as a cluster and successively merging pairs of clusters recursively
until all clusters are merged into one single group that contains all
of the points.
An overview of unsupervised approaches used in astronomy can be found in \cite{DAbrusco12}.  
        
Supervised algorithms have already yielded clear achievements in the selection of different astronomical sources. 
However, this approach only allows us to reproduce standard classifications, mostly based on optical colours, which is not optimal to extract all the relevant information from the data. 
Therefore, it is necessary to adopt unsupervised methods to
efficiently extract all the information encoded in the data.  
The applications of unsupervised machine-learning algorithms to galaxy
classification have until now mainly been applied to galaxy spectra. 
In particular, \cite{SDSS_class_2010} used an unsupervised k-means cluster analysis algorithm to classify all spectra in the final Sloan Digital Sky Survey data release~7 (SDSS/DR7). 
They identified as many as 17 different classes of galaxies. This
would have been extremely challenging using classical methods due to the huge number of spectra ($\sim$174k) to process. The classification was based on the multidimensional cuts  in the space of a mixture of features (emission/absorption lines, continuum, fluxes and errors) making use of 3849 measurements for each object.
The selected classes are well separated in the colour sequence and morphological groups. 
The spectroscopic templates obtained for each class can be used for redshift measurements ($z<0.25$) as well as to trace morphological and spectroscopic changes in cosmic time. 
        
Principal component analysis (PCA) has been used to classify astronomical data based on broadband measurements  or as a tool to clean spectra \citep[e.g.][]{marchetti13,Wild14,Marchetti17}. 
\cite{marchetti13} used a PCA algorithm to classify 27,350 optical spectra in the redshift range $0.4 < z <1.0$  collected by the VIPERS survey (Public Data Release~1, hereafter PDR1). 
The algorithm repaired parts of VIPERS spectra affected by noise or sky residuals and reconstructed gaps in the spectra. 
A classification into four main classes (early, intermediate, late and
starburst galaxies) was carried out, based on a set of orthogonal spectral templates and the three most significant components (eigen-coefficients) obtained for each galaxy.

In this paper, we introduce a new method of galaxy classification via an unsupervised learning algorithm applied to the galaxies observed by the VIMOS Public Extragalactic Redshift Survey (VIPERS). 
The VIPERS survey acquired spectra for $\sim10^{5}$ galaxies. 
For each galaxy, both spectroscopic measurements (redshift, lines, fluxes) and photometric data are provided.  
This makes VIPERS a perfect dataset for unsupervised classification; it is large enough to separate many different classes on a statistically sound level, and, at the same time, all the wealth of the spectroscopic and photometric information can be used to construct the feature space, and later for the validation process. 
Moreover, previous analyses made on the VIPERS data provide us with additional parameters such as S\'{e}rsic indices and physical properties, obtained by fitting the spectral energy distributions (SEDs) (stellar mass, star formation rate (SFR), etc.). 
All these additional measurements, even when not used for the classification itself, can serve for an a posteriori interpretation of physical properties of different classes.  
Our method is based on the multidimensional space defined by the rest-frame luminosities measured in 12 bands and, additionally, spectroscopic redshift information.

The availability of spectroscopic data for VIPERS galaxies allows us to verify how the classes obtained using the broadband rest-frame photometry are reflected in the spectral properties of galaxies. 
We demonstrate that the classification based on our automatic algorithm and confirmed by spectroscopic features  provides a homogeneous view of  different classes of galaxies which may be used as the starting point to analyse their evolutionary tracks leading to the formation of today's galaxy types. 
    
The paper is organised as follows.    
In Sect.~\ref{sec:data},  we describe the sample selection. 
Section~\ref{sec:method} gives an overview of the FisherEM  methodology. 
In Sect.~\ref{sec:results}, we present the main results and discuss their physical meaning. 
A summary is presented in Sect.~\ref{sec:summary}. 
We validate the model and the number of classes in Appendix~\ref{app:number}, and discuss the class membership probabilities in Appendix~\ref{app:A}. 
We compared FEM classification to a principal component analysis (PCA) scheme in Appendix~\ref{app:pca}, and relate FEM classes to Hubble types given by~\cite{kennicutt92} in Appendix~\ref{app:kennicutt}. 

In our analysis, we used the free statistical environment software R3\footnote{R Core Team (2013). R: A language and environment for statistical computing. R Foundation for Statistical Computing, Vienna, Austria. ISBN 3-900051-07-0, \url{http://www.R-project.org/}.}
with the FisherEM package 4 \citep{Bou2011}. 
Throughout the paper we use a cosmological framework assuming $\Omega_{m}$ = 0.30, $\Omega_{\Lambda}$ = 0.70, and $H_{0}=70$ km s$^{-1}$ Mpc$^{-1}$.

\section{Data}\label{sec:data}
        
In this paper, we make use of the final galaxy sample from the VIMOS Public Extragalactic Redshift Survey\footnote{See \url{http://vipers.inaf.it}} \citep[VIPERS,][]{scodeggio16}. 
VIPERS is a spectroscopic survey carried out with the VIMOS
spectrograph~\citep{lefevre03} on the 8.2m ESO Very Large Telescope
(VLT) aimed at measuring redshifts for $\sim$100,000 galaxies in the redshift range 0.5--1.2. 
VIPERS covered an area of $\sim$23.5 $\deg^2$ on the sky, observing galaxies brighter than $i_{AB}=22.5$ at redshifts higher than 0.5  (a pre-selection in the ($u$-$g$) and ($r$-$i$) colour-colour plane was used to remove galaxies at lower redshifts). 
A detailed description of the survey can be found in \cite{guzzo}.
The galaxy target sample was selected from optical photometric catalogues of the Canada-France-Hawaii Telescope Legacy Survey Wide \citep[CFHTLS-Wide:][]{mellier08,goranova09}.
The data reduction pipeline and redshift quality system are described by \cite{garilli14}.

\subsection{The VIPERS dataset}        
The final data release provides spectroscopic measurements and photometric properties for 86,775 galaxies \citep{scodeggio16}. 
The associated photometric catalogue consists of magnitudes from the VIPERS Multi-Lamba Survey~\citep{moutard16a}, combining CFHTLS T0007-based $u$, $g$, $r$, $i$, $z$ photometry with GALEX $FUV$/$NUV$ and WIRCam $Ks$-band observations, complemented where available by VISTA $Z$, $Y$, $J$, $H$, $K$ photometry from the VIDEO survey~\citep{jarvis2013}.    

Physical parameters including absolute magnitudes, stellar masses, and SFRs for the VIPERS sample were obtained via SED fitting with the code LePhare~\citep{arnouts2002,ilbert2006}. 
The whole multi-wavelength information available in the VIPERS fields
(from $UV$ to $NIR$) was used, applying the \cite{bruzual} models and three extinction laws. 
In addition, absolute magnitudes were computed using the nearest observed-frame band in order to minimise the dependence on models.    
The detailed description of the VIPERS data SED-fitting scheme that we adopted in the present analysis can be found in~\cite{moutard16b}. 
        
In this work, we make use of the subset of galaxies with highly secure redshift measurements \citep[with a confidence level higher than  99\%, i.e. with redshift flag 3--4 and 13--14, see][for details]{garilli14}.  
This subset contains {52,114} objects (51,522 galaxies and 592 broad-line AGNs\footnote{Broad-line AGNs were classified by VIPERS team members according to visual inspection of spectra. In the following analysis, we refer to broad-line AGNs as sources attributed with a redshift flag 13-14.}). 
They are observed in the redshift range $0.4<z<1.3$\footnote{The 1 and
  99 percentile range of redshift is given. The broad-line AGNs are
  observed up to the redshift $z\sim4.5$.}  with a mean (median) redshift of 0.7.

\subsection{The multidimensional feature data}
Data preparation is a key issue in working with learning algorithms, both supervised and unsupervised. 
In order to minimise any biases, maximise homogeneity in the input
data, and use all of the available information, 12 rest-frame
magnitudes are chosen: $FUV$, $NUV$, $u$, $g$, $r$, $i$, $z$, $B$,
$V$, $J$, $H$, and $Ks$ derived from the SED fitting (see Sect. 2, and
\citealp{moutard16b}), as well as the spectroscopic redshift~\citep{scodeggio16}. 
To avoid grouping galaxies based on differences in their
luminosities instead of differences in their SEDs, the data were standarised. 
We normalised $i$-band to unity and transformed each absolute magnitude by the normalisation factor (the redshifts were not transformed as their values are already around unity). 
This allowed us to code the data into common numerical range preventing the algorithm from splitting our sample along any direction with extended amplitudes. 

The normalised parameters, together with the spectroscopic redshift, are then used to create a multi-parameter space for the FEM algorithm. 
The spectroscopic redshift is included in the parameter space to make the classification sensitive to possible evolutionary changes with cosmic time. 
The algorithm could identify an evolving population in different cosmic epochs as belonging to physically different classes. 
Although this is not the case for the VIPERS galaxies, where all FEM classes seem to be preserved throughout the redshift range probed by the survey (see Sect.~\ref{sec:results}), we did not want to exclude this option a priori. 
However, we verified that if the spectroscopic redshift is not included in the parameter space, the FEM classification remains practically the same. 

The global picture of the classification does not suffer significantly if we reduce the feature  space by one parameter (e.g. spectroscopic redshift). 
However, excluding each single feature has an impact on the ability of the algorithm to distinguish individual classes. 
Feature importance may be statistically determined by the analysis of the orientation of the discriminative subspace. 
The x-axis of a hyperplane separating classes in a latent subspace is constructed with an 11-degree polynomial and each coefficient describes how important each feature is for the distinction of each group. 
For example, high coefficients of the hyperplane between red passive classes for $FUV$ and $NUV$ reveal their importance in distinguishing those groups.  
Therefore, excluding $FUV$ and $NUV$ will 
result in discriminating ten classes with only one large red passive class leaving the remaining classes unchanged. 
The redundancy of selected features and their importance to distinguish each group will be further discussed by~\cite{Krakowski2018}. 

We note that the redundancy of the spectroscopic redshift reveals a great potential for future photometric missions such as Euclid and LSST. 
In~\cite{siudek2018}, we explore the potential use of photometric information solely to classify galaxies and estimate their properties. 
Reliable photometric redshifts and 12 rest-frame magnitudes obtained by the SED-fitting with the photo-z scatter
 $\sigma \sim0.03$, and the outlier rate $ \mu \sim 2\%$ obtained for the VIPERS sample, were used to verify how precisely the detailed classification could be reproduced if only photometric data were available. 
The confirmed accuracy in recreating galaxy classes: 92\%, 84\%, 96\% for red, green, and blue classes, respectively , together with the ability to efficiently separate outliers (stars and broad-line AGNs) based only on photometric data, demonstrates the potential of our approach in future large cosmology missions to distinguish different galaxy classes at $z>0.5$.           

\section{Method -- Fisher EM}\label{sec:method}
Unsupervised learning algorithms are used to divide the data of {\em a priori} unknown properties into clusters. In this paper, we use the FEM \citep{Bou2011} algorithm, which is an extension of the EM algorithm. The main goal of both the EM and the FEM classifiers is to maximise the best fit of the chosen statistical model describing the data by finding the optimal parameters of this model. 
In the case of the FEM algorithm, the main assumption is that the data can be grouped into a common discriminative latent subspace which is modelled by the discriminant latent mixture (hereafter: DLM) model~\citep[]{Bou2011}. 
This discriminative latent subspace is defined by linear combinations of the input data~\citep[latent variables;][]{Bou2014}. 
It is then optimised to maximise the separation between groups and minimise their variance at the same time.
The second assumption of the FEM algorithm is that our data
can be separated into an {\em a priori} unknown number of groups, each described by a Gaussian profile in the multidimensional parameter space. The role of the FEM algorithm is to find 
the best fit of these multi-Gaussian profiles to the data, optimising both the number of the groups
and their location in the parameter space.  

\subsection{The performance of the FEM algorithm}
Unsupervised learning algorithms start by assigning initial cluster (class) centres, that is, galaxies representative of a given class. 
To select the optimal centre points, they are iteratively changed by assigning either (1) random values, or (2) pre-defined values obtained from another simpler and faster clustering algorithm. 
This is an essential step as classification algorithms yield
  different classes with each random initialisation, while we want to
  obtain final classification results that are as stable as possible. 
The randomised initialisation is fraught with the risk of finding a local probability minimum, which results in the erroneous assignment of objects to groups.
In order to avoid such a situation, a random procedure for assigning initial values of function parameters can be repeated several times, and then the model with the highest log-likelihood is selected. 
However, to achieve optimal cluster centres, the number of random values needs to be equal to the number of galaxies. 

The second approach described above is the one applied in our
analysis; in particular, for the choice of the initial values, the
k-means++ algorithm is used~\citep{Arthur2007} to obtain the optimal cluster centres. 
This algorithm starts from a random choice of cluster centres among the data points. 
It then estimates the distances of all data points from these centres, and based on a weighted probability proportional to these squared distances, it selects new centres. 
This procedure is repeated until the choice of centres does not change with the next realisation, i.e. the optimal centres are found. 
Each initialisation gives a different classification, and each run
groups similar galaxies into clusters, and so, in principle, all of them provide valuable classifications. 
The problem is then to select which classification is the best, i.e. which one should be chosen as the final classification. 
To overcome this issue, we run the k-means algorithm 15 times to find the optimal initial parameters. 
Moreover, this ensures that we obtain a representative classification, as we are able to recreate the divisions. 
As in~\cite{SDSS_class_2010}, the k-means algorithm could be used for classification purposes itself. 
However, it is not as sophisticated as FEM, as it demands a pre-defined number of clusters (classes) and it also suffers from the initialisation problem. 
Therefore, we used k-means as the first step to optimise the starting points for a more advanced tool. 

Once the starting points of the algorithm have been selected, the FEM algorithm is executed assuming that: (1)  the input parameters, magnitudes and redshift values, can be projected onto a latent discriminative subspace with a dimension lower than the dimension (K) of the observed data, and (2) this subspace (K-1) is sufficient to discriminate K classes. 
The algorithm then performs the E (expectation), F (Fisher criterion), and M (maximisation) steps described below that are repeated in each cycle.

In step E, the algorithm calculates the complete log-likelihood,
conditionally to the current value of the Gaussian mixture model. In
practice, this means the calculation of the probability of each
considered object belonging to the groups predefined by the k-means++ algorithm.

In step F, the DLM model chooses the subspace $f$ in which the
distances between groups are maximised and their internal scatter is minimised:

\begin{equation}
f=\frac{(\eta_1 - \eta_2)^2}{\sigma_1 ^2 + \sigma_2 ^2},
\end{equation}
where $\eta_1$ and $\eta_2$ are the mean values of the centres of the  analysed groups, and $\sigma_1^2$ and $\sigma_2^2$ are their variances~\citep{Fukunaga:1990}. 
The mean and variance are measured for each group in the observation space. 
The algorithm searches for a linear transformation \textit{U}, which projects the observation into a discriminative and low-dimensional subspace \textit{d,} such that the linear transformation \textit{U} of dimension \textit{$p \times d$} (where p is the dimension of the original space) aims to maximise a criterion that is large when the between-class covariance matrix (\textit{SB}) is large and when the within-covariance matrix (\textit{SW}) is small. 
Since the rank of \textit{SB} is at most equal to \textit{K -- 1}, where \textit{K} is the number of classes, the dimension \textit{d} of the discriminative subspace is therefore at most equal to \textit{K -- 1} as well. 
For details, we refer to Sects. 2.4 and 3.1 in~\cite{Bou2012}. 
 
Subsequently, in step M, the parameters of the multivariate Gaussian functions are optimised, by maximising the conditional expectations of the complete log-likelihood, based on the values obtained in the previous steps (E+F).

The algorithm then comes back to step E, now computing the probabilities for each object to belong to groups modified in the last step M.

This procedure is repeated until the algorithm converges
according to the stopping criterion which is based on the difference between the likelihoods calculated in the last two steps.

\subsection{DLM models for the FEM algorithm}\label{sec:dlm_models}

To perform the FEM analysis, it is necessary to choose a model and the number of groups.
There exist different DLM models that have been created for different applications. 
Specific models differ in the numbers of components and their parameters. 
The variety of these models then allows them to fit into various situations. 
The 12 different DLM models are considered: DkBk, DkB, DBk, DB, AkjBk, AkjB, AkBk, AkBk, AjBk, AjB, ABk and AB. The main differences between them is in the number of free parameters left to be estimated~\citep{Bou2011}.
In the primary model, DkBk, two components can be distinguished: Dk and
Bk, where Dk is responsible for modelling the variance of the actual
data (by parametrizing  the variance of each class within the latent
subspace), and Bk which models the variance of the noise (i.e. it parametrizes the variance of the class outside the latent subspace). 
The other models are in fact submodels of DkBk in which certain
parameters of the Dk and Bk components are assumed to be common between
and/or within classes. 
For example, the DBk model assumes that the variance in a latent subspace is common to all classes, whereas the DkB model assumes that the variance outside the latent subspace is common across classes. 
The combination of these two constraints (common variance inside and
outside the latent subspace to all classes) results in the DB model. 
Therefore, these submodels are characterised by a lower number of
parameters: if our thirteen-dimensional dataset is divided into 12 groups,
the "main" DkBk model would be
characterised by 1,024 free parameters, while the DkB model would be characterised by 1,013 parameters, the DBk model by 298 parameters, the DB model by  287 parameters, down to the simplest AB model with 222 free parameters. 
The number of free parameters needed is dictated by the complexity of
the input data and the mathematical equations given in~\cite{Bou2012}. 
A highly parametrised model requiring the estimation of a large number of free parameters is preferred for clustering of high-dimensional data. 
We refer the reader to~\cite{Bou2012} for a detailed description of the DLM family. 
Comparing the performance and convergence of different models, we
find that the VIPERS data are best parametrised by the DBk model with 298 free parameters.

\subsection{The selection of the optimal model and number of classes}

\begin{figure}[]
        \includegraphics[width=0.49\textwidth]{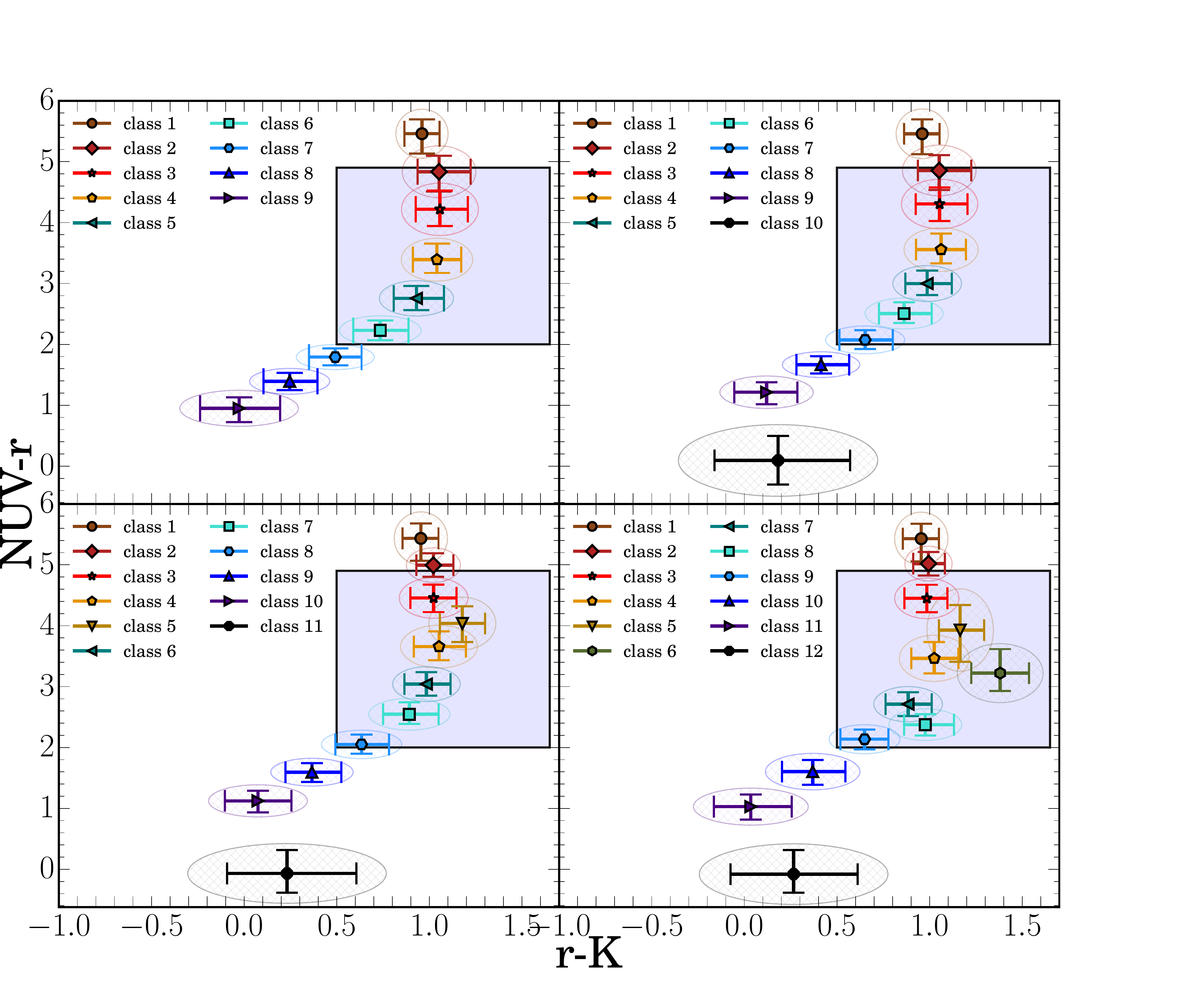}
        \caption{$NUVrK$ diagrams of FEM classes 9 -- 12. The optimal number of classes was found to be 12.  The error bars correspond to the first and the third quartile of the galaxy colour distribution, while the two half axes of the ellipses correspond to the median absolute deviation.}
        \label{fig:nuvr_9_10_11_12}
\end{figure}

The number of classes is not known a priori, which is one of the major difficulties in applying unsupervised clustering algorithms to classify astronomical sources. 
Defining the optimal number and model is not trivial. 
We do not make any a priori assumptions about galaxy separation, that is, if the data could not be described by the DLM models, for example because of the non-Gaussian nature of the datasets, the FEM algorithm simply would not converge. 
In our work, the
best DLM model and the range of possible class numbers is chosen based
on three statistical model-based criteria:  the Akaike Information
Criterion~\cite[AIC;][]{Akaike1974}, the Bayesian Information
Criterion~\citep[BIC;][]{Schwarz1978} and the Integrated Complete Likelihood~\citep[ICL;][see Appendix~\ref{app:number}]{Baudry2012}. 
These are typical criteria used to evaluate statistical models \citep[e.g.][]{souza}, which allow us to select the best model (DBk) and the approximate number of classes (9--12; see Appendix~\ref{app:number}). 
However, in order to pinpoint the diversity of physical properties among VIPERS galaxies, the final optimal number of classes is based on the flow of the galaxy distribution among a different number of classes (see Fig.~\ref{fig:flow_2_14}) and their physical properties (see Fig.~\ref{fig:nuvr_9_10_11_12}).

The analysis of the positions and properties of different classes on
the $NUVrK$ diagrams allows us to verify if the classes do indeed reveal distinct physical properties. 
Figure~\ref{fig:nuvr_9_10_11_12} shows the $NUVrK$ diagrams for the classifiers
consisting of~9,~10,~11 and 12 groups. As we can see in the figure,
the division into groups for a different number of classes differs,
especially in the region of dusty galaxies indicated by the shaded
box. We can see the emergence of three new classes (classes 5, 6, 8)
in the twelve-group division that were not distinguished by a lower number of clusters. 
The physical analysis of these classes (see Sect.~\ref{sec:results})
demonstrates that the classifier's grasp of subtle differences between
groups reveals these classes of dusty star-forming galaxies. 
Therefore, we find that division into 12 classes is physically motivated and this is also confirmed by analysis of the flow chart as all 12 classes are naturally separated from bigger groups, including separating broad-line AGNs from class 9 in the tenth iteration (see Fig.~\ref{fig:flow_2_14}). 
We also found that with 13 classes, we obtain a worse classification, as the 13th class emerges from class 11 but does not represent different physical properties with respect to the 11th class (see Appendix~\ref{app:number}). 

To summarise, using three statistical criteria: AIC, BIC, and ICL, we originally restricted the optimal number of classes to be between 9 and 12. 
After that, we checked the flow of galaxy distributions for realisations with different numbers of classes and their physical properties. 
We concluded that the optimal solution for classification of the VIPERS dataset is a DBk model with 12 classes (see Appendix~\ref{app:number}).

\section{Results}\label{sec:results}

In this section, the FEM classification of $z\sim0.4-1.3$ galaxies is presented. 
We demonstrate that the 12  classes correspond to physically different and separate galaxy categories. 
In the following analysis, different properties of our classes are investigated to show that our classes actually mirror the sequence of  galaxy types from the earliest (class 1) to the latest types (class 11) in the redshift range $0.4 < z <1.3$. 
Classes 1--11 all have very similar redshift distributions (see Table~\ref{table:properties}), centred at $z\sim 0.7$, suggesting that these classes are persistent at least over the redshift range $0.4 < z < 1.3$.        
A different median redshift is  measured within the 12th class. 
This class cannot be placed along the same sequence as the other classes. 
Class~12 mainly groups high-redshift VIPERS sources (with median redshift $z_{med}\sim2$; see Table~\ref{table:properties}). 
Members of this group are mostly identified as broad-line AGNs according to their redshift flag (see Sect.~\ref{sec:data}; $\sim95\%$, and Table~\ref{table:properties}). 
Therefore, class~12 is not part of the galaxy population at $z\sim 0.7$
that is the focus of this paper, and from now onwards only the first 11 galaxy classes will be discussed. 
The global properties of class 12 are presented in Table~\ref{table:properties} and the composite spectrum is shown in Fig.~\ref{fig:atlas2}, but it is not included in the remaining plots. 
The SED fitting procedure used for VIPERS sources does not include AGN
templates. Therefore, the AGN host properties (stellar mass and SFRs, $r$-$K$ colour, as $K$ significantly depends on models) might be wrong. 
The classification was performed on the whole sample (i.e. including broad-line AGNs, even if they are not the focus of this paper) to demonstrate the global usefulness of the FEM algorithm and its ability to separate broad-line AGNs and galaxies.
Although the algorithm was able to a separate a class of broad-line
AGNs, only $\sim50\%$ of broad-line AGNs at $z>1.3$ were assigned to
this separate class, while the other half were spread among the
star-forming classes 9--11.  The fraction of broad-line AGNs in these
classes is however negligible ($<5\%$ galaxies in a given class).    
This approach allows us to reproduce common classification schemes, which do not explicitly exclude any groups of sources.  
It should be noted that although class~12 can be expected to be
separated based on the use of spectroscopic redshift as an input parameter, even when the redshift is not included in the parameter space (i.e. classification is based only on rest-frame colours) class~12 is reproduced with an accuracy of the order of $\sim80\%$.

        \begin{sidewaystable*}

                \caption{The main physical properties of the FEM classes.}\label{table:properties}
                \centering
                \begin{tabular}{ccrcccccclclcc} 
                        \hline\hline             
                        Class & $N$ & $frac [\%]$ &  $z$ & $n$ & $NUV-r$ & $r-K$ & $U-V$ & $D4000_{n}$  & $EW(OII)$ & $log(M_{star}/M_{\odot})$ &  $log(sSFR) [yr^{-1}]$ & $N_{AGNs}$ & $frac_{AGNs}[\%]$\\
                        (1) & (2) & (3) & (4) & (5) & (6) & (7) & (8) & (9) & (10) & (11) & (12) & (13) & (14) \\
                        \hline
                        \multicolumn{14}{c}{\it Elliptical galaxies}\\ 
                        \hline
                        1 & 4476 & 9.11 & $0.67^{+0.11}_{-0.11}$ & $3.33^{+0.96}_{-1.20}$ & $5.43^{+0.37}_{-0.24}$ & $0.95^{+0.10}_{-0.10}$ & $1.99^{+0.09}_{-0.08}$ & $1.76^{+0.11}_{-0.11}$ & -- & $10.77^{+0.24}_{-0.23}$  & $-16.88^{+2.16}_{-3.55}$ & 0 & 0.00\\
                        
                        2 & 2399  & 4.88 & $0.67^{+0.10}_{-0.11}$ & $3.32^{+1.03}_{-1.30}$ & $5.04^{+0.19}_{-0.18}$ & $0.99^{+0.08}_{-0.09}$ & $1.98^{+0.09}_{-0.08}$ & $1.75^{+0.12}_{-0.10}$ & -- & $10.83^{+0.21}_{-0.22}$ & $-11.85^{+0.14}_{-0.29}$ & 0 & 0.00\\
                        
                        3 & 3558 & 7.24 &  $0.68^{+0.10}_{-0.10}$ & $3.04^{+0.99}_{-1.42}$ & $4.46^{+0.22}_{-0.21}$ & $0.98^{+0.12}_{-0.11}$  & $1.91^{+0.10}_{-0.09}$ & $1.68^{+0.14}_{-0.12}$ & -- & $10.83^{+0.23}_{-0.23}$ & $-11.28^{+0.16}_{-0.17}$ & 0 & 0.00\\    
                        
                        \hline
                        \multicolumn{14}{c}{\it Intermediate galaxies}\\ 
                        \hline                          
                        4 & 4274 & 8.70  & $0.70^{+0.11}_{-0.11}$ & $1.78^{+0.68}_{-1.11}$ & $3.47^{+0.24}_{-0.25}$ & $1.03^{+0.12}_{-0.13}$  & $1.64^{+0.14}_{-0.12}$ & $1.41^{+0.11}_{-0.14}$ & -- & $10.69^{+0.25}_{-0.23}$ & $-9.79^{+0.60}_{-0.47}$ & 4 & 0.09\\
                        
                        5 & 3375 & 6.87 & $0.63^{+0.08}_{-0.08}$ & $2.07^{+0.77}_{-1.17}$ & $3.93^{+0.51}_{-0.41}$ & $1.17^{+0.12}_{-0.13}$  & $1.82^{+0.21}_{-0.15}$ & $1.50^{+0.15}_{-0.16}$ & -- & $10.50^{+0.27}_{-0.23}$ & $-9.57^{+0.17}_{-0.33}$ & 0 & 0.00\\
                        
                        6 & 964 & 1.96 & $0.67^{+0.09}_{-0.09}$ & $1.35^{+0.51}_{-0.81}$ & $3.29^{+0.29}_{-0.38}$ & $1.41^{+0.16}_{-0.14}$  & $1.58^{+0.14}_{-0.20}$ & $1.36^{+0.10}_{-0.11}$ & $-16^{+7}_{-5}$ & $10.50^{+0.21}_{-0.22}$ & $-9.21^{+0.35}_{-0.42}$ & 2 & 0.21\\
                        
                        \hline
                        \multicolumn{14}{c}{\it Star-forming galaxies}\\ 
                        \hline   
                        
                        7 & 5099 &  10.38 & $0.67^{+0.11}_{-0.12}$ & $1.15^{+0.40}_{-0.75}$ & $2.71^{+0.19}_{-0.19}$ & $0.88^{+0.12}_{-0.12}$  & $1.35^{+0.14}_{-0.12}$ & $1.28^{+0.07}_{-0.09}$ & $-16^{+8}_{-5}$ & $10.36^{+0.30}_{-0.31}$ & $-9.29^{+0.45}_{-0.47}$ & 17 & 0.33\\

                        8 & 1755 &  3.57 & $0.72^{+0.09}_{-0.10}$ & $0.91^{+0.33}_{-0.70}$ & $2.38^{+0.16}_{-0.15}$ & $1.00^{+0.12}_{-0.14}$  & $1.15^{+0.09}_{-0.08}$ & $1.21^{+0.05}_{-0.06}$ & $-21^{+9}_{-7}$ & $10.12^{+0.23}_{-0.19}$ & $-8.76^{+0.26}_{-0.30}$ & 14 & 0.80\\
                        
                        9 & 5378 &  10.95 & $0.67^{+0.11}_{-0.13}$ & $0.92^{+0.30}_{-0.64}$ & $2.13^{+0.16}_{-0.15}$ & $0.63^{+0.12}_{-0.13}$  & $1.07^{+0.13}_{-0.12}$ & $1.21^{+0.06}_{-0.07}$ & $-24^{+8}_{-7}$ & $9.91^{+0.25}_{-0.28}$ & $-8.95^{+0.43}_{-0.39}$ & 31 & 0.58\\
                        
                        10 & 13978 & 28.45 & $0.66^{+0.10}_{-0.12}$  & $0.94^{+0.31}_{-0.63}$ & $1.60^{+0.21}_{-0.19}$& $0.36^{+0.16}_{-0.18}$  & $0.86^{+0.12}_{-0.11}$ & $1.16^{+0.06}_{-0.06}$ & $-36^{+10}_{-8}$ & $9.56^{+0.22}_{-0.23}$ & $-8.84^{+0.21}_{-0.28}$ & 123 & 0.88\\
                        
                        11 & 2699 & 5.49 & $0.71^{+0.12}_{-0.18}$ & $1.11^{+0.45}_{-0.83}$ & $1.01^{+0.21}_{-0.20}$ & $0.03^{+0.20}_{-0.22}$  & $0.59^{+0.14}_{-0.12}$ & $1.07^{+0.07}_{-0.07}$ & $-54^{+14}_{-12}$ & $9.28^{+0.21}_{-0.25}$& $-8.61^{+0.33}_{-0.32}$ & 216 & 8.00\\
                        \hline
                        \multicolumn{14}{c}{\it Broad-line AGNs}\\ 
                        \hline          
                        12 & 174 & 0.35 & $2.24^{+0.25}_{-0.56}$ & $2.80^{+1.17}_{-0.89}$ & $-0.08^{+0.31}_{-0.39}$ & *  & $0.49^{+0.49}_{-0.31}$ & $1.00^{+0.14}_{-0.16}$ & -- & * & * & 166 & 95.40\\    
                        
                        \hline
                \end{tabular}
                \tablefoot{The number of members ($N$) and fraction of whole sample ($frac[\%]$) in each class corresponds to the number and fraction in the final sample, i.e. 48,129 galaxies with high 1st-best ($>50\%$) and low 2nd-best ($<45\%$) class membership probabilities. 
                        
                        For each class, the median values of: redshift (4), S\'{e}rsic index from~\cite{krywult} (5), rest-frame colours (6-8), spectral features (9-10), and physical properties derived from SED fitting~\cite[(11-12);][]{moutard16b} are provided.
                        Errors correspond to the differences between median and 1st, and 3rd quartile, respectively. 
                        The number and fraction of broad-line AGNs (as classified by VIPERS team members) in each class are given in columns 13, and 14, respectively. 
                        
                        $EW([OII]\lambda3727)$ was not detected for the majority of galaxies (96, 91, 85, 59, 72, 98$\%$)  within classes 1--5, and 12, respectively. 
                        
                        *Stellar mass, $SFR$, $sSFR$, $r-K$ colour derived from SED-fitting are expected to be wrong, as they are estimated through the fitting of galaxy models ($BC03$), not suited for broad-line AGNs.
                        
                        Colours are given in AB system.
                        
                }
        \end{sidewaystable*}

As mentioned in Sect.~\ref{sec:intr}, standard selection methods are
powerful tools, but are however sensitive only to a few specific properties. 
We explored how such a refined classification compares with more standard two- or three-class division of galaxy population. 
The FEM classification separates VIPERS galaxies into eleven classes, which may be assigned to three wider galaxy categories: (1) red, passive, (2) green, intermediate, and (3) blue, star-forming. 
Since our classification was based on colours, the conventional nomenclature of red (classes 1--3), green (classes 4--6), and blue (classes 7--11) galaxies is applied (see Fig.~\ref{fig:uvj},~\ref{fig:nuvrrk_ID}, and~\ref{fig:nuvrrk_TM}). 
As the subsequent analysis demonstrates  (Sect.~\ref{sec:separation},~\ref{sec:properties}), the division between red (passive), green (intermediate), and blue (star-forming) galaxies is not sharp, as the intermediate groups (classes~3 and~7) are not purely passive or star-forming in terms of their global properties. 
Moreover, we note that a FEM classification into two or three main groups is not entirely unequivocal. 

We compared our final eleven-class classification with a two-class FEM separation. 
The simple separation into two main clouds (red and blue) is able to distinguish a separate group of blue star-forming galaxies: $97\%$ of galaxies from classes 7--11 are assigned to the blue cloud and only $3\%$ of green galaxies (classes 4--6) were found in the blue cloud. 
At the same time, red and green galaxies are indistinguishable in the red cloud: $100\%$ of red galaxies (classes 1--3) were assigned to the red cloud, as were $97\%$ of green galaxies (classes 4--6).

In the subsequent step, the standard three-class (red/green/blue) division is compared  with the FEM 11-class classification. 
As in the case of the two-class division, we are also not able  to separate a red passive population from green galaxies. 
Almost all red galaxies assigned to classes 1--3 ($99\%$) were found in a red group. 
However, this group is strongly contaminated by green galaxies: $43\%$ of intermediate galaxies (classes 4--6) were found in the red cloud. 
The distinction between green  and blue galaxies is also not obvious. 
Only $67\%$ of blue star-forming galaxies (classes 7--11) were assigned to the blue cloud, while the remaining $33\%$ were found within the green population. 

For the two-class separation, red and green galaxies go together to
form one group only, while for the three-class division, the green population is split between red and blue galaxies. 
This implies that the borderlines between green/blue and red/green
populations are much less sharp than that for the eleven-class division. 
Only a more detailed classification can appropriately yield the division between red, green, and blue populations. 

The FEM classification yielded distinct clusters in the thirteen-dimensional space, although the separation between classes is smooth. 
Some galaxies are close to the borders of different classes, and this is reflected in their lower posterior probabilities of being members of the class to which they are assigned. 
The posterior probability is correlated with the distance of the sources from the centre of the group in multidimensional space. 
There is no correlation of probabilities with the properties of the input data, that is, no dependence of the probability on the redshift measurement accuracy or luminosity was found. 
We assume that the classification, which assigns a probability of being a member of the class instead of a single class membership,  should be a  better approximation of the galaxy evolution, as a continuous transition between different groups (even if they are well separated in the feature space) is expected. 
Therefore, each group contains its core representative population and a (usually small) number of galaxies that are more loosely mapped to them. 
A detailed description of the class membership probabilities is given in Appendix~\ref{app:A}.
In the following analysis, we focus on the representative galaxies, for which the class membership is not questionable.  
Our initial sample of 52,114 galaxies was therefore cleaned by excluding
objects located in between adjacent classes, and outliers, based on their probability.
In particular, 2,947 galaxies with low probabilities ($< 50\%$) of
being class members, and 1,038 objects with high ($> 45\%$) probabilities of belonging to a second group were removed.
However, it is worth noting that this leads to the rejection of only $8\%$ of the sample, therefore demonstrating the robustness of the clustering process performed with the FEM algorithm.

The resultant final catalogue consists of 47,556 galaxies (and 573 broad-line AGNs). 
The number of sources in each class, as well as the basic properties of the FEM classes, are summarised in Table~\ref{table:properties}.  

\subsection{Multidimensional galaxy separation versus standard methods}\label{sec:separation}

\begin{figure*}[ht]

                \begin{subfigure}[a]{0.49\textwidth}
                        \includegraphics[width=\textwidth]{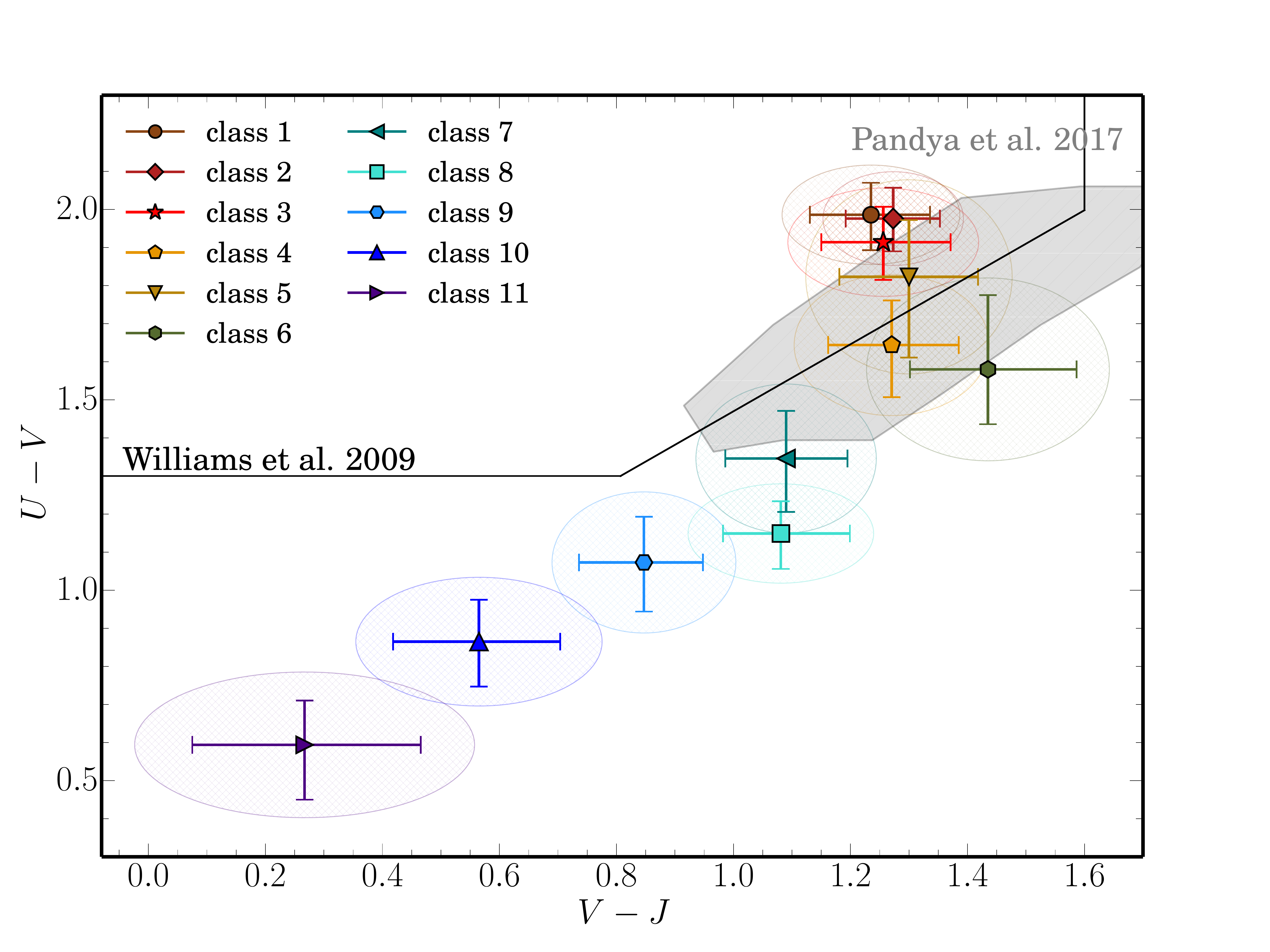}
                        \caption{$UVJ$ diagram;}
                        \label{fig:uvj}
                \end{subfigure}
                \begin{subfigure}[a]{0.49\textwidth}
                        \includegraphics[width=\textwidth]{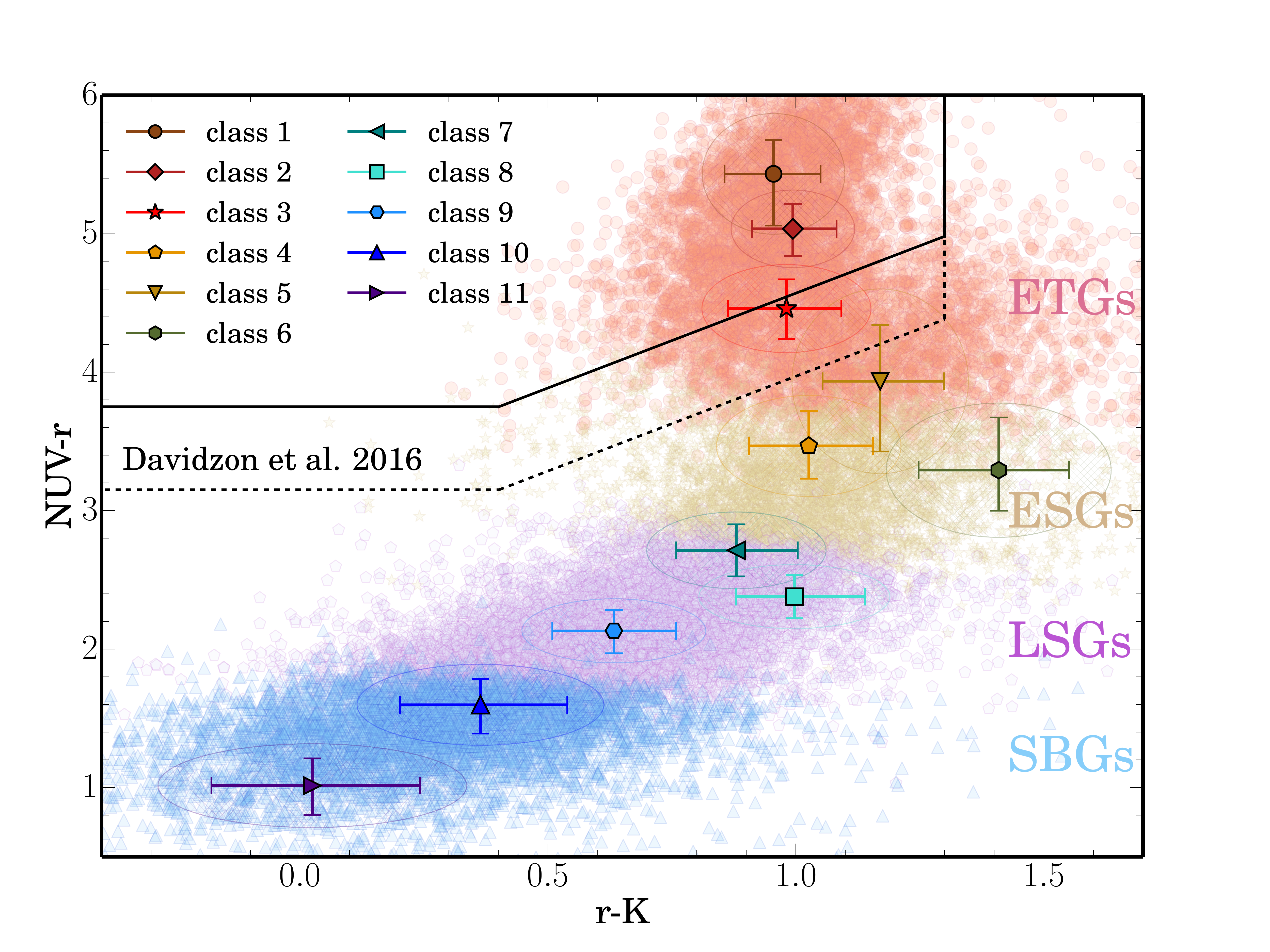}
                        \caption{$NUVrK$ diagram;}
                        \label{fig:nuvrrk_ID}
                \end{subfigure}
                \begin{subfigure}[a]{0.49\textwidth}
                        \includegraphics[width=\textwidth]{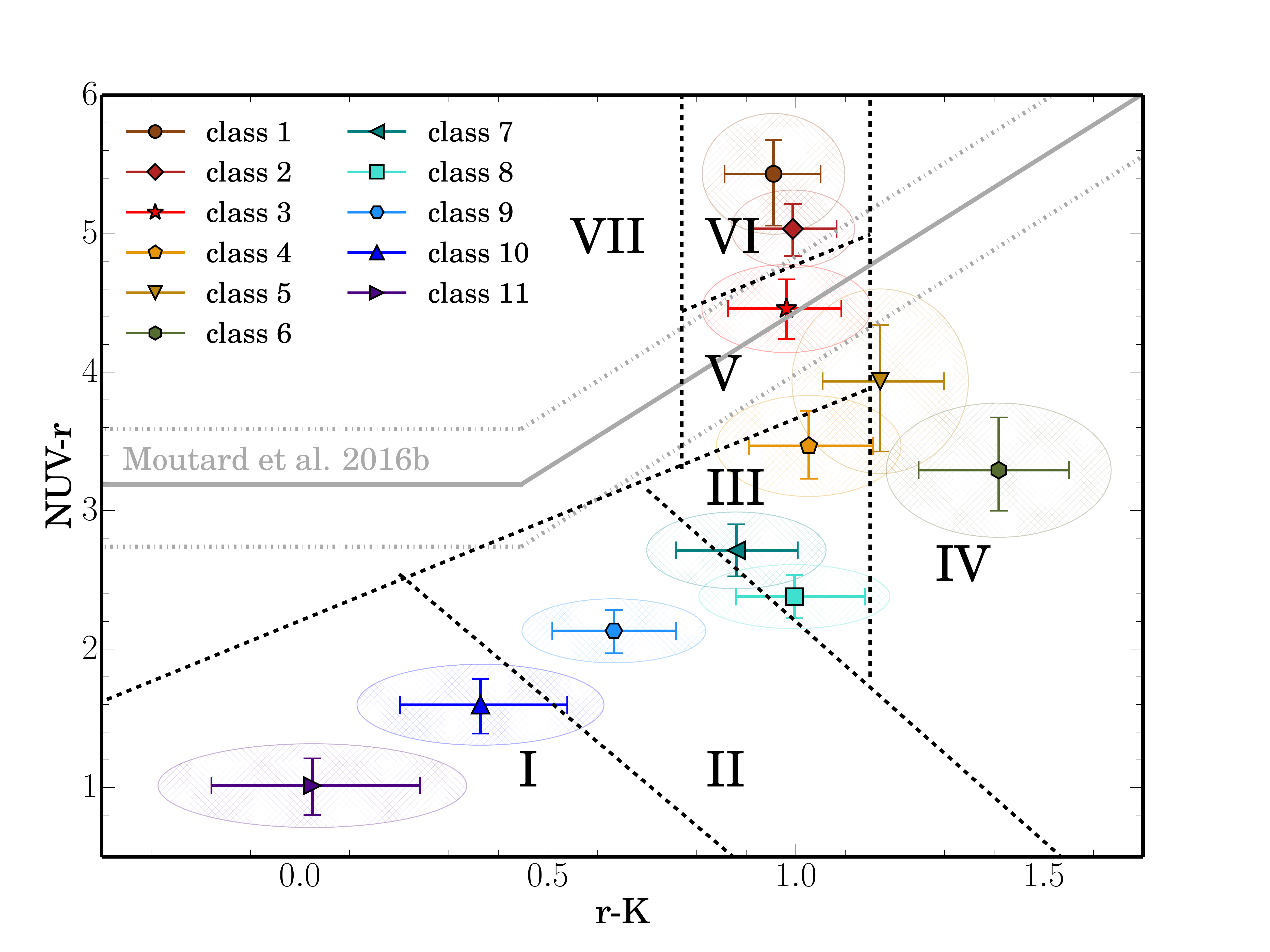}
                        \caption{$NUVrK$ diagram;}
                        \label{fig:nuvrrk_TM}
                \end{subfigure}  
                \begin{subfigure}[a]{0.49\textwidth}
                        \includegraphics[width=\textwidth]{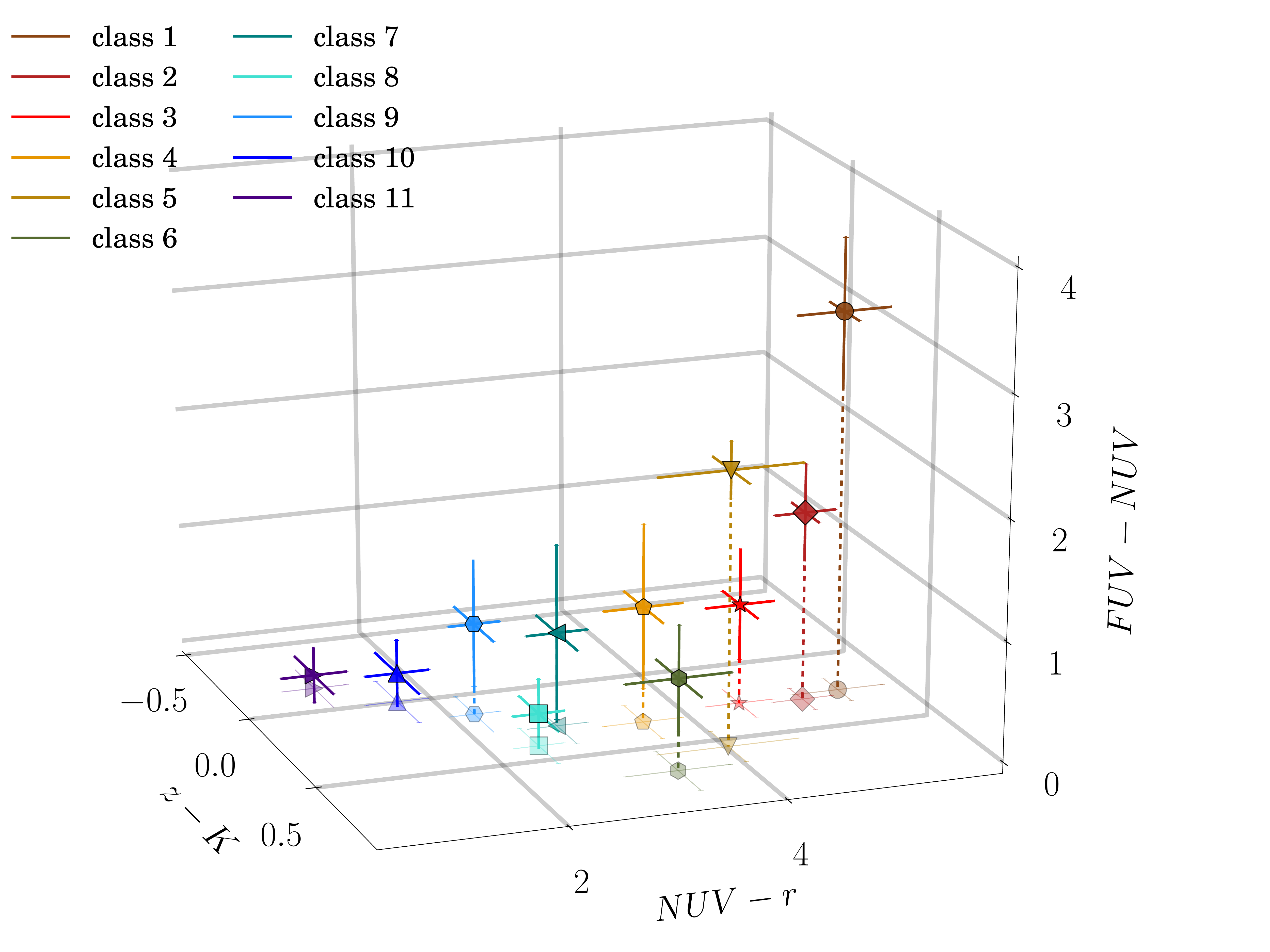}
                        \caption{3D colour diagram;}
                        \label{fig:3d}
                \end{subfigure}

        \caption{Colour-colour(-colour) diagrams of the VIPERS galaxies classified into 11 classes with the FEM algorithm. The error bars correspond to the first and the third quartile of the galaxy colour distribution, while the two half axes of the ellipses correspond to the median absolute deviation.  (a) The $UVJ$ diagram. The solid line corresponds to the standard separation between quiescent and star-forming galaxies. The area occupied by CANDELS transition galaxies is  shown as a grey shaded area~\citep{pandya}.  (b) The $NUVrK$ diagram. The black solid line corresponds to the separation of red passive galaxies, and the dotted line separates additionally galaxies located in green valley. Galaxies photometrically classified by their SED type as: (1) early-type (red E and Sa; ETGs), (2) early spiral (ESGs), (3) late spiral (LSGs), and (4) irregular or starburst (SBGs) following the prescription given in~\cite{fritz14} are marked with light salmon, gold, violet, and blue, respectively. (c) The $NUVrK$ diagram. The black dashed lines corresponds to the division of CFHTLS galaxies into seven groups proposed by~\cite{moutard16a}.  The grey solid line corresponds to the separation of red passive galaxies. The grey dash-dotted lines correspond to upper (lower) limits of the green valley galaxies proposed by~\cite{moutard16b}.  (d) The 3D diagram. The dotted lines indicate the projection of FEM classes on the bottom plane ($z-K$ vs. $NUV-r$).  }
        \label{fig:color}
\end{figure*}

The FEM classification allows for a more sophisticated galaxy separation than the standard two-dimensional (2D) colour-colour diagrams. 
The typical classification schemes are mostly based on tight and linear cuts in the 2D space, while an unsupervised approach associates each object to the group based on its location in the multidimensional space describing galaxy properties. 
Colour-colour diagrams, including $NUVrK$~\citep{arnouts} and  $U-V$ versus $V-J$~\citep[$UVJ$;][]{Williams}, are coarser classifications than the ones obtained with the multidimensional approach, even if the trends are continuous. 
At the same time, the FEM classes correspond well to the classification schemes based on all these colour-colour diagrams. 
FEM classes are able to reproduce the standard colour-colour separation into passive and star-forming galaxies. 
Unsupervised classification further introduces the division into subclasses, which monotonously change their physical, spectroscopic, and morphological properties from class to class. 
This reveals the differences within passive, intermediate, and star-forming galaxy populations. 
The FEM classification creates a multidimensional separation cut. The
advantage of this approach is that it is sensitive to a larger number
of galaxy properties with respect to standard classification
techniques. For example, as shown in a subsequent analysis, the three
red passive classes are indistinguishable in the $r-K$, $U-V$, and $V-J$
colours, but have different $FUV$ and $NUV$ properties.         
Figure~\ref{fig:color} presents colour-colour diagrams: $FUV-NUV$ versus $z-K$ versus $NUV-r$, $NUVrK$, and $UVJ$, where the median colours for the 11 FEM classes are shown. 
The error bars correspond to the first and third quartiles of the galaxy
colour distribution, while the semi-major and minor axes of the ellipses correspond to the normalised median absolute deviation (NMAD) defined by~\cite{mad}, as $NMAD = 1.4826\cdot median(|P-median(P)|)$, where P corresponds to {the measured colour reported on each axis.
Classes are labelled according to their $NUVrK$ colour, from the reddest, class~1, to the bluest, class~11.  
We note that green galaxies (classes 4--6) are labelled to follow their $r-K$ colour change rather than $NUV-r$, which is more sensitive to dust obscuration.

The FEM classes may overlap with each other on 2D diagrams, and the
clear separation may only be revealed when an additional parameter is added. 
This is especially relevant for red passive galaxies (classes 1--3) which are not distinguishable in the $UVJ$ diagrams (see Fig.~\ref{fig:uvj}), and are only partially separated in the  $NUVrK$ diagram (classes 1--2 overlaps, see Fig.~\ref{fig:nuvrrk_TM} and~\ref{fig:nuvrrk_ID}). 
Only when an additional parameter is added to the diagram (see
Fig.~\ref{fig:3d}) is the clear separation between three classes of red passive galaxies achieved, and the inhomogeneity of red galaxies becomes visible.

\subsubsection{The $UVJ$ diagram}\label{sec:uvj}

As proposed by~\cite{Williams} and confirmed by many others~\citep[e.g.][]{2011Whitaker, patel12,2015Dokkum}, passive and star-forming galaxies occupy two distinct regions on the $UVJ$ diagram. 
Figure~\ref{fig:uvj} shows the distribution of the 11 FEM classes on the $UVJ$ diagram, with the standard division between passive and star-forming galaxies marked with a black solid line. 
Passive galaxies (classes 1--3) are redder in $U-V$ and bluer in $V-J$ relative to galaxies that are young and dusty, which are red in both $U-V$ and $V-J$ colours (class 6). 
Galaxies classified as green intermediate (classes 4--6) are not as red in $U-V$, which may indicate that they still have some active star formation. 
Galaxies within classes~4 and~5 reproduce remarkably well the CANDELS sample of 1,745 massive ($>10^{10}$ $M_{\odot}$) transition galaxies observed at $0.5 < z < 1.0$ on the $UVJ$ diagram~\citep[see Fig.~A1 in][]{pandya}. 
Star-forming and transition CANDELS galaxies are not well separated on the $UVJ$ plane; the region occupied by the FEM classes~6 and~7 is already strongly occupied by the star-forming sample; therefore, we do not connect them with CANDELS transition population. 
Moreover, class~4 is placed in the region of dust-free CANDELS
transition galaxies, whereas class~5 corresponds to the more dusty galaxies~\citep[see the distribution of the optical attenuation in Fig.~A1 in][]{pandya}. 
These galaxies tend to occupy a transition region populated by  galaxies with a variety of  morphologies~\citep{moutard16a}. 
Therefore, we conclude that classes~4 and~5 consist of green
intermediate galaxies, representing a mixed population  in the
transition phase between passive and star-forming  categories.

Intermediate galaxies  are located in the green valley, a wide region in the ultraviolet-optical colour magnitude diagram between the blue and red peaks, 
and usually they are hard to distinguish, as the classical selection criteria are not well defined~\citep[e.g.][and references therein]{salim}. 
However, \cite{schawinski} have already shown the existence of two different populations of green galaxies with respect to their gas content, separating intermediate galaxies into green spirals and green elliptical populations. 
The three intermediate FEM classes (4--6) confirm that the green valley population is not a homogeneous category of galaxies. 
Star-forming galaxies (classes 7--11) are well separated on the $UVJ$
diagram, showing bluer $U-V$ and $V-J$ colours with increasing class number.  
The median $U-V$ and $V-J$ colours for 11 FEM classes are given in Fig.~\ref{fig:colorproperties} and Table~\ref{table:properties}. 

\subsubsection{The $NUVrK$ diagram}\label{sec:nuvrrk}

Figures~\ref{fig:nuvrrk_ID} and~\ref{fig:nuvrrk_TM} present the
distribution of the 11 FEM classes in the $NUVrK$ diagram~\citep{arnouts}. 
The $NUVrK$ diagram is similar to the $UVJ$ plane (see Fig.~\ref{fig:uvj} and Sect.~\ref{sec:uvj}), but allows for a better separation between passive and active galaxies. 
The $NUVrK$ diagram is also a better indicator of dust obscuration and current versus past star formation activity.
Old, quiescent galaxies exhibit redder $NUV-r$ colours, while galaxies with a younger stellar content are bluer. 
However, the $NUV-r$ colour is highly sensitive to dust attenuation,
meaning that dusty star-forming galaxies may also show reddened $NUV-r$ colours~\citep{arnouts07,martin07}. 
The vector for increasing dust reddening acts perpendicularly to the vector of decreasing specific SFR (defined as the SFR per stellar mass unit, hereafter: $sSFR$), enabling the degeneracy to be broken. 
Therefore, the $NUVrK$ diagram is extensively used to separate different galaxy types~\citep[e.g.][]{arnouts,fritz14,moutard16b,davidzon16}.

\cite{davidzon16} proposed criteria for the selection of passive
and intermediate objects in the $NUVrK$ diagram (black solid and black
dashed lines in Fig.~\ref{fig:nuvrrk_ID}, respectively) based on
VIPERS PDR1 galaxy sample. \cite{moutard16b} defined a slightly different division between quiescent and star-forming galaxies (black solid line in Fig.~\ref{fig:nuvrrk_TM}), as absolute magnitudes were derived through SED-fitting with other assumptions. 
In particular, the slope of the line separating active and passive galaxies in the $NUVrK$ diagram found by~\cite{davidzon16} is flatter than the one presented in~\citealp{moutard16b} (slopes are $S=1.37$, $S=2.25$, respectively). 
Both criteria show a similar behaviour with respect to the FEM classes.   
Classes 1--2  perfectly match the area occupied by red passive galaxies, while class~3 is close to the separation line between red passive and the  green valley region as defined by~\cite{moutard16b}. 
As previously mentioned, class~3 is not purely passive and may represent the population of red galaxies that have just joined the passive evolutionary path.  

There is a clear path in the $NUVrK$ diagram along which the FEM classes are distributed. 
Figures~\ref{fig:nuvrrk_TM} and~\ref{fig:nuvrrk_ID} show that classes 1--3 are placed at the top of the diagram, while classes 7--11 occupy its bottom part with the intermediate area reserved for classes 4--6. 
The FEM classification also very closely follows the  photometric selection based on the SED fitting by~\citealp{fritz14} (see points in Fig.~\ref{fig:nuvrrk_ID}, colour-coded according to SED type). 
Almost all FEM red passive galaxies (classes 1--3; $\sim98\%$) are
defined as ETGs (red E/Sa) by the SED classification (ETGs are marked with salmon circles in Fig.~\ref{fig:nuvrrk_ID}), and most star-forming galaxies (classes 10--11) are classified as irregular or starburst types ($\sim97\%$; SBGs marked with blue triangles in Fig.~\ref{fig:nuvrrk_ID}).  
The intermediate (4--6) and star-forming (7--9) classes match
reasonably well ($\sim70\%$) with the early- and late-type spiral
galaxies classified based on their SEDs (ESGs and LSGs are marked with yellow stars and purple pentagons, respectively).

The FEM  classes (Fig.~\ref{fig:nuvrrk_TM}) also follow very well the classification of CFHTLS galaxies proposed by~\cite{moutard16a}. 
The region of dusty star-forming galaxies mainly corresponds to classes 5--6, whereas  classes 7--11 are found in the star-forming area~\citep{moutard16a}.
Galaxies become bluer (both in $NUV-r$ and $r-K$; except the $r-K$ colour for intermediate galaxies) with increasing class number, that is, classes 7--11 contain the bluest galaxies. 
When the stellar populations become older or the amount of dust in galaxies increases, the $r-K$ colour becomes redder.
The green galaxies, members of classes 4--6, are characterised by redder $r-K$ and $NUV-r$ colours relative to the star-forming cloud  (classes 7-11). 
Only edge-on galaxies may have the reddest $r-K$ colours~\citep{arnouts,moutard16a}.   
Therefore, as FEM class 6 shows the reddest $r-K$ colours, we conclude that its colours may be a consequence of dust within the disks or their high inclinations.  
The area of the $NUVrK$ diagram occupied by classes 4 and 5 is placed in the region  where~\cite{moutard16a} located  a morphologically inhomogeneous class of galaxies,  which in our classification may be divided into more homogeneous classes. 
\cite{moutard16b} found these galaxies to be most likely transiting from the star-forming to the passive population. 
Class 4 has similar $r-K$ colours to classes 1--3, showing that this class, as already mentioned, is close to passive galaxies.  
The top of the diagram is reserved for classes 1--3, which show the
reddest $NUV-r$ colours in the FEM classification.

Besides the clear differences between the three main classes (red/green/blue) on the $NUVrK$ diagram, the difference is visible also within subclasses.
The red subclasses show the progressive reddening in $NUV-r$ colour starting from class~3, and ending in class~1, as shown in Figs.~\ref{fig:nuvrrk_ID} and~\ref{fig:nuvrrk_TM}. 
The clear separation of three red passive classes is clearly visible in the $FUV$-$NUV$ colour (see Fig.~\ref{fig:3d}). 
At the same time, there is no significant change in their $r-K$ colour. 
Red passive galaxies  are populated by old stellar populations and have little dust, and therefore we do not expect to distinguish different red passive populations in $r-K$ colour, which is sensitive to dust obscuration.  
At the same time, these subclasses show only small differences in the
strengths of their $D4000_{n}$ (see Fig.~\ref{fig:lineproperties}, and Table~\ref{table:properties}), suggesting only small differences in their stellar ages. 
However, classes 1--3 show significant changes in $sSFR$ (see Fig.~\ref{fig:sedproperties}, and Tab.~\ref{table:properties}), which may indicate that star formation contributes more to class~3 than to the first and second classes. 

Figure~\ref{fig:nuvrrk_z} shows the $NUVrK$ diagram in six redshift bins spanning the redshift range $0.4 < z < 1.0$.
The colour evolution of the galaxy populations with redshift is clearly visible.  
\citet[][and references therein]{madau2014} have already shown that galaxy properties such as $SFR$ and colour change significantly within a galaxy population as a function of redshift.  
Figure~\ref{fig:nuvrrk_z} shows that properties of galaxy types indeed vary with cosmic time.

Red passive galaxies (classes 1--3) form three different, well-separated clusters in the $NUVrK$ diagram at $z\sim 0.4$.
When we move back with cosmic time, classes 1 and 2 tend to progressively merge up to $z\sim 1$. 
At $z\sim1,$ the separation between classes 1 and 2 is less evident. 
This could be a consequence of the colour-colour pre-selection sample bias, as at $z\sim1$ VIPERS observed only the most massive and the brightest galaxies, but may also imply that the population of red passive galaxies was more homogeneous at earlier epochs. 
Red passive galaxies achieve their final morphology at $z\sim1$, whereas at higher redshifts ($1< z < 2$) the peak of their evolution is expected~\citep[e.g.][]{Bundy10}. 
The homogeneity of classes~1 and~2 at $z\sim1,$ at least in $NUV-r$ and $r-K$ colours, may therefore indicate that these groups of red galaxies were inseparable at that epoch with respect to some of their physical properties, when they still attain their final form~\citep[e.g.][]{Cimatti2004,Glazebrook2004}. 
The detailed analysis of the physical processes leading to the separation of three different red passive galaxy classes will be presented in a forthcoming paper.   

\begin{figure*}[]
        \includegraphics[width=0.49\textwidth]{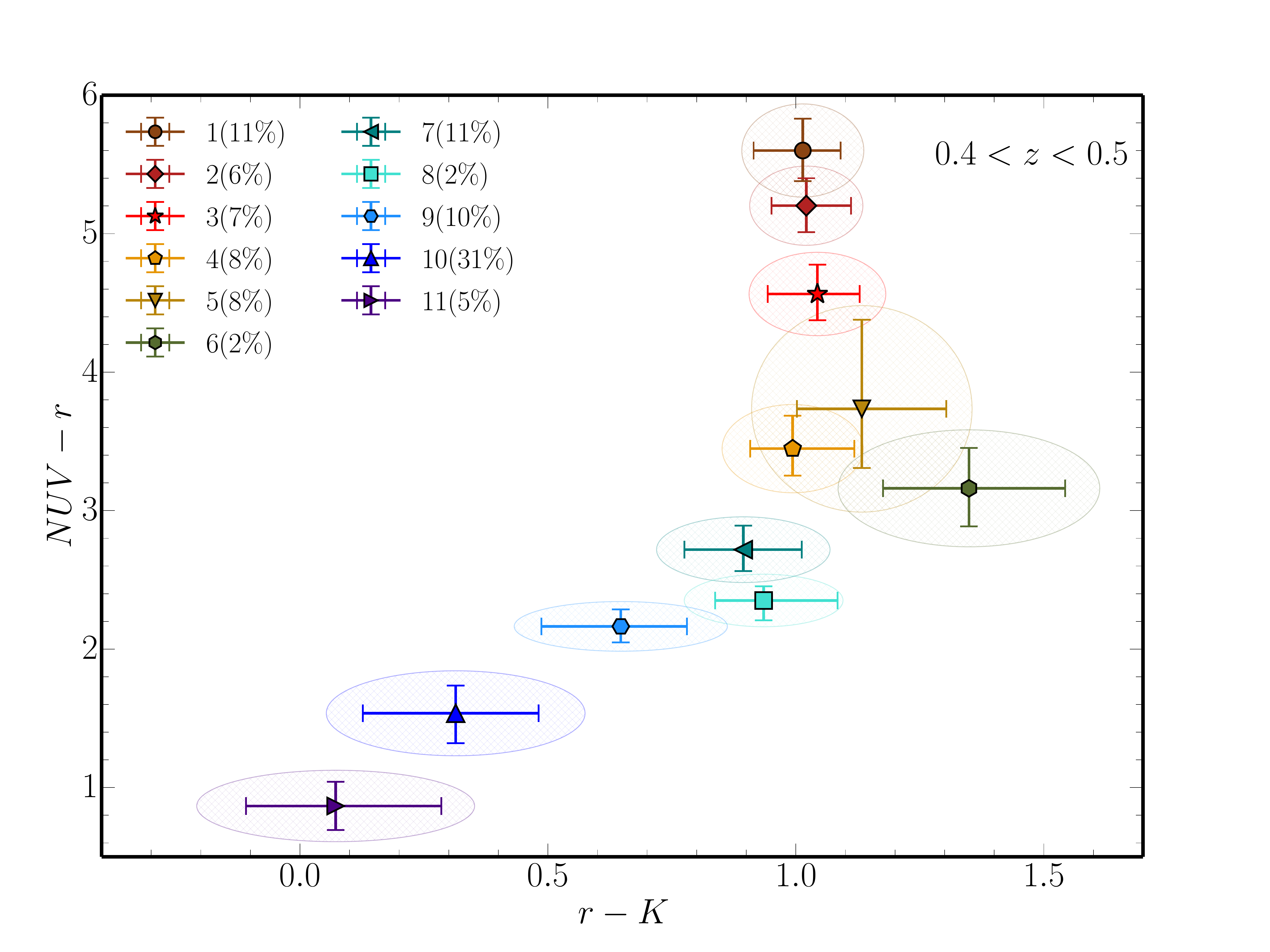} 
        \includegraphics[width=0.49\textwidth]{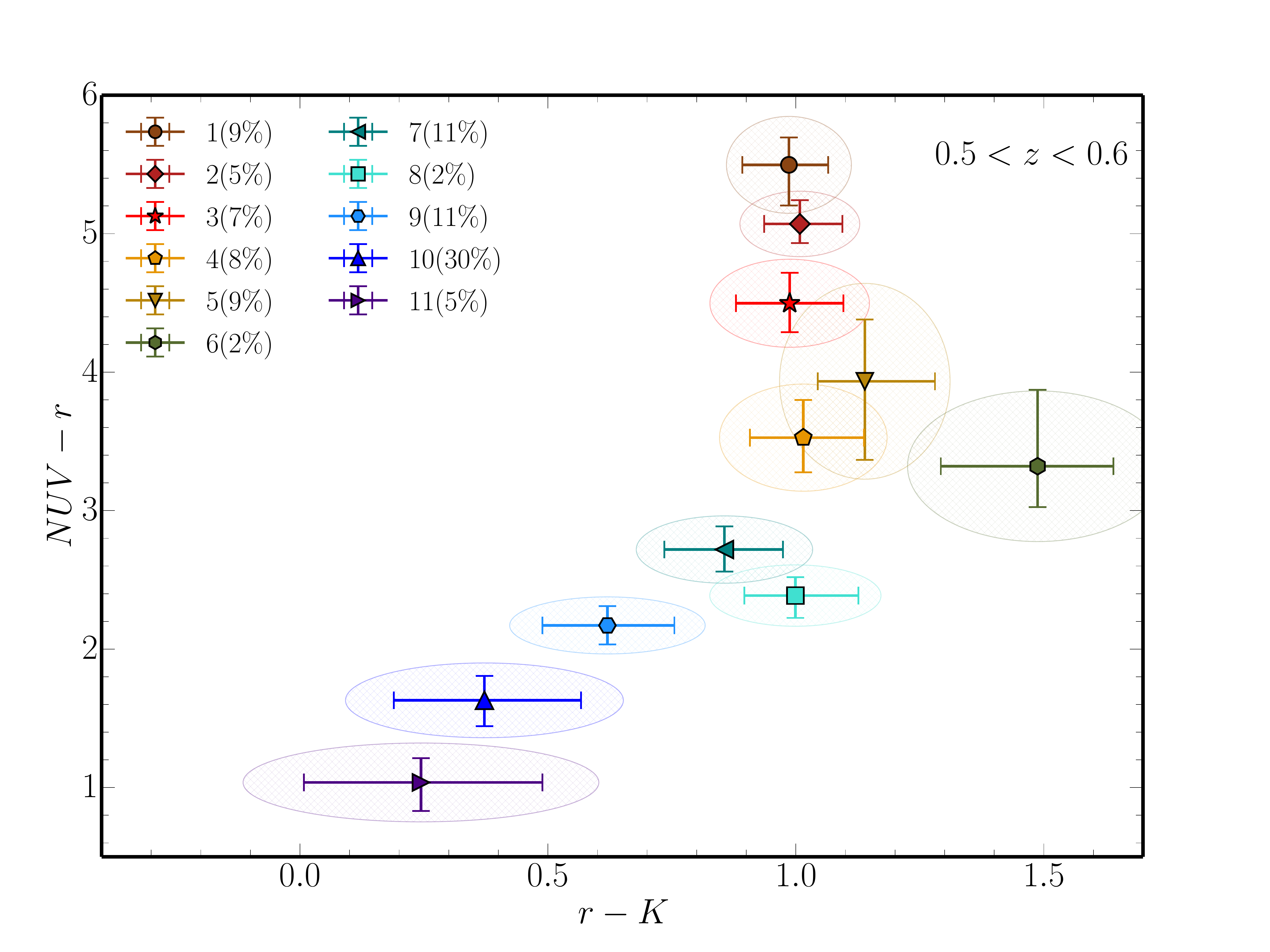}                            
        \includegraphics[width=0.49\textwidth]{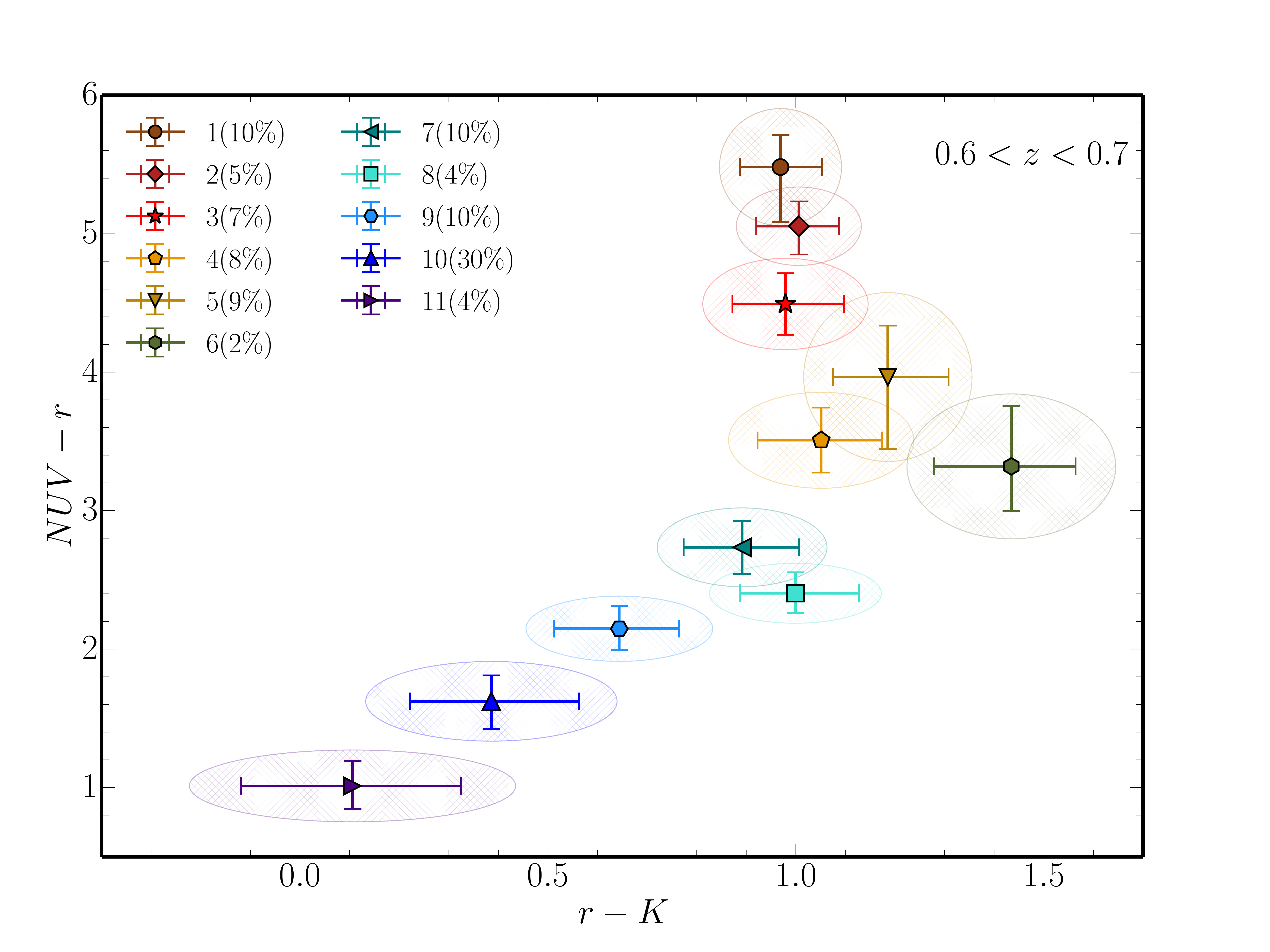} 
        \includegraphics[width=0.49\textwidth]{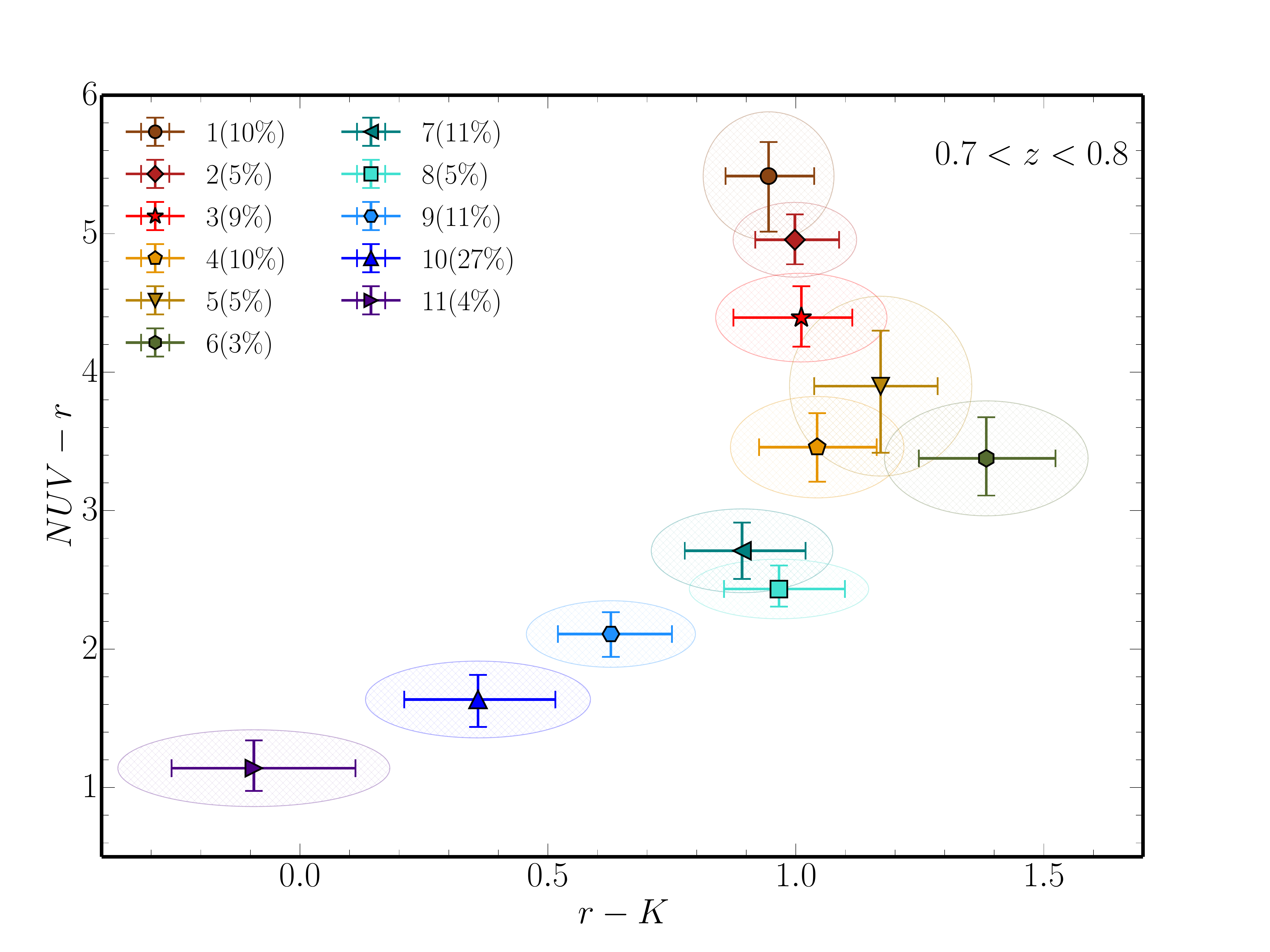}  
        \includegraphics[width=0.49\textwidth]{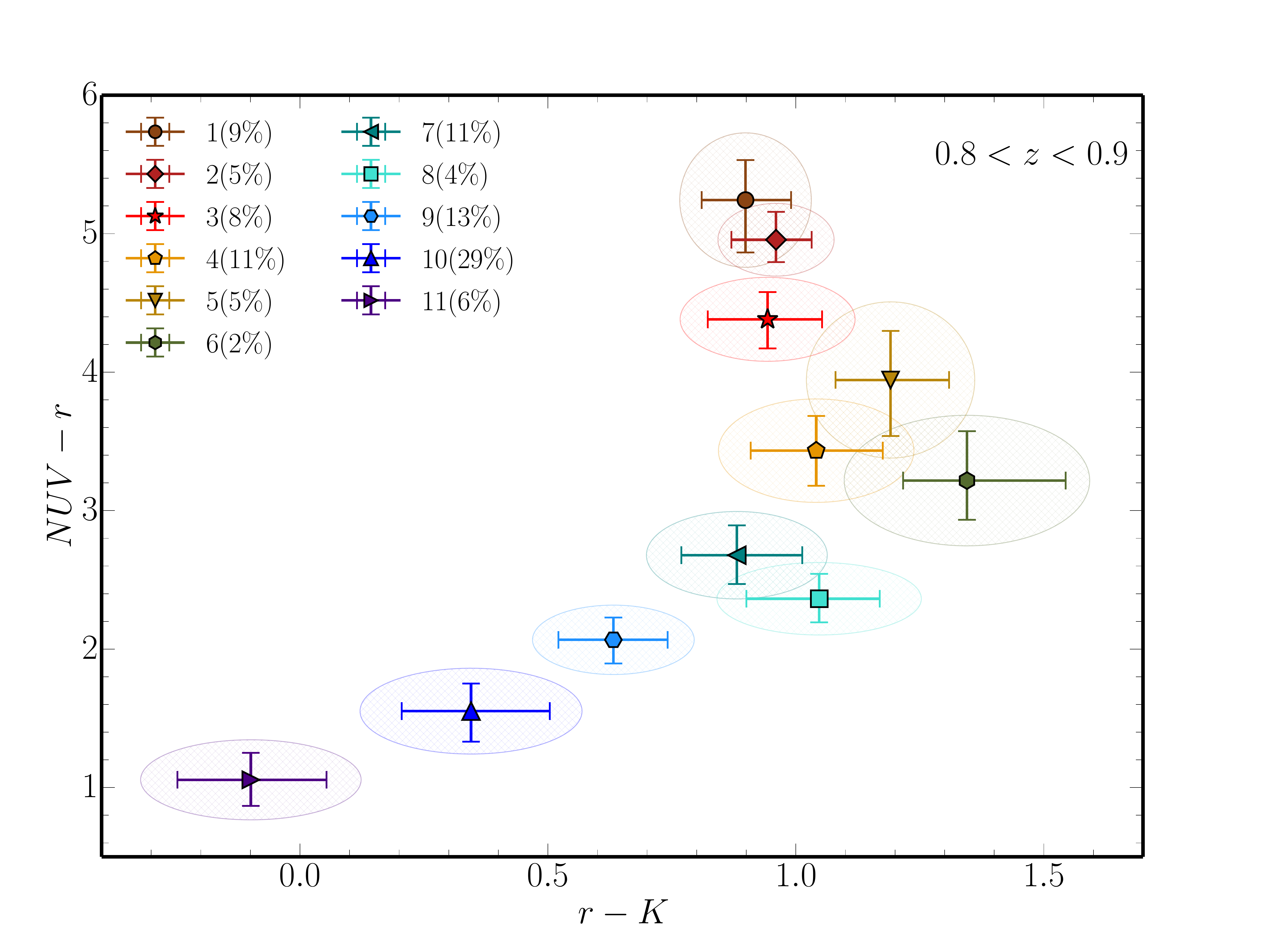}      
        \includegraphics[width=0.49\textwidth]{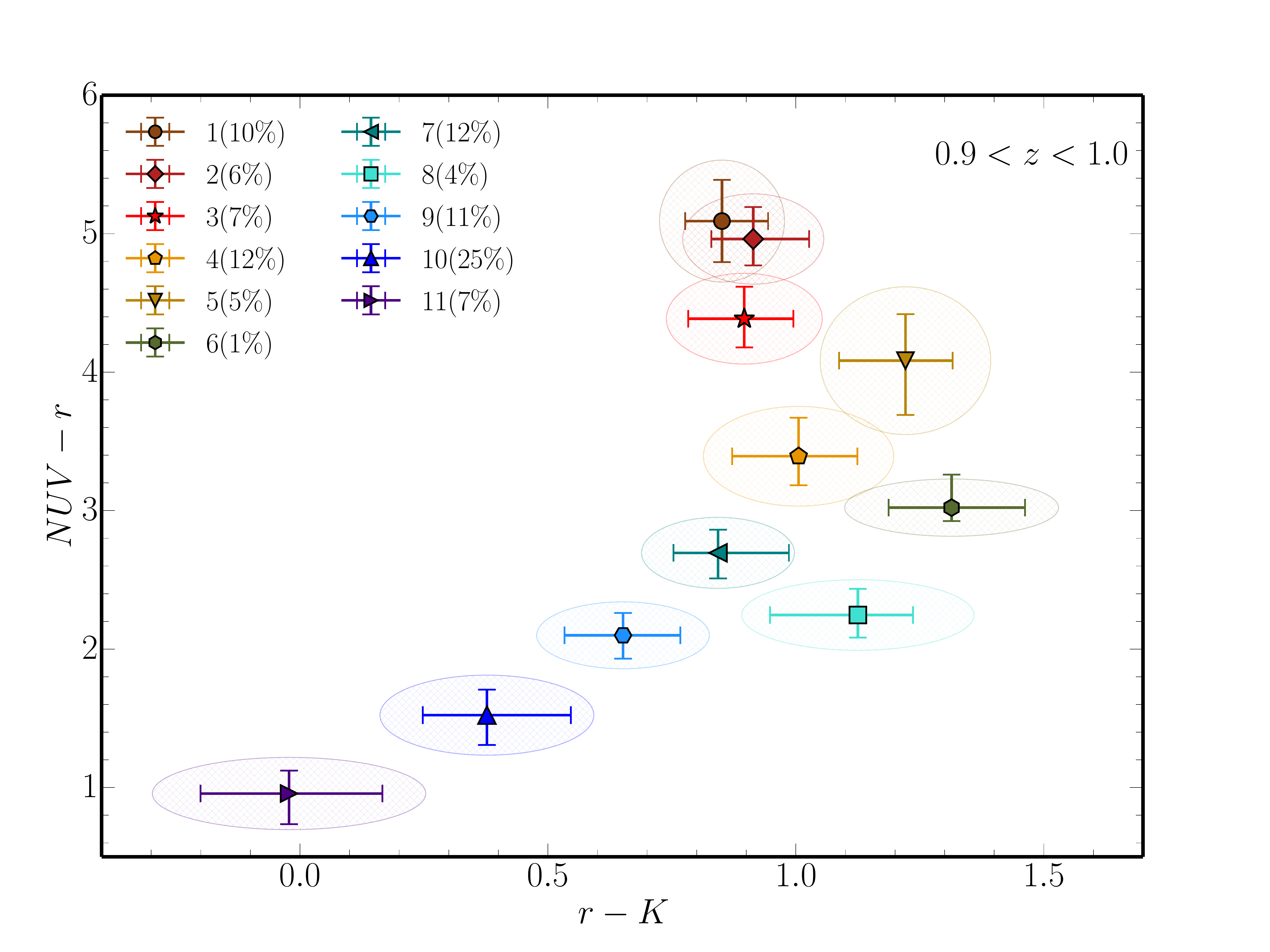}                 
        \caption{$NUVrK$ diagrams of the 11 FEM classes in six different
          redshift bins spanning the redshift range $0.4 <z <
          1.0$. The error bars correspond to the first and the third
          quartiles of the galaxy colour distribution, while the two
          axes of the ellipses correspond to the median absolute
          deviations. The fraction of galaxies in each class is given in the legend. 
        }
        \label{fig:nuvrrk_z}
\end{figure*}

\subsection{Global properties of FEM classes}\label{sec:properties}

A visible separation of 11 classes in the 3D and 2D colour-colour diagrams may be expected, as the FEM classification is based on the normalised absolute magnitudes and, therefore, colours. 
In this section, we examine properties that were not included in the parameter space used for the automatic classification.  
Below we investigate morphological, spectral, mass, and star formation
properties of the different FEM classes to examine whether or not there is a correspondence between our classification and these properties.

The distributions of main properties along the 11 FEM classes are shown in Figs.~\ref{fig:lineproperties} and~\ref{fig:sedproperties}, and summarised in Table~\ref{table:properties}. 
In particular,  the following features were derived for VIPERS galaxies: S\'{e}rsic index \cite[$n$; calculated for VIPERS sample by][]{krywult}, equivalent widths of $[OII]\lambda3727$, the strength of the $4000\AA$ break~\citep[$D4000_{n}$, as defined by][]{balogh1999}, and physical properties derived from SED fitting: stellar masses, and $sSFR$~\citep[calculated by][]{moutard16b}.     
The following analysis is based on the median values of these parameters  derived for each class. 
The error bars correspond to the first and third quartiles of the galaxy property distribution. 

\begin{figure}[]
        \includegraphics[width=0.49\textwidth]{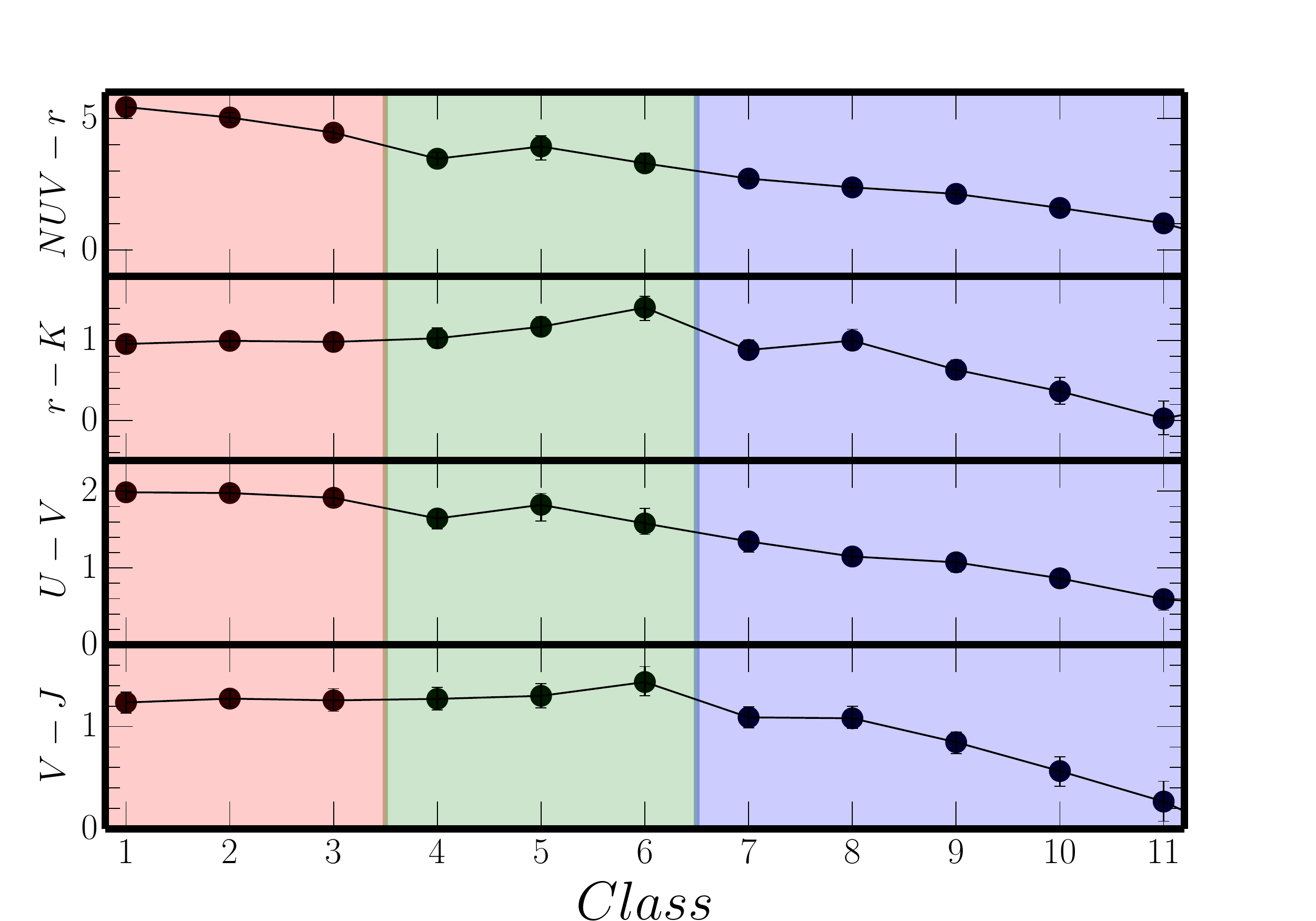}
        \caption{Colours: $NUV-r$, $r-K$, $U-V$ of the 11 FEM classes as a function of class number. The median values of parameters for red (classes 1--3), green (classes 4--6), and blue (classes 7--11)  galaxies are shown in red, green, and blue areas, respectively. }
        \label{fig:colorproperties}
\end{figure}

\begin{figure}[]
        \includegraphics[width=0.49\textwidth]{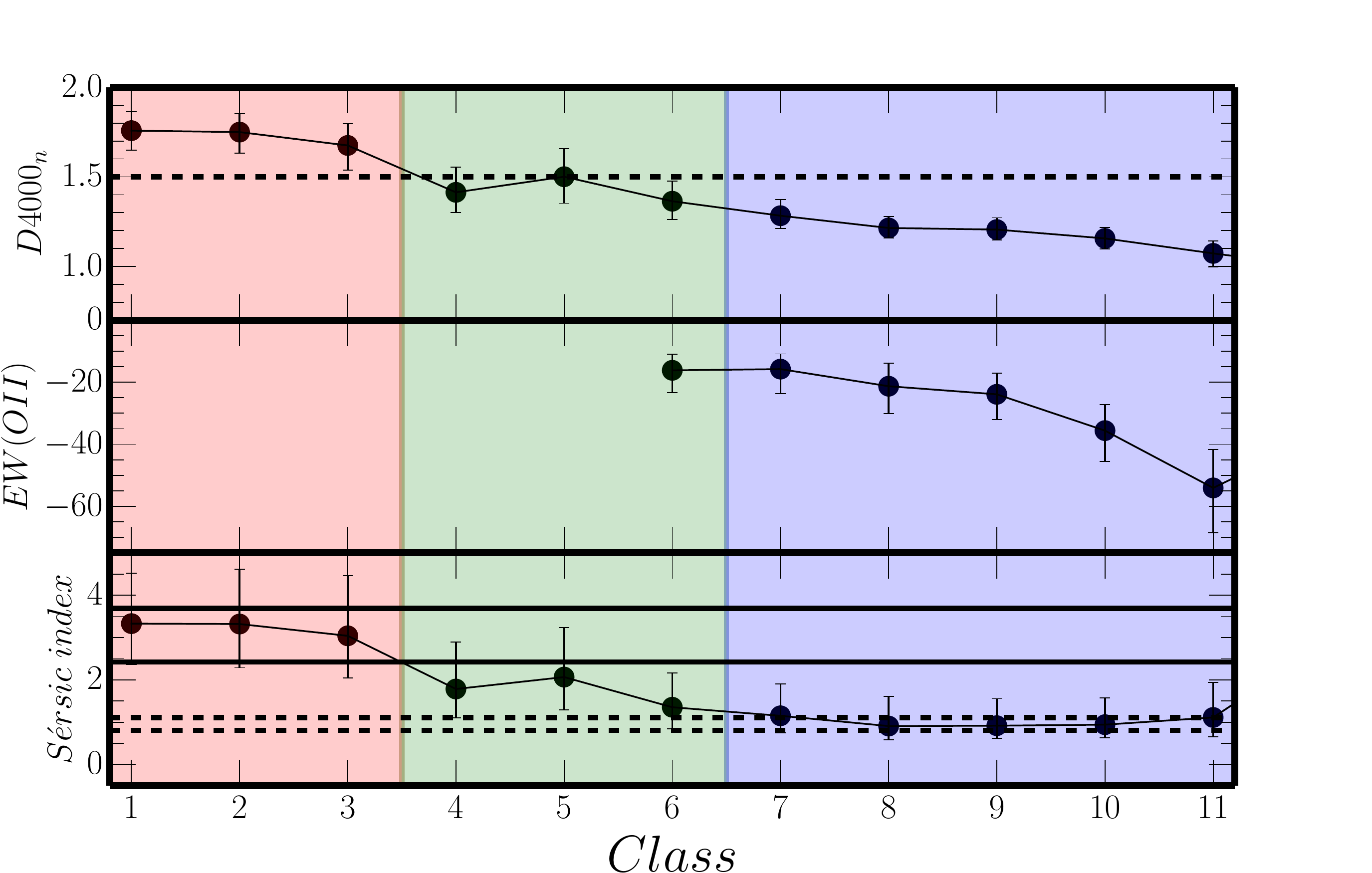}
        \caption{Spectral and morphological properties of the 11 FEM classes: $D4000_{n}$, $EW([OII]\lambda3727)$, and S\'{e}rsic index as a function of class number. The division between red passive and blue active based on $D4000_{n}$ according to~\cite{kauffmann03} is marked with a black dashed line. The range of mean values of S\'{e}rsic index for VIPER red passive and blue star-forming galaxies obtained by~\cite{krywult} are marked with black solid and dashed lines, respectively. The $[OII]\lambda3727$ line has not been detected in the majority of galaxies within classes 1--5 (for 96, 91, 85, 59, and 72 $\%$, respectively). }
        \label{fig:lineproperties}
\end{figure}

\begin{figure}[]
        \includegraphics[width=0.49\textwidth]{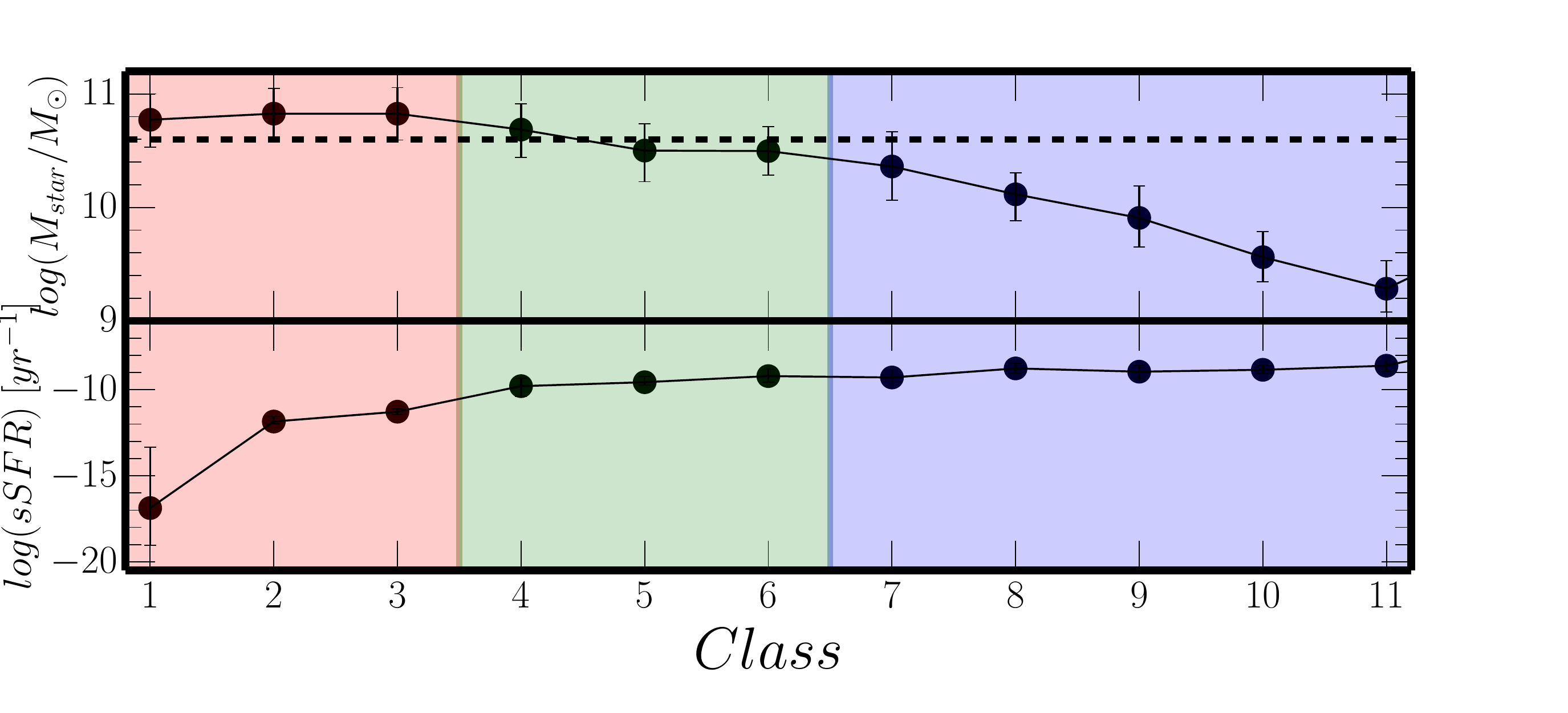}
        \caption{SED-dependent properties of the 11 FEM classes: stellar mass ($log(M_{star}/M_{\odot})$), and $log(sSFR)$ $[yr^-1]$ as a function of class number. The transition mass found for VIPERS galaxies at $z\sim0.7$ by~\cite{davidzon13} is shown with a dashed line.}
        \label{fig:sedproperties}
\end{figure}

To trace the change of spectral properties along the FEM classes, the strength of the $4000\AA$ break and equivalent width of $[OII]\lambda$3727 of individual galaxies in each FEM class is measured. 
Figure~\ref{fig:lineproperties} shows the weakening of the median $4000\AA$ break, and the  increasing of the median EW($[OII]\lambda$3727) with increasing class number.
Galaxies within classes 1--3 have $D4000_{n}$ greater than 1.5 (dashed
line in Fig.~\ref{fig:lineproperties}), and simultaneously display
negligible emission in $[OII]\lambda$3727, while galaxies within
classes 7--11 have strong emission in the $[OII]\lambda$3727 line, and
a $4000\AA$ break lower than 1.5.
The threshold for $D4000_{n}$ at 1.5, dividing actively star-forming and passive galaxy populations, has been found by~\cite{kauffmann03} for local Universe and extended to higher redshift by~\cite{vergani2008}.
This cut allows us to associate galaxies hosting old stellar populations with no sign of star formation activity to classes 1--3, and younger objects with stronger on-going star formation to classes 7--11. 
The more detailed description of the spectral properties of the 11 FEM classes is presented in Sect.~\ref{sec:spectral}.

The reflection of our classification on different galaxy properties indicates the robustness of our approach and the fact that the proposed classification may be able to trace the evolutionary stages  from blue and active to red passive types. 

\subsubsection{Morphological properties}\label{sec:morpho}

One way to define the type of a galaxy is to analyse its structure. 
In the local Universe, passive galaxies are usually spheroidal, while star-forming galaxies are irregular or disk shaped~\citep[e.g.][]{bell2012}.
\cite{krywult} showed that this is also the case for the whole
mass distribution ($8 \lesssim log(M_{star}/M_{\odot}) \lesssim 12$)
and redshift range ($0.4 \lesssim z \lesssim 1.3$) of VIPERS galaxies.
To describe the shapes of the light profiles of VIPERS galaxies, the S\'{e}rsic index is used~\citep[$n$,][]{sersic63}.
The index  has low values ($n\sim1$) for spiral galaxies whose disks have surface brightnesses with a shallow inner profile, and high values ($n\sim3-4$) for elliptical galaxies which have surface brightnesses with a steep inner profile~\citep[e.g.][]{simard, bell2012, krywult}.

\cite{krywult} showed that VIPERS disk-shaped galaxies have S\'{e}rsic index mean values in the range  $n\sim0.81-1.11$, whereas spheroid galaxies are characterised with average S\'{e}rsic indices in the range $n\sim2.42-3.69$.  
As shown in the lower right panel of Fig.~\ref{fig:lineproperties}, there is a very good correlation between the FEM galaxy class and S\'{e}rsic index.
FEM red passive galaxies (classes 1--3) have a median S\'{e}rsic index $n>3$, indicating a spheroidal shape, while classes 7--11 show a  significantly lower median S\'{e}rsic index $n\lesssim1$, typical for disk galaxies. 
For intermediate classes, the median S\'{e}rsic index is $n\sim1.7$, confirming that classes 4--6 are mainly composed of intermediate galaxies also in terms of this structure.
\cite{krywult} demonstrated the strong correlation between morphology and galaxy colour, which is also reflected in our studies. 

\subsubsection{Physical properties}\label{sec:physical}

The top panel of Fig.~\ref{fig:sedproperties} shows the median stellar
masses obtained for the 11 FEM groups. 
The stellar mass decreases with class number. 
Galaxies assigned to classes 7--11 are less massive (with median stellar mass $\sim10^{9.7\pm 0.3} M_{\odot}$) than galaxies within classes  1--3  (median stellar mass $\sim10^{10.8 \pm 0.2} M_{\odot}$).
The stellar mass change is much more rapid for star-forming classes
(0.3~dex per class), whereas for red passive classes the median stellar mass is almost constant (0.05~dex).   
Our classification follows well the location of passive and active galaxy types with respect to the transition mass.
The transition mass separates blue star-forming and red passive populations, since above the transition mass, red passive galaxies dominate, and below that mass, star-forming galaxies are the most numerous population~\citep[e.g.][]{kauffmann03,vergani2008,pannella,davidzon13}.
Based on the VIPERS dataset,~\cite{davidzon13} determined the transition mass to be $log(M_{star}/M_{\odot}) = 10.6$ for galaxies at $z\sim0.7$.
Our classification is consistent with this result. 
Median stellar masses of galaxies within  classes 1--3 are above the transition mass (marked with the dashed black line in Fig.~\ref{fig:sedproperties}), while classes 7--11 are located below the transition mass consistent with the fact that these galaxies are still forming stars. 
The intermediate galaxies within class~4 have the median stellar mass which matches the transition mass perfectly. 
This confirms that this is the group of sources that are just entering  the passive evolutionary path. 
Classes 5--7 have stellar masses just below the transition mass ($10^{10.5}$~$M_{\odot}$) between the red and blue populations.

Finally, the bottom panel of Fig.~\ref{fig:sedproperties} shows the change of $sSFR$ as a function of class number. 
The FEM classes are well separated in $sSFR$, with red passive
galaxies (classes 1--3) showing the lowest star formation activity,
whereas sources from the blue classes (7--11) have the highest sSFRs. 
At the same time, from Fig.~\ref{fig:lineproperties} we can see that classes 7--11 have high $EW([OII]\lambda3727)$, which is typical for blue star-forming galaxies~\citep[e.g.][]{Cimatti2002}.  
The $sSFR$ obtained for the intermediate galaxies ($log(sSFR)\sim-9$ $[yr^{-1}]$) is in agreement with the results derived for 1,745 CANDELS transition galaxies observed at $0.5 < z < 1.0$~\citep[$log(sSFR)\sim-9$,][]{pandya}.       
Summarising, the distributions of the physical properties (see Figs.~\ref{fig:colorproperties},~\ref{fig:lineproperties}, and~\ref{fig:sedproperties}, and Table~\ref{table:properties}) show the trends of global and systematic changes along the FEM classes. 
The main spectral, morphological and physical properties correlate well within and among the groups, that is, the most massive spheroidal galaxies populated by old stellar populations are the reddest in comparison to the disk-shaped bluer galaxies hosting younger stellar contents.  
This demonstrates that our classification traces the evolutionary phases and galaxy types.

\subsection{The $SFR-M_{*}$ relation}\label{sec:sfr}

Galaxies show a correlation between their $SFR$ and stellar mass at redshifts at least up to $z\sim6$~\citep[e.g.][]{brinchmann,noeske,whitaker12,Speagle2014, 2015Salmon}. 
This correlation, often called the galaxy main sequence (MS), is likely connected with the physical mechanisms responsible for galaxy growth, regulated by the accretion of gas from cosmic web and gas feedback~\citep[e.g.][]{bouche10}.

\begin{figure}[]
        \includegraphics[width=0.49\textwidth]{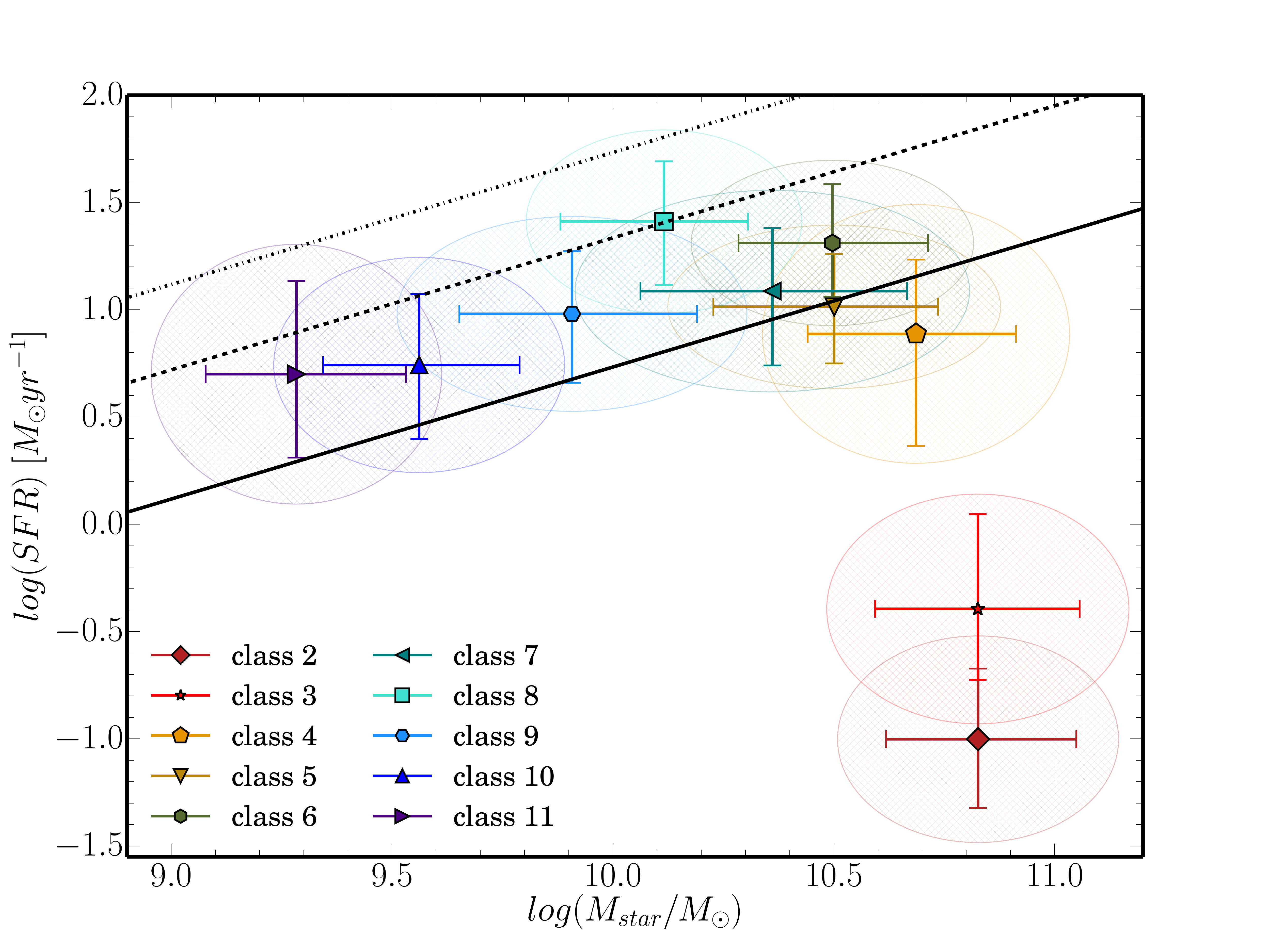}
        \caption{$SFR$--stellar mass relation for FEM
          classification. The median $log(SFR)$ vs.  median
          $log(M_{star}/M_{\odot})$ for classes 2--11 are shown. The
          error bars correspond to the first and third quartile of the
          galaxy $SFR$-stellar mass distribution, while the area of
          ellipses correspond to the median absolute deviations.  The
          colours are given as in Fig.~\ref{fig:color}. The
            first class is not plotted due to its very low median
          $SFR$. The black solid line corresponds to the MS trend at $z=0.7$ found by~\cite{whitaker12}, while dashed and dashed-dotted lines correspond to $4\times MS$, and $10\times MS$ to represent active star-forming and starburst galaxies, respectively, following~\cite{rodighiero11}. }
        \label{fig:sfr}
\end{figure}

The $SFR$ dependence on stellar mass for the different FEM classes is shown in Fig.~\ref{fig:sfr}.
The black solid line corresponds to the MS at $z=0.7$ according to~\cite{whitaker12}.
\cite{whitaker12} have established the slope and the normalisation of
the $SFR(M_{*})$ as a function of redshift allowing us to reproduce the MS trend at $z=0.7$, the median redshift of VIPERS galaxies.  
Passive galaxies within classes 2--3 (class~1 is not presented in
Fig.~\ref{fig:sfr} due to its very low SFR;
$log(SFR)=-6.1$ $[M_{\odot}yr^{-1}]$) occupy an area well below the MS line. 
The star-forming galaxies assigned to classes 7--11 instead follow the
tight MS trend, showing a steady increase in $SFR$ with stellar mass as expected for the MS at this redshift. 
Therefore, this confirms classes 7--11 to be representative clusters of star-forming MS galaxies. 
However, we note that most of these median values are above the solid line. 
The global offset for star-forming galaxies could be due to the extinction law and SFH used for SED fitting. 
The~\cite{calzetti} extinction law is characterised by larger attenuations at longer wavelengths which results in lower stellar masses compared to other recipes such as \cite{charlot2000} or  \citealp{lofaro2017} (for more detailed discussions we refer to \citealp{lofaro2017} and \citealp{Malek2018}). 
Therefore, we relate the offset in the $SFR$ to the method used to calculate $SFR$. 
\cite{whitaker12} used the \cite{Kennicutt1998} relation which assumes a constant $SFR$. 
This assumption leads to the overestimation of the $SFR$ with respect to the other SFHs in the literature (and with respect to the delayed SFH used for the SED fitting; e.g. \citealp{lofaro2017}). 
To summarise, the different models used for VIPERS SED fitting and to
obtain the MS relation have influence on the observed offset in Fig.~\ref{fig:sfr}.    
Galaxies assigned to class 8 show a $SFR-M_*$ relation slightly above MS. 
However, we stress that within uncertainties this class is still
consistent with the trend defined by the other classes. 
The median $SFR$ of galaxies in class~8 is located at $4\times MS$
(dashed line), which is attributed to galaxies with enhanced star formation~\citep{rodighiero11}. 
This class is also characterised by redder $r-K$ colours than, for example, class~7, and a strong $H_{\beta}$ line, but not one stronger than the $H_{\beta}$ line for class~10 (see Fig.~\ref{fig:stacked_spectra2}).

\subsection{Spectral properties}\label{sec:spectral}

In this section, the spectral properties of the photometrically motivated classes are presented.  
To compare the spectral properties to the classification scheme, the stacked spectra for each of the 11 FEM classes were derived. 
The spectra were co-added in narrow redshift bins ($\delta z=0.1$ from 0.4 to 1.0) in the same way as described in~\cite{siudek17}. 
Firstly, the rest-frame spectra were re-sampled to a common wavelength grid. 
Individual spectra were normalised by dividing the flux at all wavelengths by the scaling factor derived using median flux computed in the wavelength region $4010 < \lambda(\AA) < 4600$. 
The stacked spectra were then obtained by computing the mean flux from all individual spectra at all wavelengths in the common wavelength grid, and rescaled by multiplying the flux at all wavelengths by an average value of scaling factors of the individual spectra. 
Given the large sample of VIPERS galaxies, the constructed stacked
spectra are characterised by a signal-to-noise ratio (S/N) high
enough to detect absorption lines that are undetectable on typical, individual spectra~\citep[e.g. the $H\delta$ line; see details in][]{siudek17}.

\begin{figure*}[]
        \includegraphics[width=0.99\textwidth]{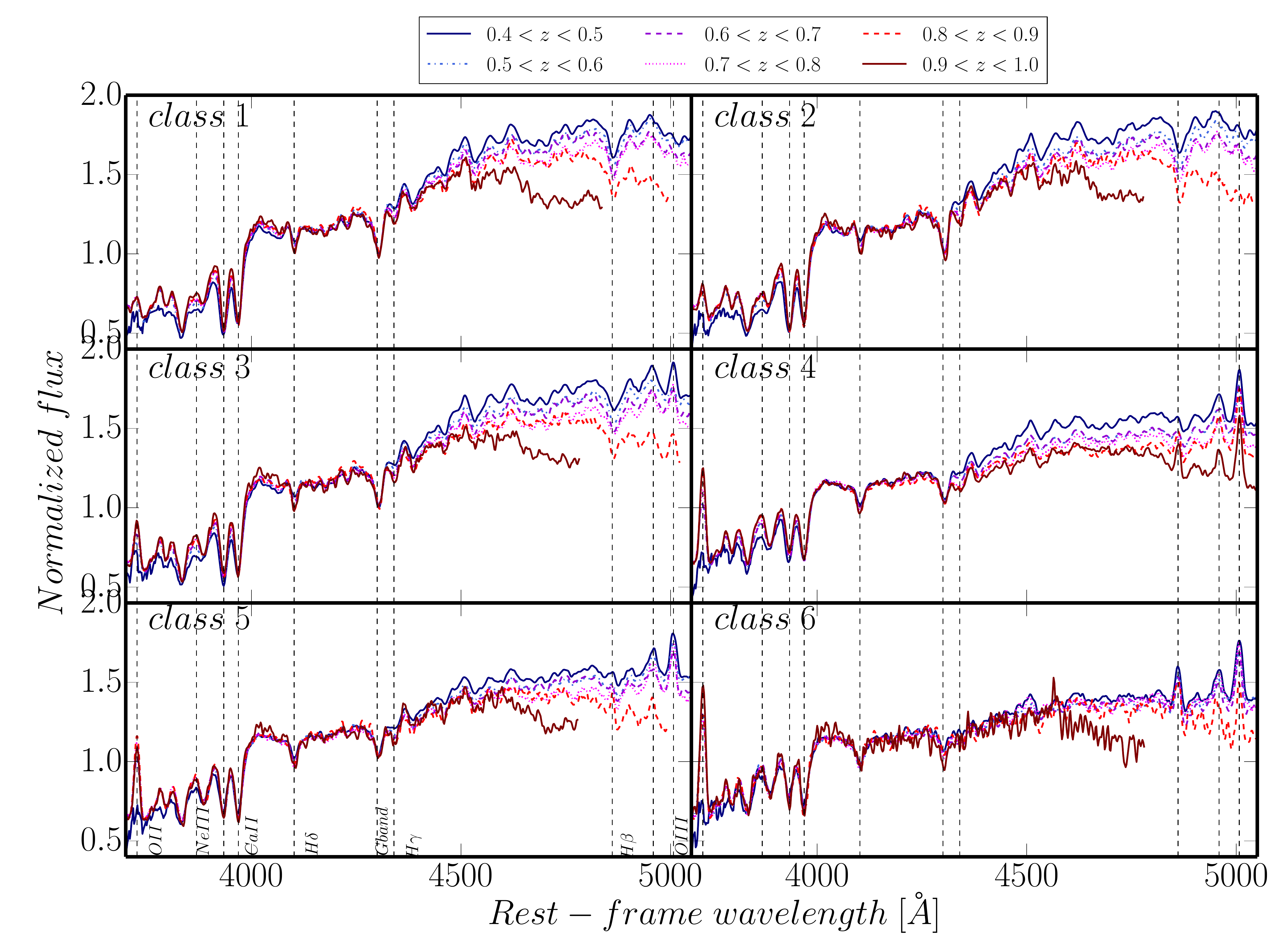}
        \caption{Stacked spectra of VIPERS galaxies among FEM classes 1--6 in different redshift bins.
                Rest-frame composite spectra were normalised in the region $3600 < \lambda < 4500 $ $\AA$. 
                The most prominent spectral lines are marked with vertical solid lines with labels. 
        }
        \label{fig:stacked_spectra}
\end{figure*}

\begin{figure*}[]
        \includegraphics[width=0.99\textwidth]{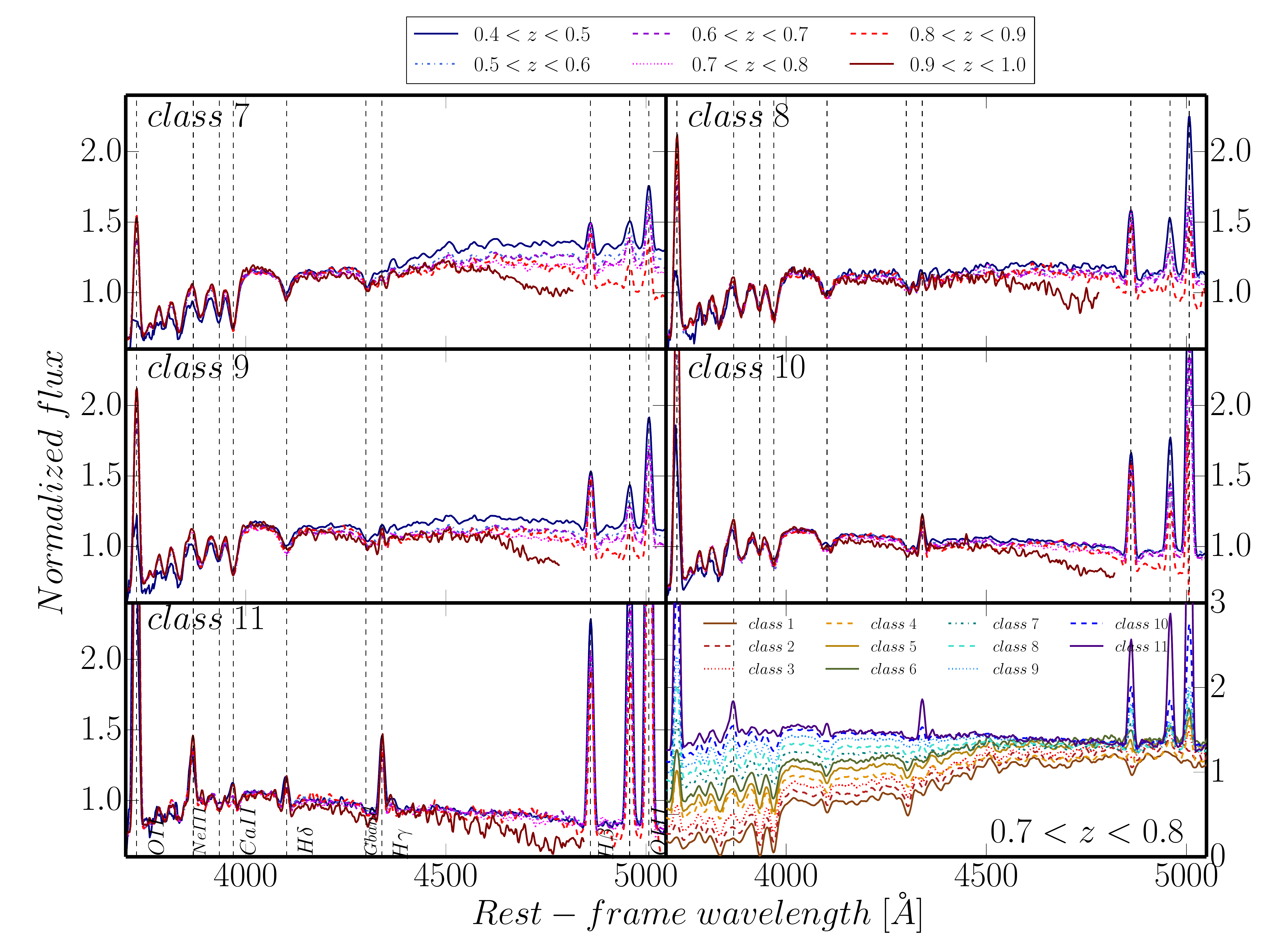}
        \caption{Stacked spectra of VIPERS galaxies among FEM classes 7--11 in different redshift bins.
                Rest-frame composite spectra were normalised in the region $3600 < \lambda < 4500 $ $\AA$. 
                The most prominent spectral lines are marked with vertical solid lines with labels. The last panel shows stacked spectra of FEM classes 1--11 in redshift bin $0.7 < z<0.8$.
        }
        \label{fig:stacked_spectra2}
\end{figure*}

Figures~\ref{fig:stacked_spectra} and ~\ref{fig:stacked_spectra2}
show the stacked spectra of the 11 FEM classes in six redshift bins spanning the redshift range $0.4 < z < 1.0$. 
The stacked spectra show that there is a gradual change as a function of class number. 
The  lines go from absorption (in the first class) to strong emission (in the eleventh class). 

All composite spectra of galaxies assigned to classes 1--3 are dominated by absorption lines and show weak emission lines.
We can clearly see the strong $4000\AA$ break, G-band $(4304\AA)$, and Balmer lines over most of the redshift range, even if some of these features are not observed at $z>0.8$ because of the 
 wavelength range $5500-9500\AA$ of VIPERS spectra; see ~\cite{scodeggio16} for details.
The strong absorption lines  for these features are typical for early-type galaxies~\citep[e.g.][]{ wortheya, worthey, gallazzi2014,siudek17}.  
Therefore, we conclude that the spectral properties indicate that galaxies in classes 1--3 consist of old stellar populations. 
From Fig.~\ref{fig:stacked_spectra}, we can see that the $H\delta$ line is getting stronger with redshift for all three red passive galaxy classes, which may be simply indicating that stellar populations are getting older as time passes. 
There is also a change in the relative strength of the CaII H ($3969\AA$) and CaII K ($3934\AA$) lines, as the CaII K line dominates at $z\sim 1$, while the CaII H line dominates at lower redshifts, especially for galaxies in class~3. 
The CaII K line dominates in galaxies with old stellar populations, whereas CaII H dominates when the younger stars appear. 

Spectra of the green group (classes 4--6)  show properties in-between the red and blue populations~\citep[see also][]{vergani2017}. 
The representative stacked spectra of classes~4 and~5 are characterised by strong emission in the $[OIII]\lambda\lambda4959,5007$ doublet with no or little sign of the recombination line $H\beta$ at redshift range $0.4 < z < 0.7$. 
Since a high ratio of $[OIII]\lambda\lambda4959,5007$ to $H\beta$ lines is an indication of AGN photo-ionisation, this suggests that a non-negligible fraction of galaxies in these classes may host a Seyfert nucleus. 
However, this is not confirmed by the localisation of classes 4 and~5 on the BPT diagram (see Sect.~\ref{sec:bpt}), even if only galaxies within redshift range $0.4 <z <0.7$ are considered. 
Therefore, we are not able to conclude whether those galaxies host a
Seyfert nucleus or not. 
The stacked spectra of intermediate galaxies within class 6 show diagnostic lines (e.g. $[OII]\lambda3727$, $[NeIII]\lambda3869$, $H\beta$) in emission. 
There is also a hint of star formation activity in  the intermediate
classes (4--6) revealed by detectable emission in the
$[NeIII]\lambda3869$ line in all redshift ranges~\citep{ho}.

The stacked spectra of galaxies in classes 7--11 show that they are undergoing a significant level of star formation, indicated by prominent emission lines, like the $[OII]\lambda3727$ or $H\beta$ lines, and a weak $4000\AA$ break~\citep[e.g.][]{mingoli,haines16}. 
The emission lines are getting stronger with increasing class number of star-forming galaxies. 
The possibility of AGNs is further discussed in Sect.~\ref{sec:bpt}. 
In this paper, we focus on general properties of the whole classification scheme.
The detailed properties and evolutionary trends of the FEM classes will be discussed in future papers.

\subsubsection{The comparison of FEM classes with Kennicutt's Atlas}
To better define the morphological and spectral types of each of the 11 FEM classes, we compare their representative stacked spectra with those of galaxies of different Hubble types as given by~\cite{kennicutt92}. 
Kennicutt's Atlas consists of 55 integrated spectra of nearby
galaxies, covering the wavelength range $3650 <\lambda[\AA]<7100$ with a resolution of $5-8\AA$, grouped according to their morphological and spectral types. 
~\cite{kennicutt92} provides a set of individual normal and peculiar galaxies following the Hubble sequence, from giant ellipticals ($NGC1275$) to dwarf irregulars ($Mrk$35). 
We compared the 11 FEM classes with the Atlas by assigning to each FEM class the best spectrum in the Atlas based on the $\chi^2$ minimisation. The 11 FEM classes tend to follow the Hubble sequence as classes 1--3
show morphologically earlier types than the other classes. 
Stacked spectra of galaxies in classes 7--11 are quite well reproduced
by the spiral, irregular, and emission-line galaxies (Sc, Im), whereas
spectra of Sb galaxies best fit the stacked spectra of intermediate galaxies (classes 4--6), and the template spectrum of Sab galaxy fits the representative stacked spectra of classes 1--3.  
The detailed comparison of spectral properties of the 11 FEM classes to the spectral Atlas of~\cite{kennicutt92} is discussed in Appendix~\ref{app:kennicutt}.

\subsubsection{Comparison to principal component analysis (PCA) classification of VIPERS galaxies}\label{sec:pca}      

In this section, we compare the FEM classification to a classification
scheme used within the VIPERS survey by~\cite{marchetti13}, based on the PCA technique applied to spectra of VIPERS galaxies. 
The PCA-based algorithm divided VIPERS galaxies into 15 different clusters based on the first three eigen coefficients ($\theta-\phi$ diagram).
The PCA classification distinguished eight groups among the red and intermediate galaxy types from E to Sc, and seven classes of more active starburst galaxies. 
We find that our classification follows the track found by~\cite{marchetti13}, since the reddest, early-type galaxies fall in the region of the bottom left edge of the $\phi-\theta$ diagram, and with increasing $\theta$ and $\phi,$ the number of the FEM class is increasing, which implies that galaxies are bluer (see Fig.~\ref{fig:pca_z}).  

We find  that $\sim70\%$ of early-type galaxies selected with PCA (PCA classes 1--2 contain E and Sa galaxies) are distributed in the FEM classes 1--3. 
This indicates the similarities in the  capability of separation of ETGs, especially the oldest ones,  in the VIPERS dataset by both methods. 
The dusty spiral galaxies, $\rm{Sb_{4,6}}$~\cite[with $E(B-V)>0.4$; ][]{kinney96}, assigned to PCA classes 3--6 are spread among various FEM classes, with the majority of them ($\sim70\%$) being located in the FEM classes 7--11.
Almost all Sc galaxies ($\sim95\%$) selected by PCA (PCA classes 7-8) are assigned to the FEM classes 9-11. 
The spiral galaxies with smaller amounts of dust,
$\rm{Sb_{1,2}}$~\cite[with $E(B-V)<0.2$;][]{kinney96}, within PCA
classes 9--13, are also found among FEM classes 10--11  ($\sim80\%$ of them). 

This shows that there is a global agreement between the FEM and PCA classification schemes. 
However, it should be noted that these two classification schemes,
being based on different input data  (photometric data for FEM, and
spectroscopic for PCA), are not fully coherent with each other and therefore do not show precisely the same patterns. 
In Appendix~\ref{app:pca} it is shown how well, using the derived eigenvalues for VIPERS PDR1, the FEM classes are separated in the $\theta - \phi$ PCA diagram.
 
\subsubsection{The Baldwin, Phillips \& Terlevich diagram}\label{sec:bpt}

To differentiate star-forming galaxies from AGNs, we checked the distributions of the intermediate and star-forming galaxies (classes 4--11) on the diagnostic diagram for emission-line galaxies.
The distribution of VIPERS galaxies in the BPT~\citep{Baldwin81} diagram is shown in Fig.~\ref{fig:bpt}. 
We are able to separate  LINERS and Seyferts based on their emission line ratios.  
We measured emission lines on individual spectra within the redshift range $0.4 < z < 1.3$ assigned to classes 4--11, and the $H\beta$ measurements were corrected for an average absorption component. 
The distribution of VIPERS galaxies assigned to classes 4--10 indicates that those galaxies are star-forming galaxies. 
Class 11 is placed in the composite area (SF/Sy2 in Fig.~\ref{fig:bpt}), which may indicate that it contains AGNs. 
The contamination by broad-line AGNs has no influence on our result,
as only line measurements of galaxies within redshift range $0.4 < z <
1.0$ and redshift flag 3--4 are included (i.e. excluding the flags
corresponding to broad-line AGN). 
However, AGNs are very rare among low-mass galaxies (the median stellar mass of galaxies within class 11 is $\sim10^{9}$ $M_{\odot}$), therefore, we suspect that these might be low-metallicity galaxies.   
However, both these options (AGN contributions and low-metallicity galaxies) are consistent with the spectroscopic properties of this group (see Sect.~\ref{sec:spectral}), as the spectra show strong emission lines.

\begin{figure}[]
        \includegraphics[width=0.49\textwidth]{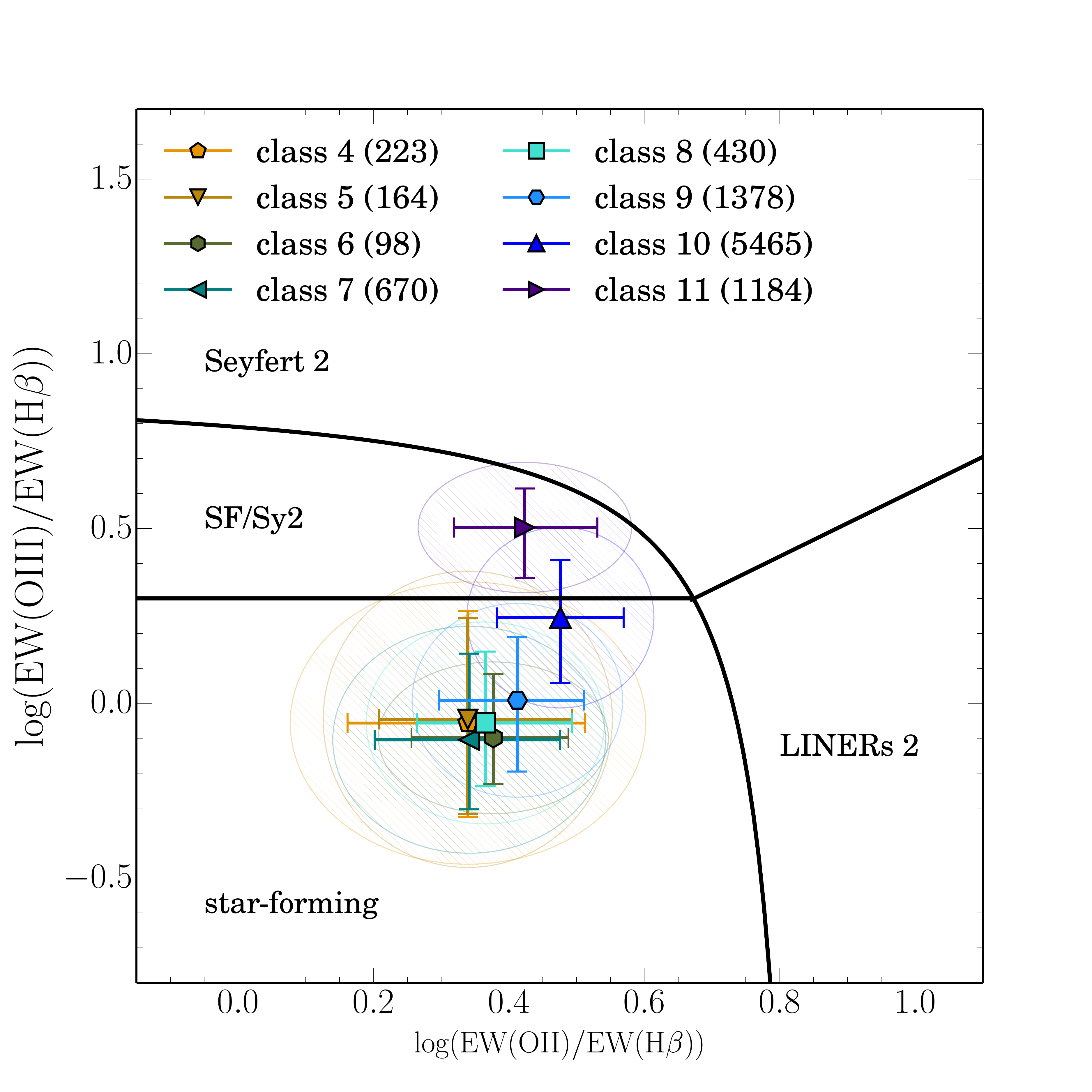}
        \caption{The distributions of FEM classes 4--11 on the "blue" BPT diagram introduced by~\cite{2010Lamareille}. The number of spectra in each class for which lines were measured in the redshift range $0.4 < z < 1.3$ are given in the legend. The error bars correspond to the first and third quartile of the line measurements distribution, while the area of ellipses correspond to the median absolute deviations.  }
        \label{fig:bpt}
\end{figure}

\section{Summary}\label{sec:summary}

In this paper, a new approach to galaxy classification is introduced,
based on the thirteen-dimensional parameter space built from 12 absolute
magnitudes and the spectroscopic redshift. 
An unsupervised classifier based on the FEM algorithm blindly separated 52,114 VIPERS galaxies into 12 classes. 
The model selection (DBk) and the determination of the optimal number of classes were based on statistical criteria (BIC, AIC and ICL; see Appendix~\ref{app:A})  and found to be in the range 9--12. 
Subsequently, the final class number (12) was decided based on the analysis of the galaxy flow with a changing number of groups (see Fig.~\ref{fig:flow_2_14}), and the interpretation of physical properties of classes in different realisations (see Fig.~\ref{fig:nuvr_9_10_11_12}). 
All these techniques resulted in the same model and an optimal number
of 12 classes in the VIPERS dataset.  
These classes follow a well-defined sequence from the earliest to the latest types, separating galaxies into three major groups: red, green, and blue. 
The FEM classification automatically finds groups that share physical and spectral properties, beyond the features used for classification purposes. 
Galaxies are not unequivocally assigned to a single class, but the probability of belonging to each group is given. 
Such an approach is more realistic as the transition between  classes can be continuous. 
In spite of this, a majority of galaxies ($92\%$) in the sample have high ($>50\%$, with $<45\%$ second best probability) probabilities of belonging to the selected group.
We obtain three main classes: red, green, and blue, which can be further separated into subclasses: three red, three green, and five blue, and an additional class 12, which consists of outliers.
For class 12, $95\%$ of its members are broad-line AGNs according to
the visual classifications by the VIPERS team~\citep{garilli14}.
Their median redshift is $z_{med}\sim2$, which removes this class from the global picture of VIPERS galaxy types observed up to $z\sim1$. 

We demonstrated that our approach leads to a new classification scheme allowing us to track  galaxy evolutionary paths. 
The main advantage of this approach is the ability to distinguish 11 galaxy types, which share physical and spectral properties not used in the classification procedure. 
The presented separation between different galaxy types differs from traditional selection methods based mainly on the bimodal distribution in colours~\citep[e.g.][]{Bell2004, balogh2004,Franzetti07}, spectral properties (e.g., H$\alpha$~\citep{balogh}), $[OII]\lambda3727$ emission~\citep{mingoli}, 4000$\AA{}$ break~\citep{kauffmann03, vergani2008}, and SFH~\citep{brinchmann}.      

Our main results are as follows:

We present a new unsupervised approach to galaxy
          classification based on the multidimensional space of
          absolute magnitudes and the spectroscopic redshift, which does not introduce any a priori defined cuts. 

We find three red, three green, and five blue classes which are distributed along a well-defined path in multidimensional space.  

The borders between classes are not sharp; the probability of belonging to a given class is associated to each galaxy.  However, the probabilities of belonging to a given class are high ($\sim80\%$) and, in spite of the presence of outliers, the classes are well separated in the feature space and are therefore more faithfully representative of the full complexity of the galaxy population at these redshifts.

We show the evolution of the 11 classes over the redshift range $0.4 < z < 1.0$.

We demonstrate that there are significant differences in physical and spectral properties between galaxies classified as red/green/blue FEM classes and their subclasses.

We find  a very good correlation between the FEM classes and spectroscopic classes in the Atlas of~\cite{kennicutt92}. The 11 FEM groups follow the path from the earliest to the latest galaxy types.

In particular, the following FEM class properties were found:

Classes 1--3 host the reddest spheroidal-shape galaxies showing no sign of star formation activity and dominated by old stellar populations (as testified by their strong $4000\AA$ breaks). 

Classes 4--6 host intermediate galaxies whose physical properties, such as colours, sSFR, stellar masses, and shapes, are intermediate relative to red, passive, and blue, active galaxies. 
        These intermediate galaxies have more concentrated light profiles and lower gas contents than star-forming galaxies (as indicated by the S\'{e}rsic index, and $EW(OII)$).
        This tendency is also observed for intermediate galaxies observed in the local Universe~\citep{schiminovich07, schawinski}. 

Classes 7--11 contain the star-forming galaxies.
        The blue cloud of disk-shaped galaxies is actively forming new stars and are populated by young stellar populations (as indicated by the weak $4000\AA$ break).       
        Class 11 may consist of low-metallicity galaxies, or AGNs according to its localisation on the BPT diagram. 

Automatic unsupervised classifications are becoming an invaluable tool
in the current era of information deluge. 
The FEM algorithm can also be applied to photometric samples with comparable efficiency in distinguishing a full panoply of galaxy types~\citep{siudek2018}.    
With the increasing number of deep surveys, such as Euclid and LSST,
such algorithms may allow us to study galaxy formation and evolution across the lifetime of the Universe. 
The presented classification scheme has great potential, as we can ascertain the class to which a galaxy or a galaxy region belongs. 
Based on defined classes, different stellar populations can be traced and galaxies within structures can be classified. 

\begin{acknowledgements}
The authors wish to thank the referee for useful and constructive comments. 
The authors wish to thank Didier Fraix-Burnet  and Charles Bouveyron for useful and constructive discussion.  
We acknowledge the crucial contribution of the ESO staff for the management of service observations. In particular, we are deeply grateful to M. Hilker for his constant help and support of this program. Italian participation in VIPERS has been funded by INAF through PRIN 2008, 2010, and 2014 programs. LG and BRG acknowledge support of the European Research Council through the Darklight ERC Advanced Research Grant (\# 291521). OLF acknowledges support of the European Research Council through the EARLY ERC Advanced Research Grant (\# 268107). KM, TK, JK, MS have been supported by the National Science Centre (grant UMO-2013/09/D/ST9/04030). MS also acknowledges financial support from UMO-2016/23/N/ST9/02963 by the National Science Centre. RT acknowledge financial support from the European Research Council under the European Community's Seventh Framework Programme (FP7/2007-2013)/ERC grant agreement n. 202686. EB, FM, and LM acknowledge the support from grants ASI-INAF I/023/12/0 and PRIN MIUR 2010-2011. LM also acknowledges financial support from PRIN INAF 2012. 
 \end{acknowledgements}
        
\bibliographystyle{aa}
\bibliography{vipers}

\appendix

\section{Determination and validation of the number of classes}\label{app:number}

The AIC, BIC, and ICL are standard criteria widely used in machine learning algorithms to evaluate the statistical model.

\begin{itemize}
        \item AIC - The Akaike information criterion penalises the
          log-likelihood by $\gamma(M)$, where $M$ is the used model
          and $\gamma$ is the number of parameters in this model
          \citep{Akaike1974}. The AIC is used to select the best model from the available pool. Using this criterion, we search for the model closest to reality with a minimum number of parameters. It was first introduced to cosmology by \cite{Takeuchi2000}. 
        \item BIC - The Bayesian information criterion is the most popular criterion which penalises likelihood by $\frac{\gamma(M)}{2}log(n)$, where $M$ is the used model and $n$ is the number of observations \citep{Schwarz1978}. BIC is analogous to AIC \citep{Bou2011}, but is derived from Bayesian statistics. 
        \item ICL - The integrated complete likelihood penalises the log-likelihood by $\sum^n_{i=1}\sum^K_{k=1}t_{ik}\log(t_{ik})$ in order to favour well-separated models, where $k$ is the number of mixture components, $i$ is the number of observables, and $t$ is the posterior probability~\citep{Baudry2012}.
\end{itemize}

In this paper, all three criteria are applied. They allow us to compare different DLM models and define the optimal number of groups  in the data. 
The AIC and BIC are widely used penalised likelihood criteria, while
ICL is an alternative approach, which starts with the BIC criterion
and adds a so-called entropy component (the sum of posterior probability memberships given by $\sum_{i}\sum_{j}pp_{ij}ln(pp_{ij})$, where $pp_{ij}$ is the $i$'s posterior probability membership in $j$-group). 
Therefore, the scoring between AIC/BIC cannot be directly compared to ICL scoring.

\begin{figure}
        \includegraphics[width=0.5\textwidth]{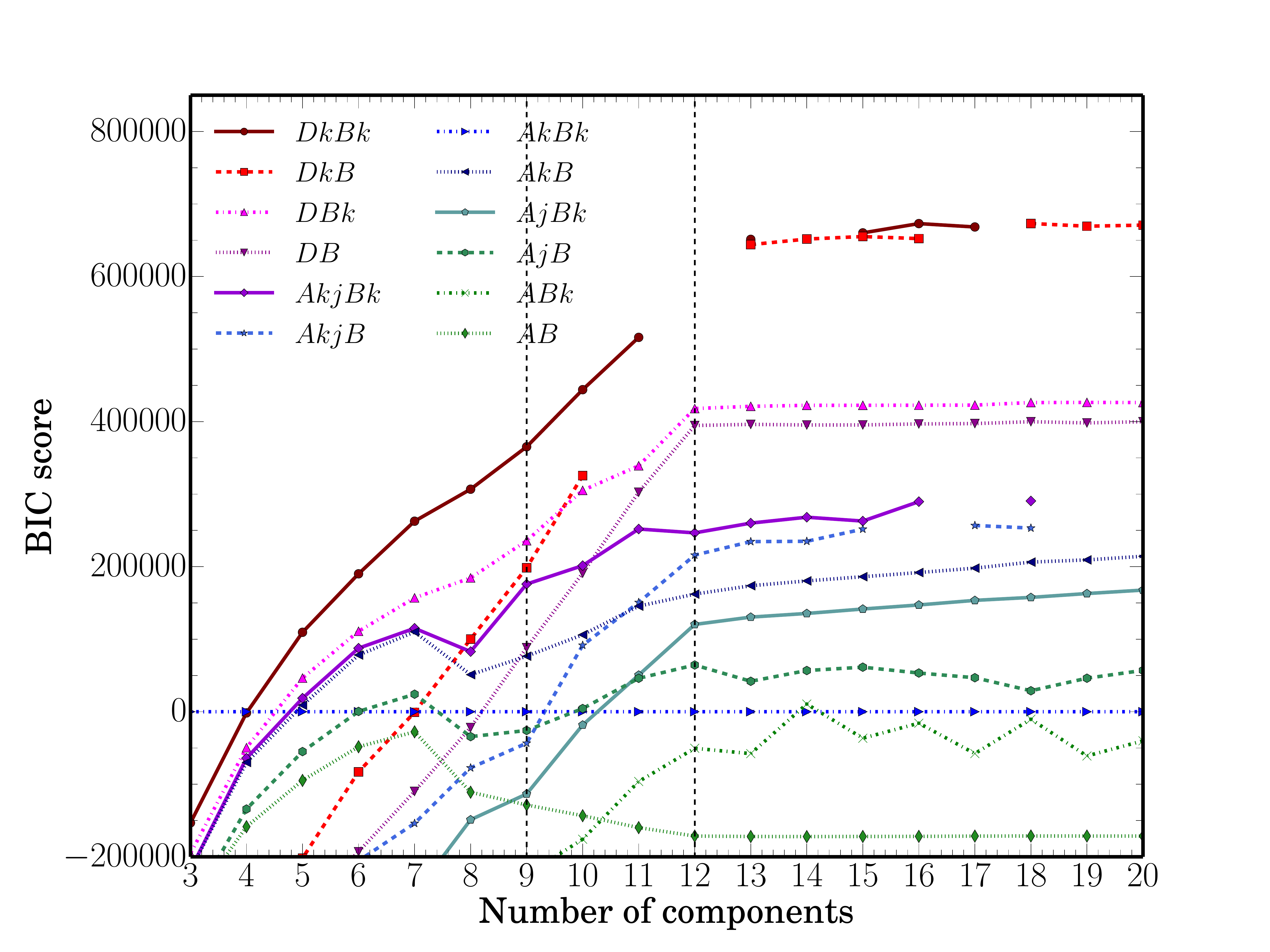}        
        \includegraphics[width=0.5\textwidth]{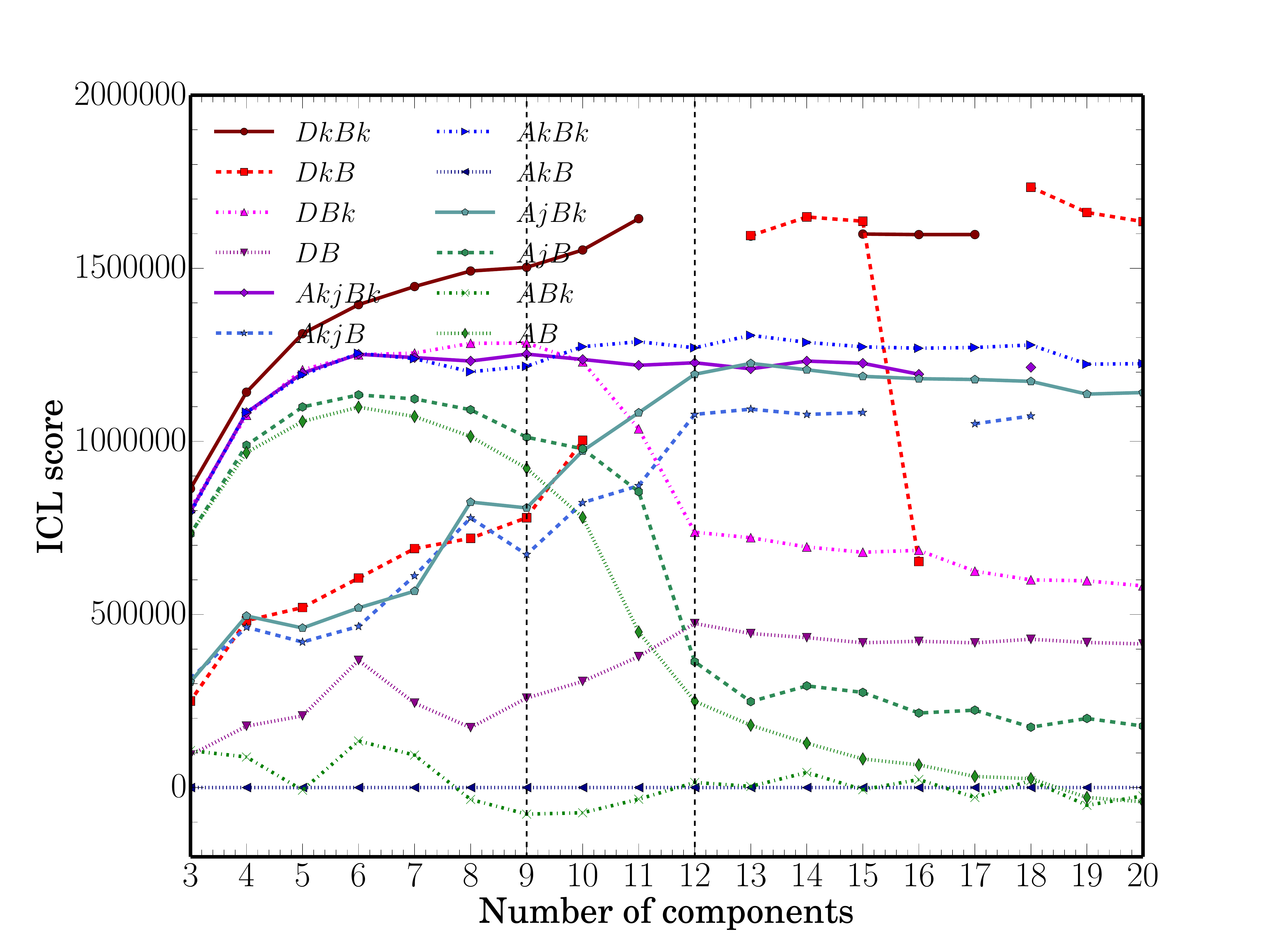}
        \caption{Results of model and number of classes  chosen based on the BIC and ICL criterion. Number of components corresponds to number of classes. }
        \label{fig:bic_icl_score}
\end{figure}

Each criterion (AIC, BIC and ICL) consists of two parts: the former increases the score for models with increasingly well separated groups, and the other is responsible
for penalising an excessive number of groups. 
Otherwise, the division which receives the highest score would be the one in which every object was a separate group \citep{Bou2011}. Based on the received scores the optimal number of groups can be chosen.

The change of BIC (AIC is not shown as it mostly gives the same
scoring as the BIC criterion) and ICL scoring for different numbers of classes and models is shown in Fig.~\ref{fig:bic_icl_score}. 
The algorithms were unable to converge (and therefore unable to
calculate the scoring) for some combinations of numbers of classes and
models, resulting in discontinuities in the DkBk and DkB curves. 
Although the DkB and DkBk models have higher AIC/BIC scores, we did not decide to use them. 
The DkBk model is unstable for more than 11 components, where it is expected to achieve a maximum (i.e. the most preferable number of classes) and therefore the optimum number of classes cannot be specified. 
The similar situation is observed for the DkB model, which is unstable for steps 11 and 12 and for this reason it is impossible to state its behaviour and find the maximum. 
Based on Fig.~\ref{fig:bic_icl_score} we can see that the AIC/BIC
scores for the DBk and DB models are very similar. 
However, when we consider the ICL scores, we see significant differences in favour of the DBk model. 
Therefore, we decided to use the DBk model which gives the best scoring among all the three criteria. 
For a detailed description of the DBk model see~\citealt{Bou2011}.
According to Fig.~\ref{fig:bic_icl_score}, the BIC scores increase
steadily as the number of classes is increased, up to a limit of 12
classes. The scores do not continue to increase for models with more
than 12 classes, and so we consider 12 classes to be the optimal choice.
The ICL criterion achieves its maximum score for
nine classes. The ICL scores decrease rapidly from step 9 to step 10,
and declines further to 12 classes. Above this number, the model score (as in case the of BIC criterion) practically does not change. 
Based on the analysis of the BIC, AIC, and ICL criteria, we conclude
that the optimum model is the DBk model with approximately nine
(according to the ICL criterion) or twelve (according to the AIC and
BIC criteria) classes. 

\begin{figure*}[h]
        \centering
                        \includegraphics[width=0.99\textwidth]{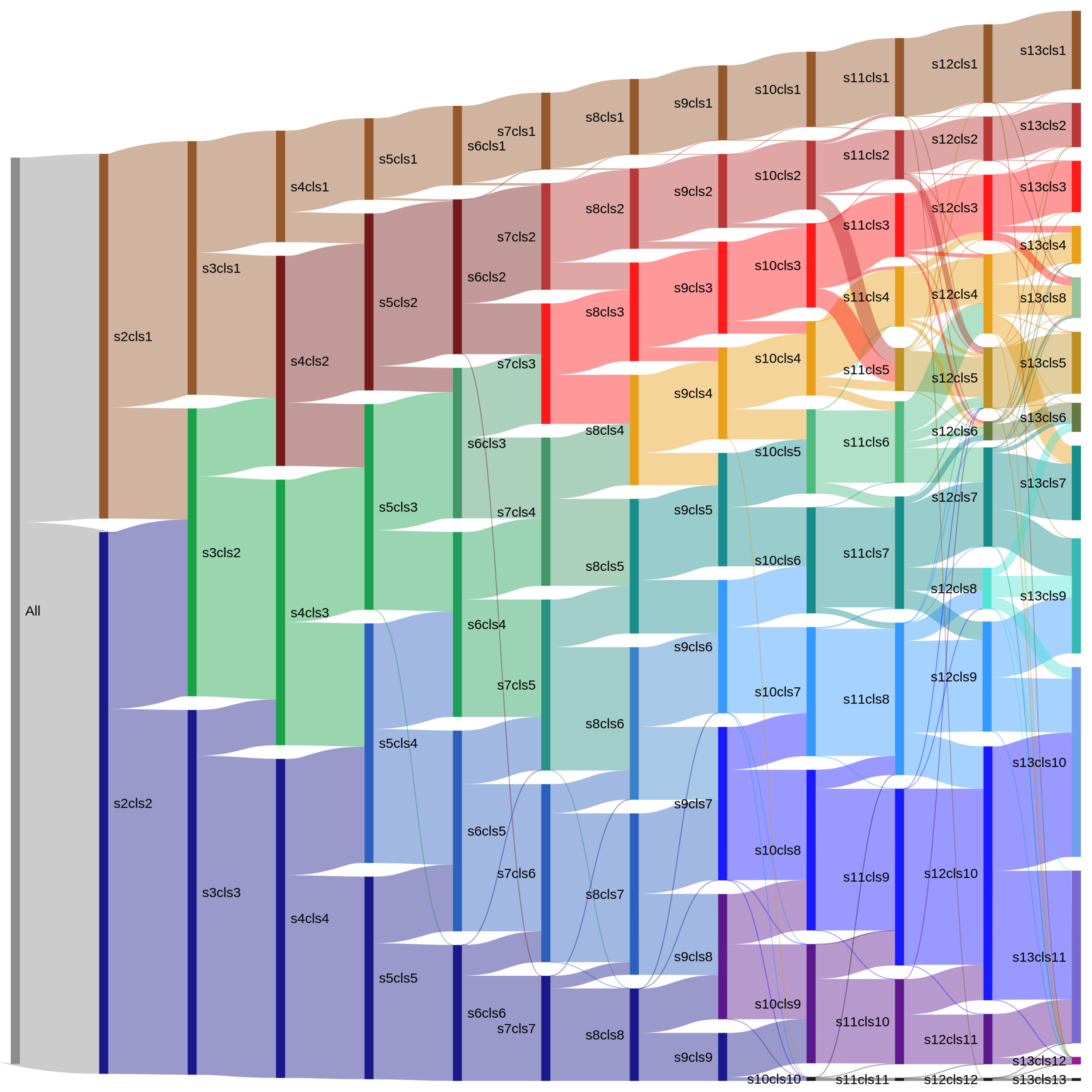}
                        \caption{The dependence of galaxy distribution
                          on the number of classification steps. The
                          galaxy flow beginning from one single class
                          up to the classification with thirteen classes is
                          shown. The step number (equal to the number
                          of classes) followed by the class number is given (s2cls1, where s2 corresponds to the second step, division into two classes, while cls1 corresponds to the first class). }
                        \label{fig:flow_2_14}

        \end{figure*}
        
In Fig.~\ref{fig:flow_2_14} we present the flow chart of 52,114 VIPERS
galaxies. This includes those objects with low probabilities of being
members of any class, which are less than 1\% of our sample at the
first step and less than 10\% of our sample at the twelfth step). These are visible as thin lines that correspond to several dozen objects, which
in all cases are a small percentage (3\%) of all galaxies in a given group. 
In order to select the optimal number of classes, the galaxy flow
between classifications was checked at each step, from one single class up to thirteen classes. 
Each step corresponds to the number of classes into which the VIPERS galaxies were classified (i.e. notation s2 corresponds to the second step and division into two classes).
Figure~\ref{fig:flow_2_14} shows that some classes (e.g. cls1, where 1 corresponds to the first class) are well separated in the very early steps (cls1 is basically unchanged for s4--13), while others (e.g. s12cls7) are formed from a mixture of galaxies from different classes from the previous steps. 
This is also mirrored by their posterior membership probabilities, as well-defined classes (e.g. s12cls1) have high membership probabilities, while "flowing" classes (e.g. s12cls7) are characterised by lower probabilities. 
The new classes distinguished between steps 10 and 12 (12cls6, 12cls8,
12cls12)  are characterised by relatively small numbers of galaxies
(see Fig.~\ref{fig:probability}) and tend to separate beside the main linear trend formed by s9cls1-s9cls9 (see the first panel in Fig.~\ref{fig:nuvr_9_10_11_12}). 
In particular, class 12cls12 is separated in step 10 (s10cls10), while
in steps 11 and 12 the classes containing dusty star-forming galaxies (s12cls5, s12cls6, and s12cls8) are distinguished. 
Based on the physical properties of these newly separated classes in
s11-12 (see Fig.~\ref{fig:nuvr_9_10_11_12}), we found each of them to be representative of distinct galaxy types, and therefore, we found 12~classes as the optimal number of galaxy types reflecting a full panoply of VIPERS dataset. 
We also verified that forcing the algorithm to separate one more
additional group (step 13) does not lead to formation of a well-defined (with respect to physical galaxy properties, see Fig.~\ref{fig:nuvrrk_13}) new class (s13cls12), which emerges from 12cls11, however do not reveal distinct properties at least in its colours.    

\begin{figure}
        \includegraphics[width=0.5\textwidth]{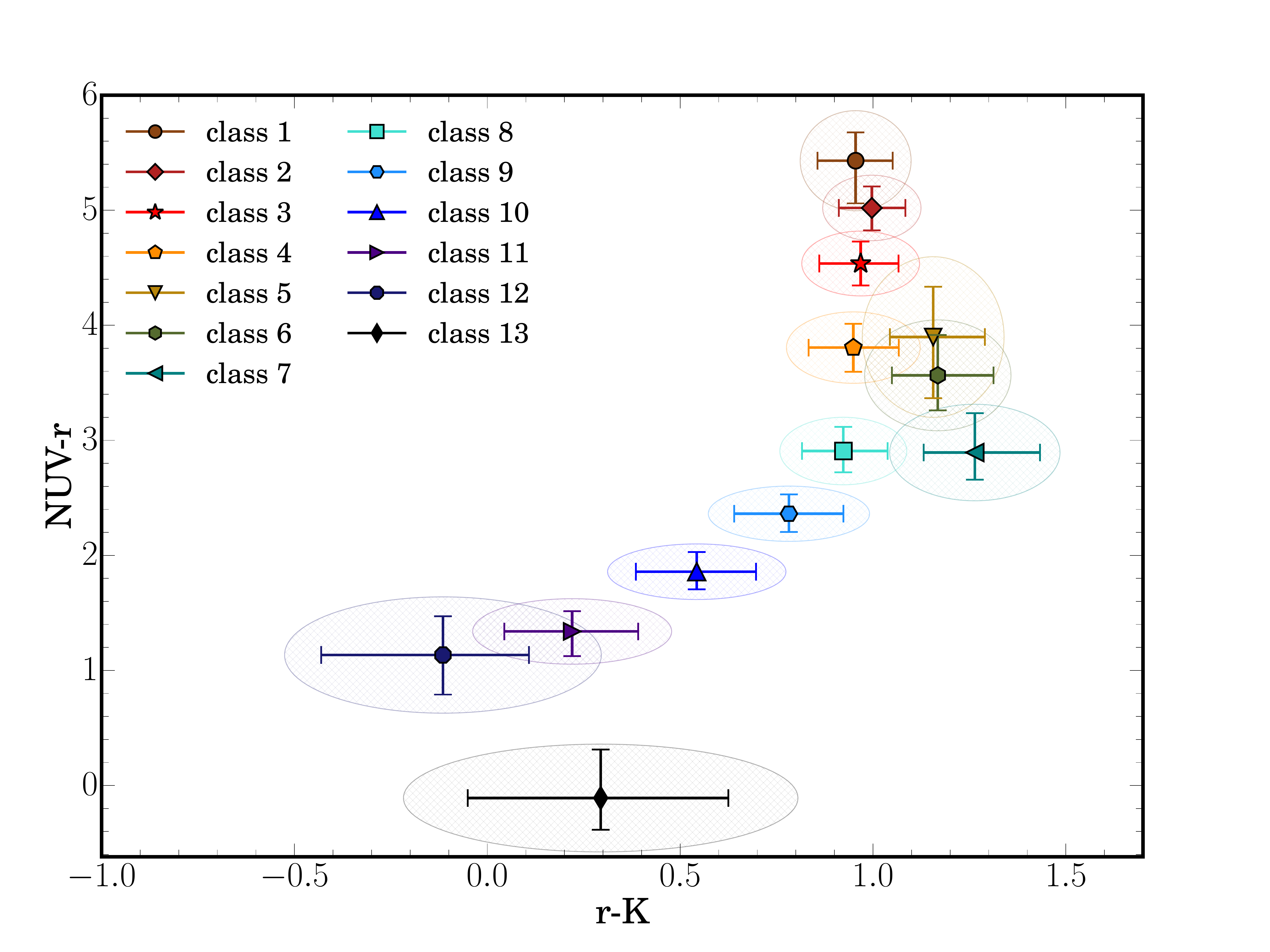}        

        \caption{$NUVrK$ diagrams of FEM classes 1--13. The error bars correspond to the first and the third quartile of the galaxy colour distribution, while the two half axes of the ellipses correspond to the median absolute deviation }
        \label{fig:nuvrrk_13}
\end{figure}

A detailed description of the diagram ~\ref{fig:flow_2_14} will be discussed in \citealp{Krakowski2018}.

\section{Class membership probabilities}\label{app:A}
The FEM algorithm assigns each VIPERS galaxy to one of the 12 classes. 
However, the classification does not provide sharp borders between classes. 
The class membership probability is based on the distance of an object
from the centre of the class in the multidimensional feature space.
Therefore, each galaxy is characterised by the posterior probability of it belonging to a particular class. 
This has allowed us to quantify the number of galaxies with
problematic classifications that could belong to two or more classes with roughly similar probabilities. 

In our case, the classes found by the FEM algorithm are well defined,
as the majority of their members are well separated from the
neighbouring    classes. This is reflected by the majority of galaxies having a high probability of being a member of the class to which they are assigned. 
Therefore, almost all galaxies ($94\%$) have high probabilities ($>50\%$) of belonging to their assigned class. 
The remaining $6\%$ are outliers, which do not fall into any class. 
       
Among the $94\%$ of "well classified" galaxies (with $>50\%$
probability of class membership) $2\%$ also have $> 45\%$ probability of belonging to the second-best class, implying that those galaxies are midway between two classes. 
Although the numbers of sources at the borders of the classes (1,038)
and other outliers (2,947) is very small (in total $8\%$ of the sample), we exclude them from the final set used to analyse properties of the FEM classes.
          
The distributions of class membership probabilities for each class are presented in Fig.~\ref{fig:probability} (marked with dark blue). 
The probability distribution for the final sample, including only
objects with a high first-best probability ($>50\%$) and a low second-best probability ($<45\%$) is marked with light blue. 
Following~\cite{Krakowski16} and~\cite{Kurcz16}, the influence of increasingly severe probability cuts on the quality of the estimated global properties of each class is checked. 
No significant improvement in purity or derived properties was found when adopting more severe cuts. 
Even for the purest sample, including only galaxies with class
membership probabilities higher than $80\%$,  the global properties of
each class (reaching the highest deviation for colours, but not
greater than 0.3$\sigma$) remain in a broad agreement with those
obtained when less severe cuts are applied. 
As more severe cuts (with class membership probabilities higher than $80\%$) do not change our results, but reduce significantly the number of objects (to $56\%$ of the sample), in this paper, we applied less severe cuts (high first-best and low second-best class membership probabilities).

        \begin{figure*}[]
                \centering
                \centering
                \includegraphics[width=0.99\textwidth]{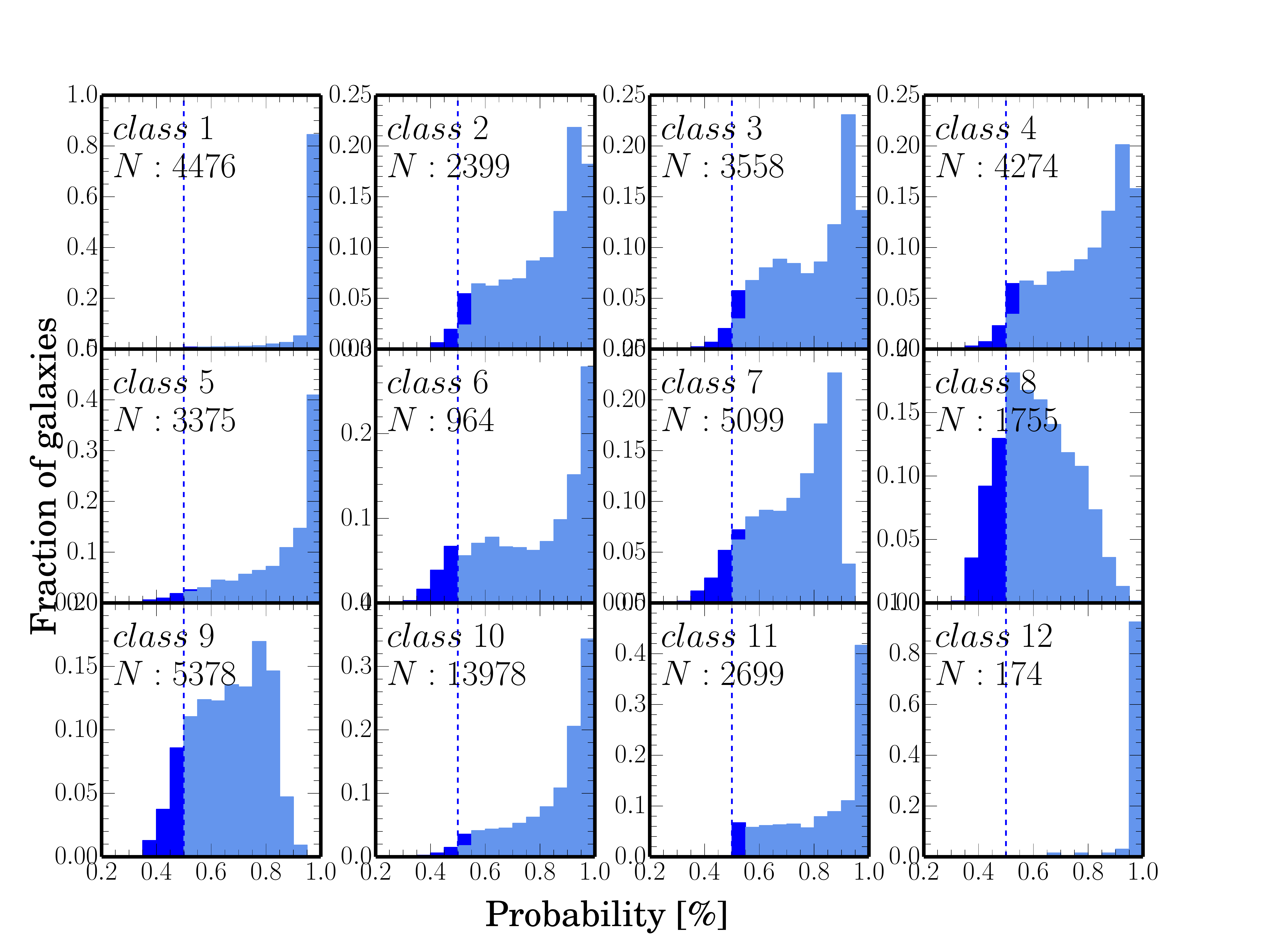}
                \caption{Distribution of the class membership probabilities for all sources (dark blue), and for the final sample of galaxies (with first-best probability $>50\%$ and second-best $<45\%$; in light blue) for each FEM class. 
                        The threshold ($50\%$) used to remove outliers
                        in the subsequent analysis  is marked with a blue dashed line. 
                        The final number of class members is given in each panel.}
                \label{fig:probability}
        \end{figure*}

\section{Comparison of the FEM-based and PCA-based classifications of the VIPERS galaxies}\label{app:pca}

        To date galaxy classification has been mostly based on their
        colours as determined from broad-band photometry~\citep[e.g.][]{Bell2004,fritz14,siudek17} or from their spectra~\citep[e.g.][]{balogh,marchetti13,souza}. 
        One of the most common methods used to distinguish different galaxy populations from their spectra is the PCA method. 
        In this method, each spectrum is decomposed into a set of representative templates, which reproduce the most important galaxy features (spectral slope and strong emission lines). 
        This transformation is characterised by orthogonal vectors (eigenvectors), which describe the original spectra. 
        \cite{marchetti13} have used the first three eigenvalues
        ($a1$, $a2$, $a3$), which have the highest importance in
        effectively representing the data, and provide an optimal input for spectroscopic classification. 
        The parameter space was further reduced to Karhunen-Lo\'{e}ve angles~\citep[$\theta$,$\phi$;][]{karhunen1947,Connolly1995}.
        The spread of galaxies on the  $\theta-\phi$ plane allows us to identify different galaxy types, as redder galaxies have smaller  $\theta$, and $\phi$ values, while bluer galaxies  are characterised by higher  values~\citep{marchetti13}.  
        In order to verify how our FEM classification based on galaxy
        colours is relevant to spectroscopic galaxy classification we
        investigate the eigencoefficients for FEM classes and compare the FEM-based and the PCA-based classes.  
                        
        The distribution of galaxies belonging to the different FEM classes in the $\theta-\phi$ diagram is shown in Fig.~\ref{fig:pca_z}. 
        The values of $\theta$ and $\phi$ were obtained by~\cite{marchetti13}, who made a classification of the spectra of the VIPERS galaxies from the PDR-1 making use of the Kinney--Calzetti templates~\citep{calzetti94,kinney96}, which we also use for comparison (see the last panel in Fig.~\ref{fig:pca_z}).
        
        We examine the FEM classification  of the PCA-based galaxy
        types from early types (E) at the bottom, through the
        star-forming (Sa and Sb) populations, and up to the Sc galaxies in the top of the diagram.
        Lower-$z$ red passive galaxies (classes 1--3) are tightly
        located in the bottom edge of the $\theta-\phi$ diagram, in the
        locus of the early-type family of galaxies ($\theta<1.6$ and $\phi$<0). 
        With increasing redshift, the galaxies of classes 1--3 tend to
        move to the  region of Sa, almost reaching the area where the Sb6 are placed. 
        This shows how passive galaxies evolve with cosmic time, showing a hint of star formation activity in earlier epochs and quenching with cosmic time. 
        Green intermediate galaxies (classes 4--6) tend to have a
        similar distribution to sources within classes 1--3, although
        they show a larger scatter, especially in $\phi$, revealing
        extended star formation activity. This may be attributed to them being in the intermediate stage between red passive and blue star-forming galaxies.
        The star-forming galaxies (classes 7--11) follow a sequence on
        the $\theta-\phi$ diagram ending with galaxies assigned to
        class 11 being distributed in the top-left corner.
        This area is dominated by the templates of galaxies which are actively forming new stars (Sb, and Sc).
        This is especially significant for classes 10 and 11, since galaxies form a tail beginning in the area of Sb and Sc templates with a sharp cut in redshift. 
        
        The presented good correlations with eigenvalues show that we
        can use FEM classification based on rest-frame magnitudes when galaxy spectra of good quality are not available. 
        
        \begin{figure*}[]
                \centering{     
                        \includegraphics[width=0.8\textwidth]{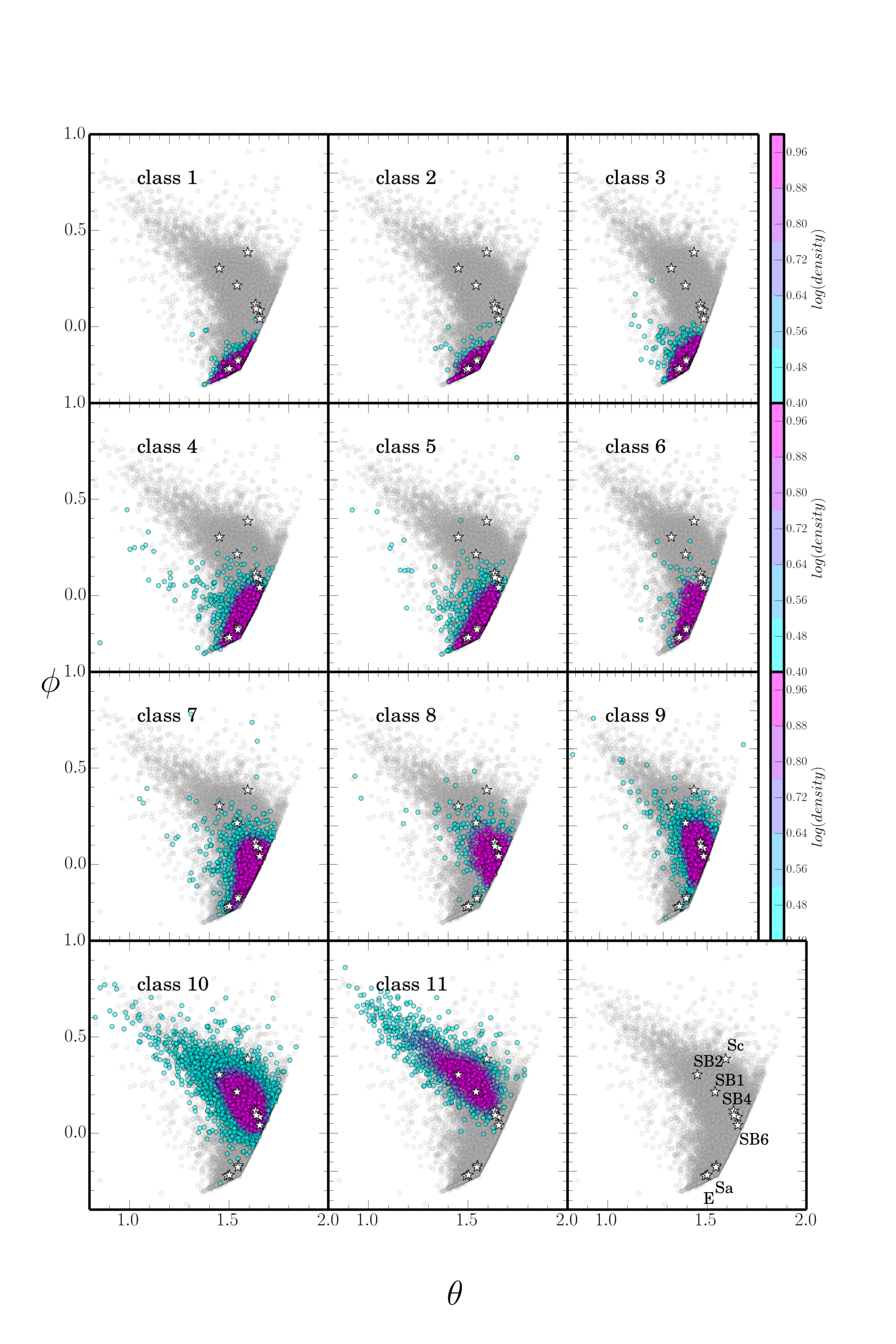}}                                                          
                \caption{PCA components derived from VIPERS spectra
                  by~\cite{marchetti13} for the 11 FEM classes. VIPERS
                  PDR1 galaxies are marked with grey dots.   The
                  distribution of 2688, 1428, 2185, 2674, 2201, 598,
                  3300, 1096, 3630, 7885, 1395 eigen coefficients for
                  the 11 FEM classes are colour-coded according to their redshift values.  The location of the Kinney--Calzetti spectra~\citep{calzetti94,kinney96} for different galaxy families are marked with white stars following~\cite{marchetti13}.}
                \label{fig:pca_z}
        \end{figure*}      
        
        \section{Relation between FEM classes and Hubble types}\label{app:kennicutt}
        
        One of the methods to classify galaxy optical spectra is to examine their structures and compare them to the Hubble sequence, or its variations. 
        Spectral types may be derived through spectral features or SED fitting~\citep[e.g.][]{SDSS_class_2013, conselice2011}.
        In this paper, we adopt the approach
        of~\cite{SDSS_class_2013}, and compare optical composite
        spectra of the 11 FEM classes with the spectral types defined in the Atlas of~\cite{kennicutt92}.
        To fit the observed spectra against models, we first rest-framed the spectra, downgraded their resolution to $14\AA$~\cite[corresponding to the typical resolution of VIPERS spectra as shown by][]{siudek17}, and normalised the template spectra over the wavelength range $3800 < \lambda(\AA)< 5500$.
        The observed spectra were stacked within each FEM class in redshift range $0.5 < z < 0.6$ and normalised in the same wavelength range as the templates. 
        For each of the 11 representative stacked spectra we performed a $\chi^2$ minimisation over the entire Kennicutt Atlas. 
        Based on the $\chi^2$ we determined the best template for each FEM class.
           
        \begin{figure*}[]
                \includegraphics[width=0.49\textwidth]{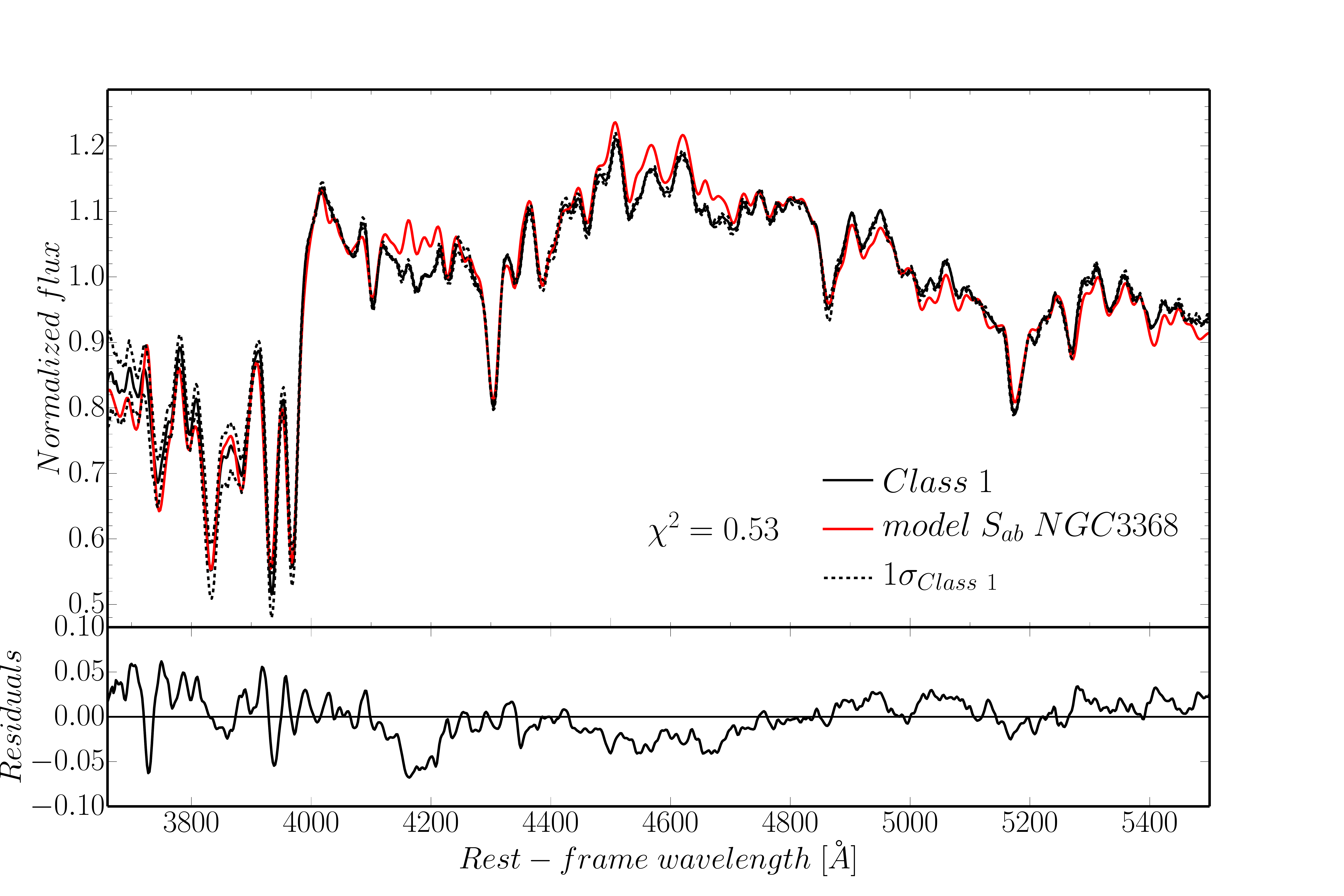}
                \includegraphics[width=0.49\textwidth]{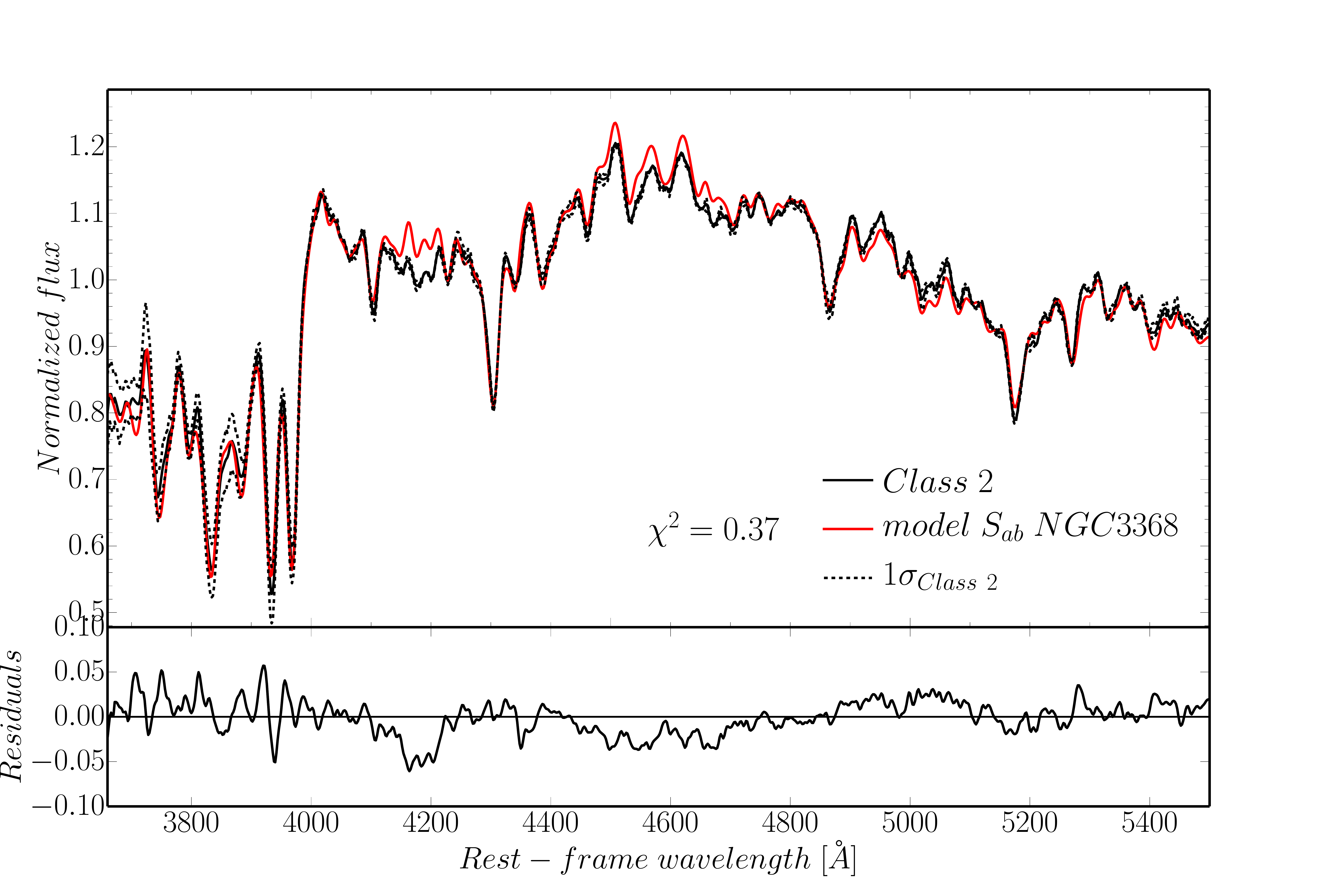}
                \includegraphics[width=0.49\textwidth]{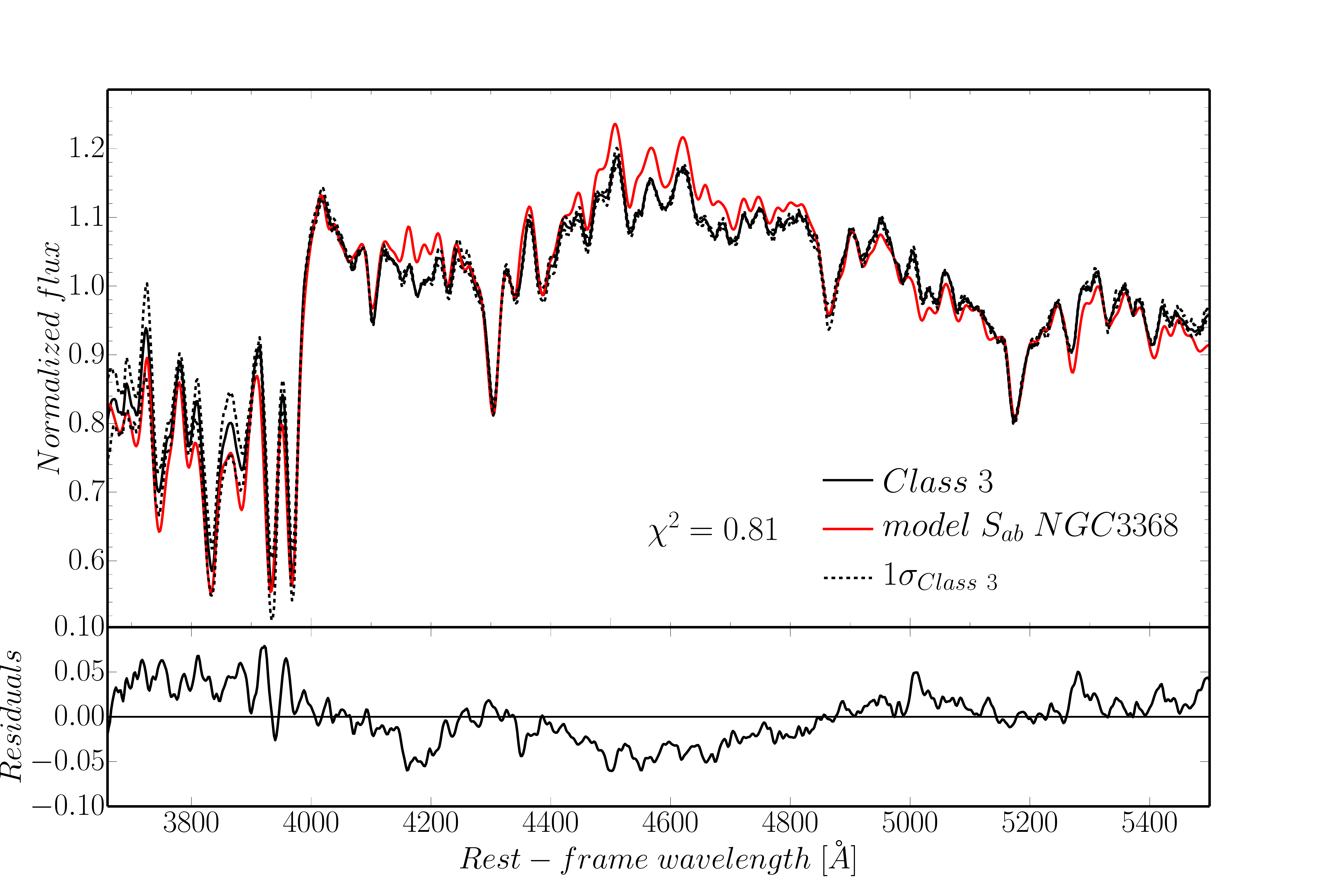}
                \includegraphics[width=0.49\textwidth]{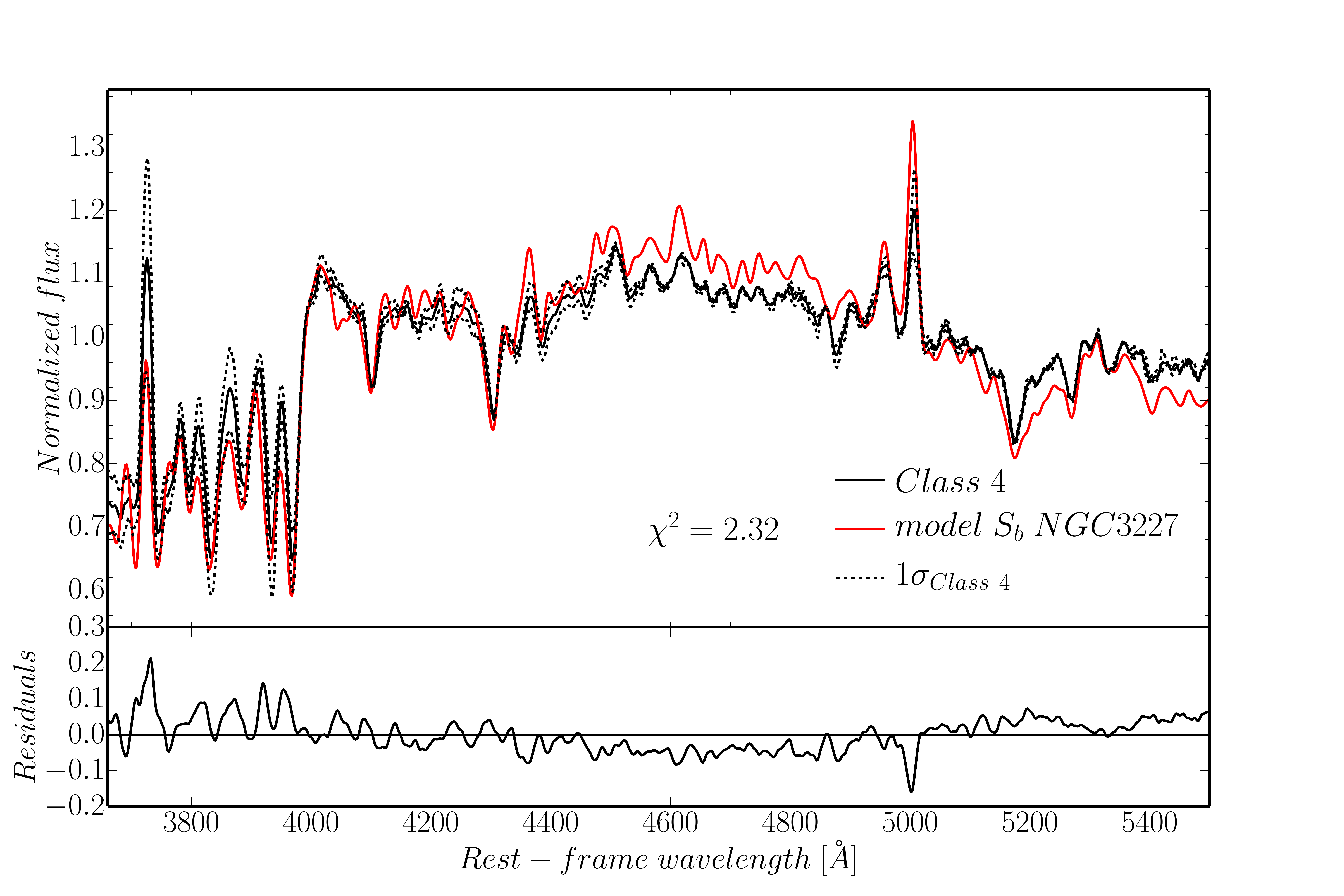}
                \includegraphics[width=0.49\textwidth]{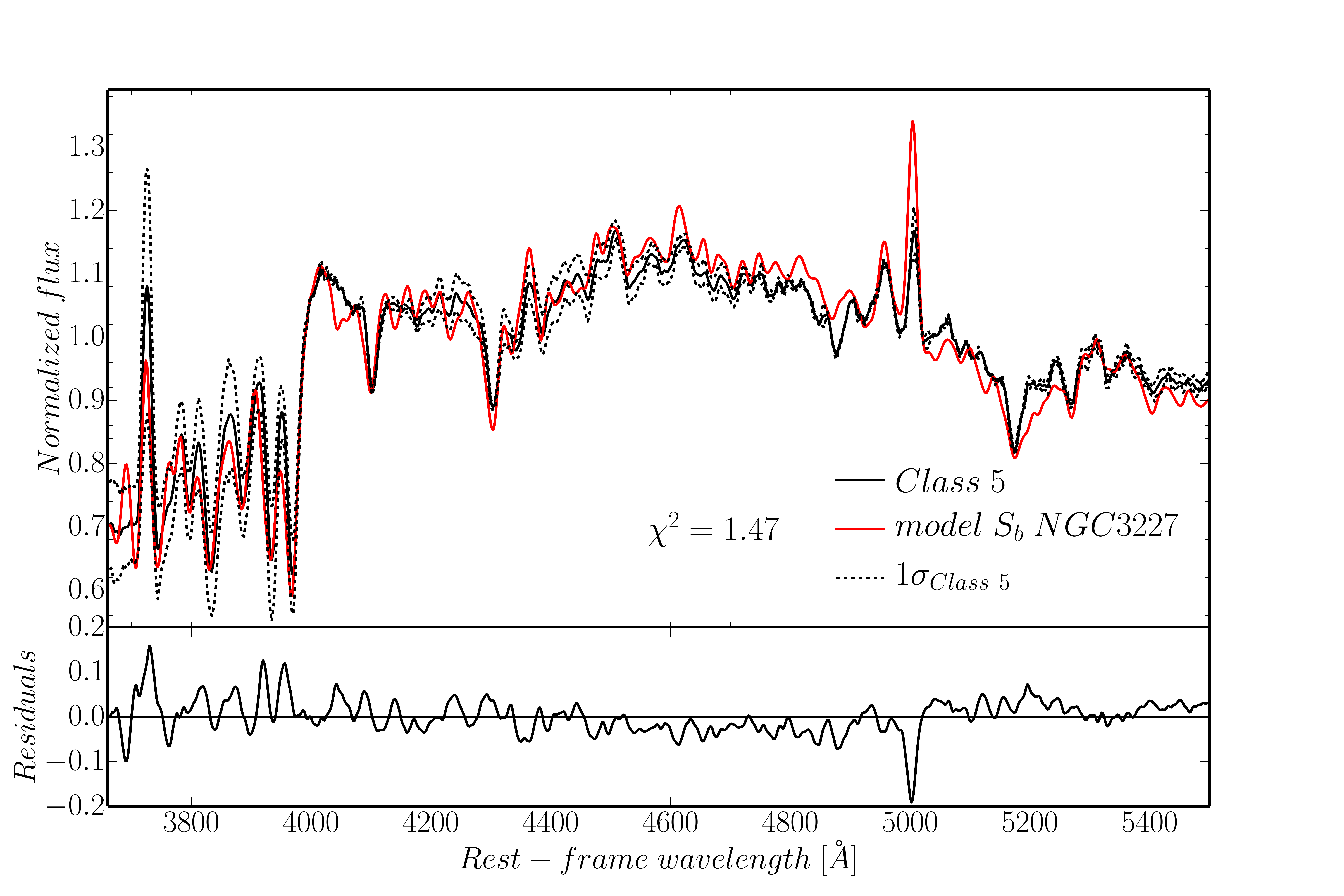}
                \includegraphics[width=0.49\textwidth]{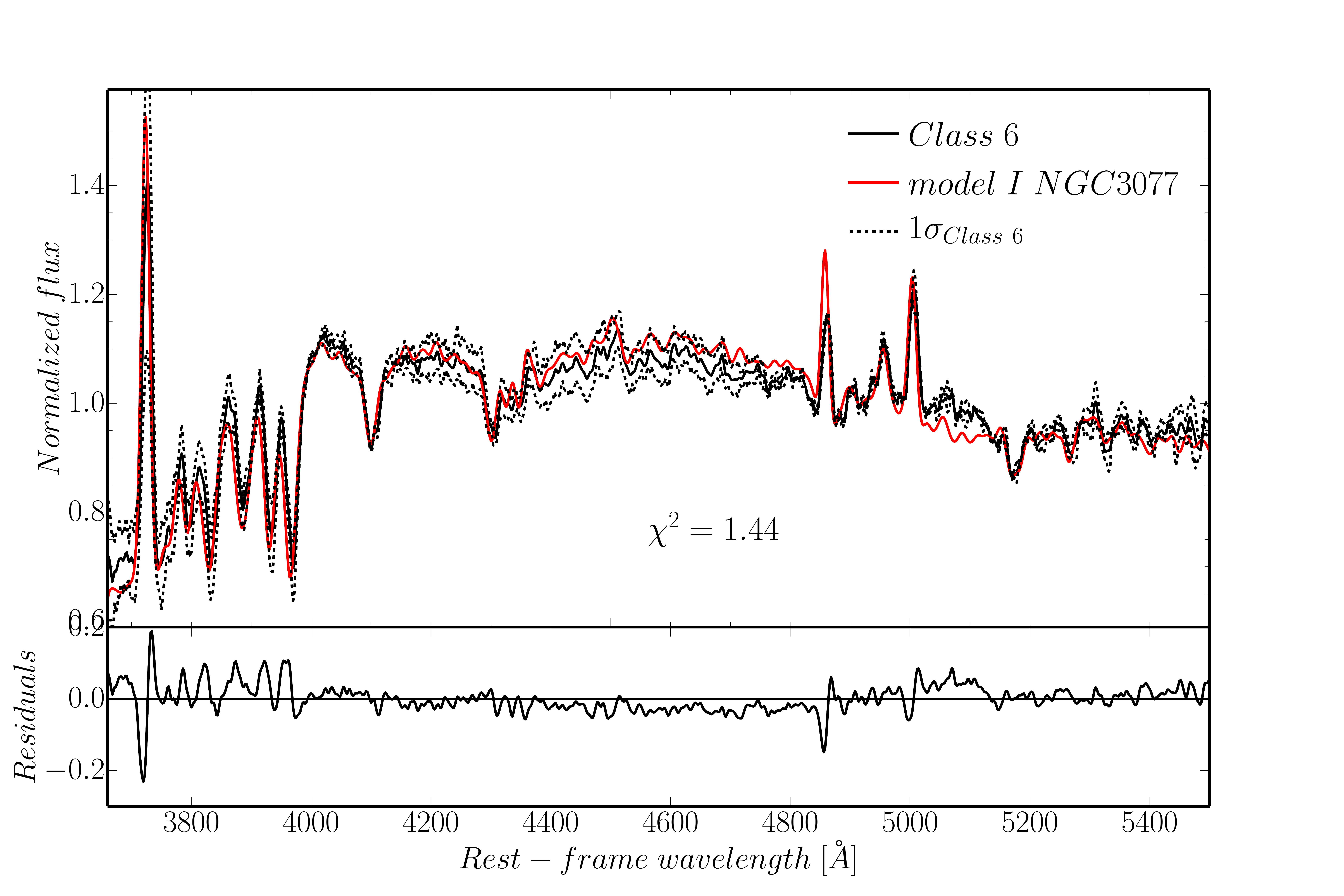}
                \includegraphics[width=0.49\textwidth]{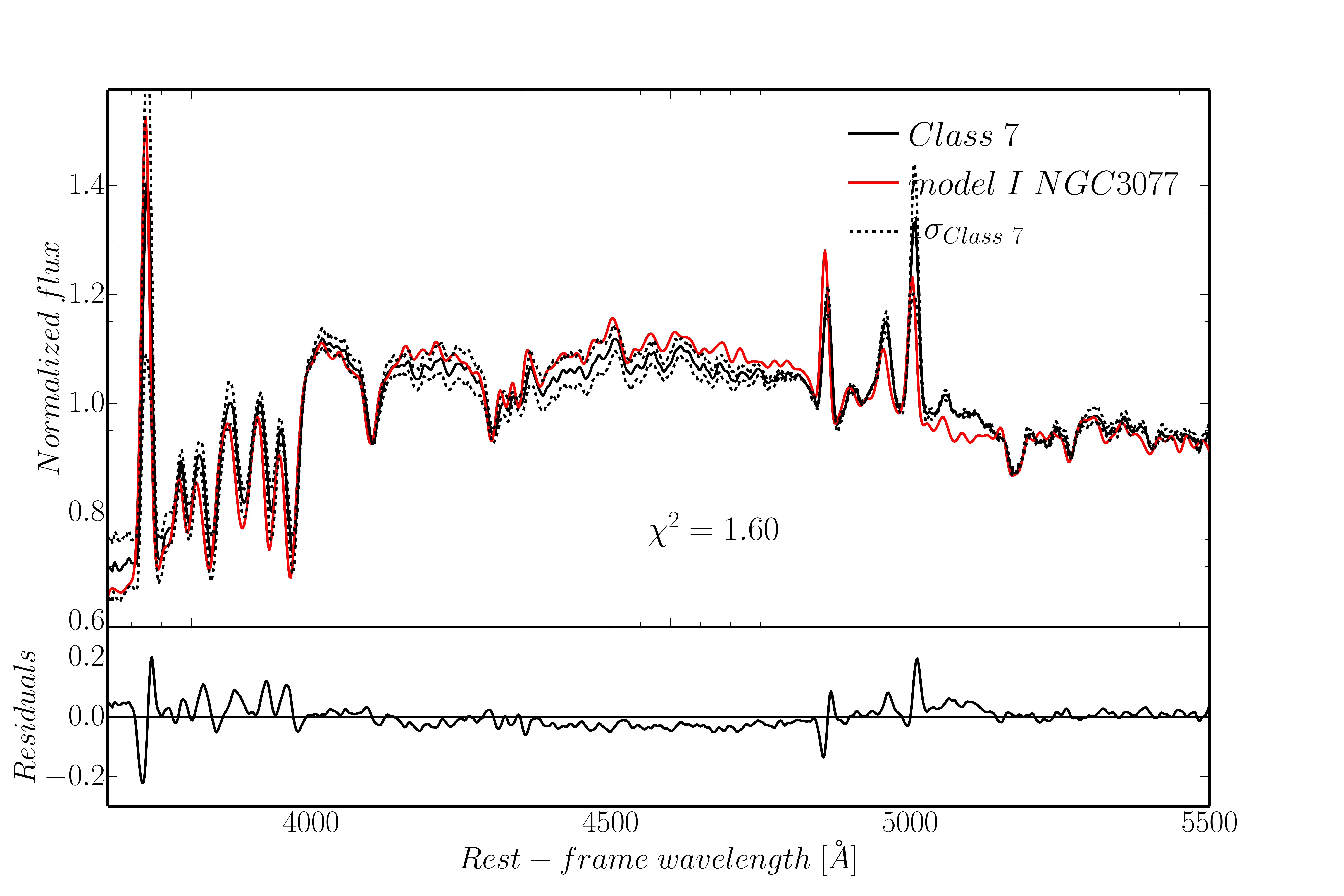}
                \includegraphics[width=0.49\textwidth]{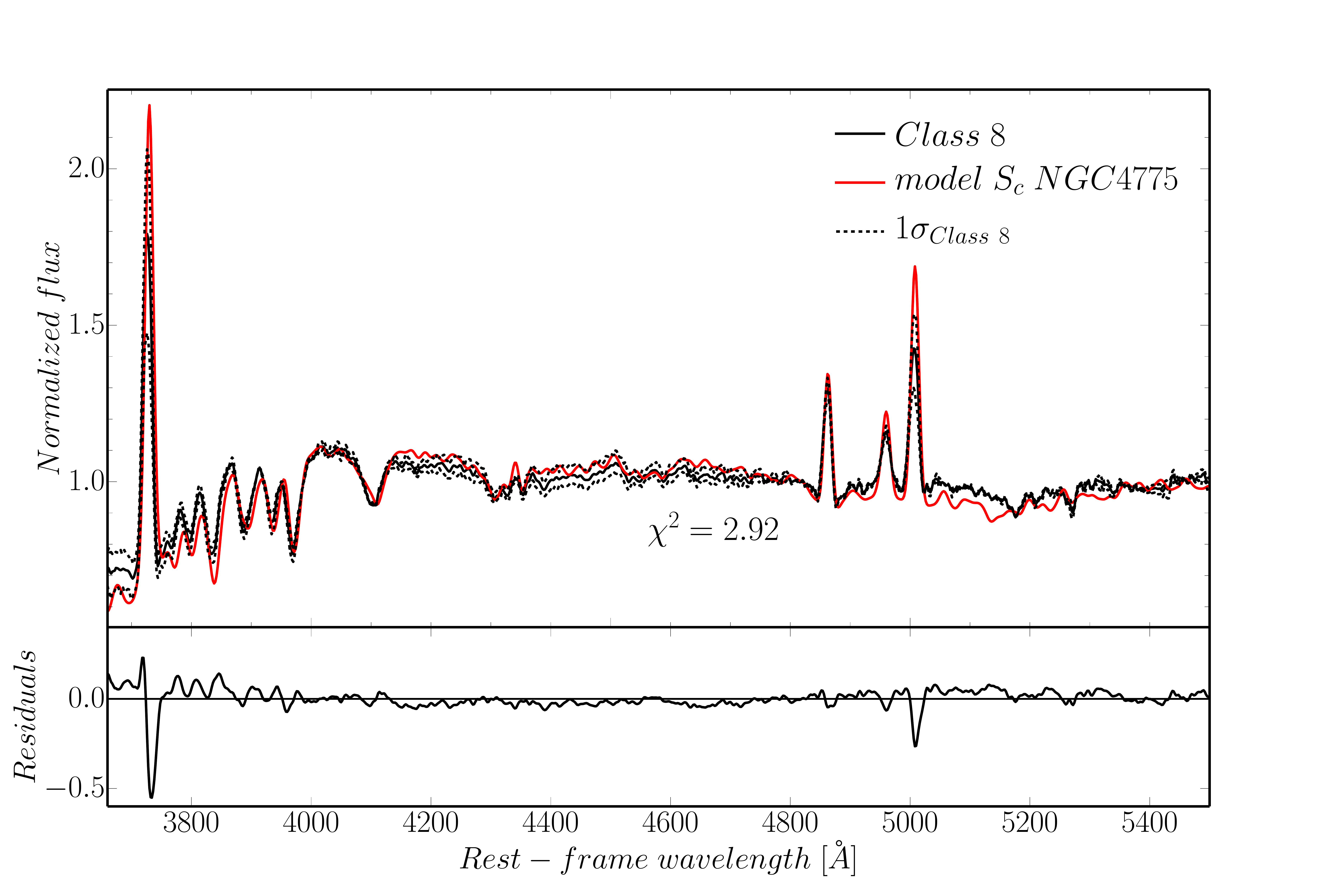}
                \caption{The comparison of the 11 FEM stacked spectra
                  in redshift bin $0.5< z < 0.6$ with the best-fit
                  template spectra from the spectral Atlas of~\cite{kennicutt92}. }
                \label{fig:atlas}
        \end{figure*}
        
        \begin{figure*}[]               
                
                \includegraphics[width=0.49\textwidth]{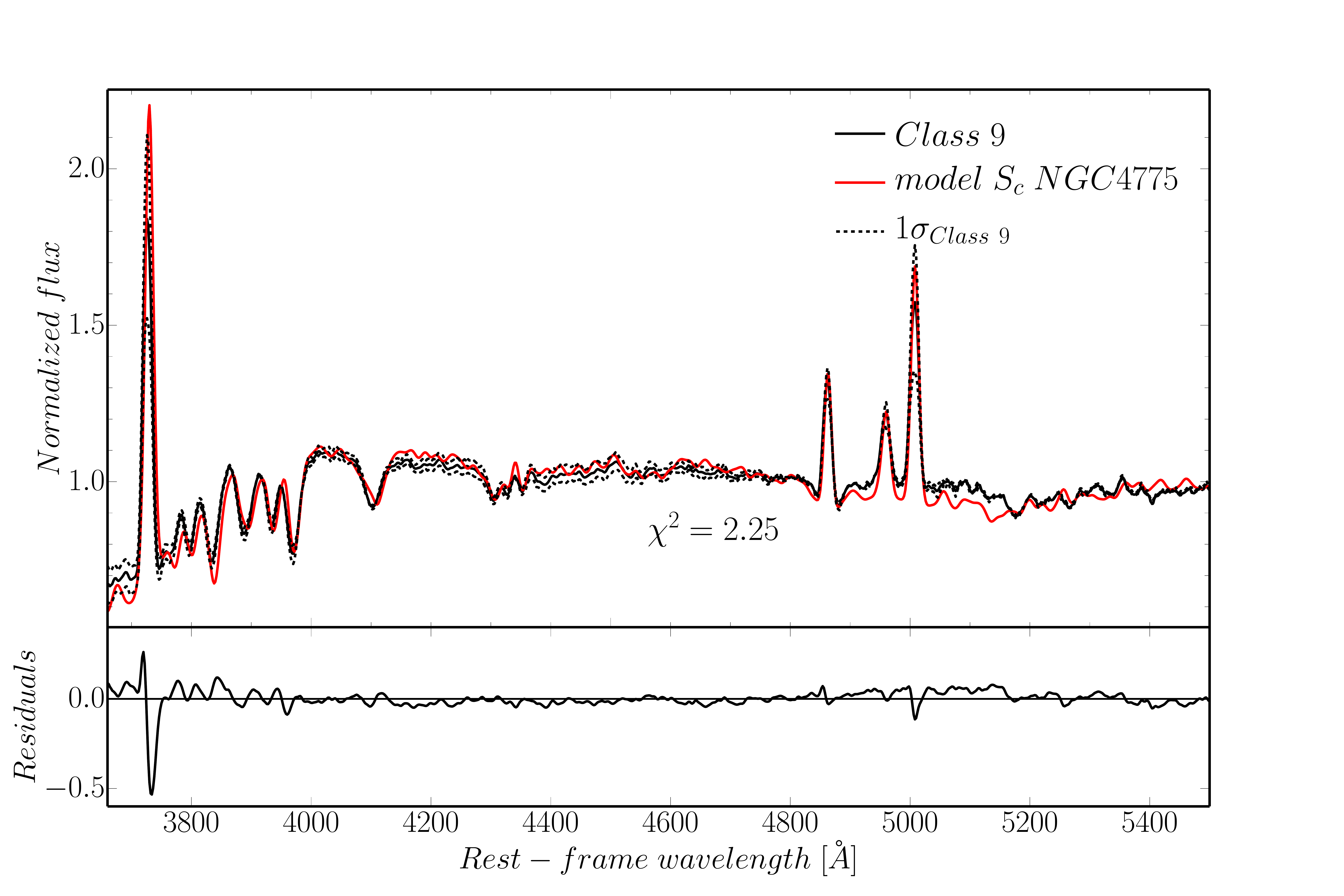} 
                \includegraphics[width=0.49\textwidth]{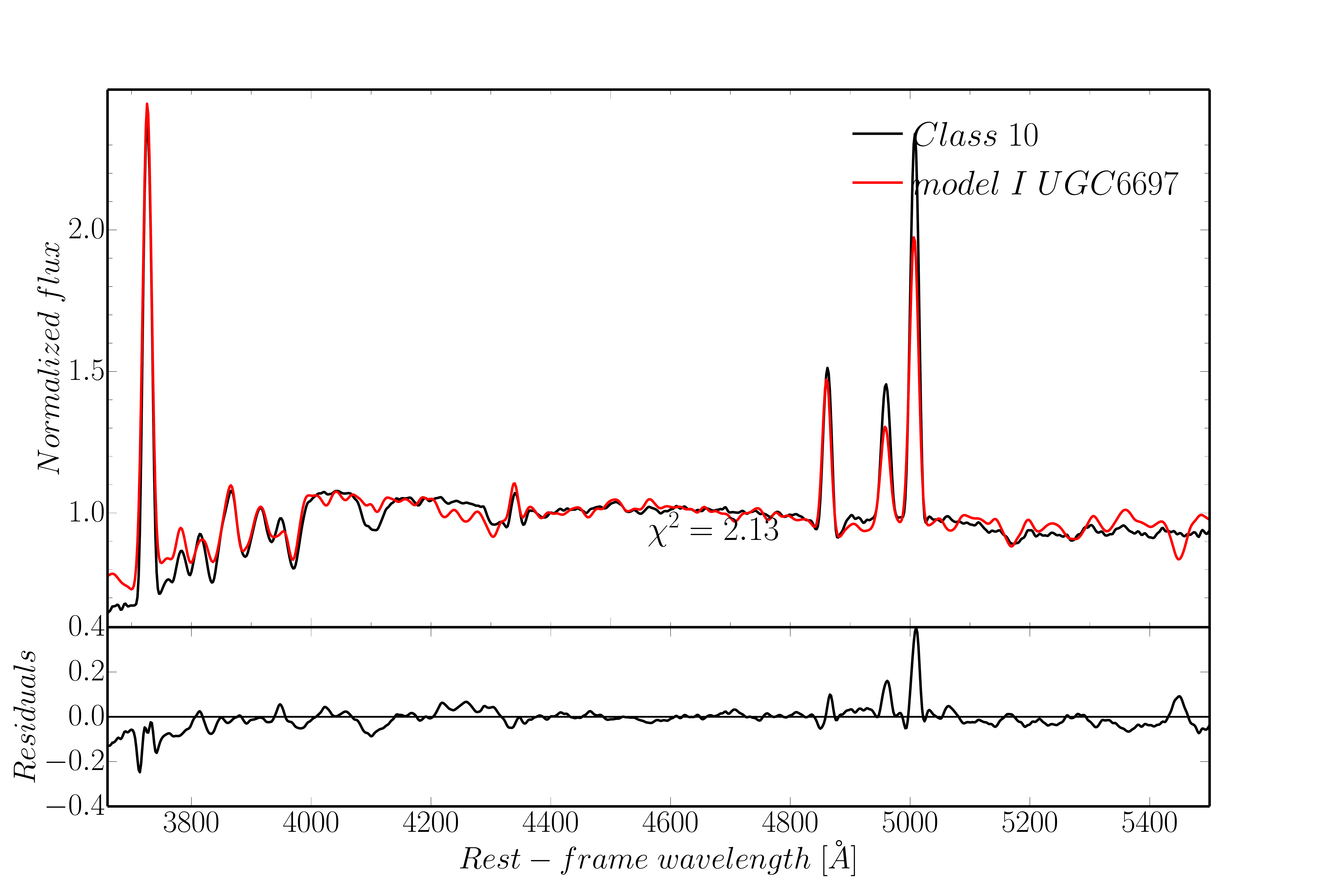} \includegraphics[width=0.49\textwidth]{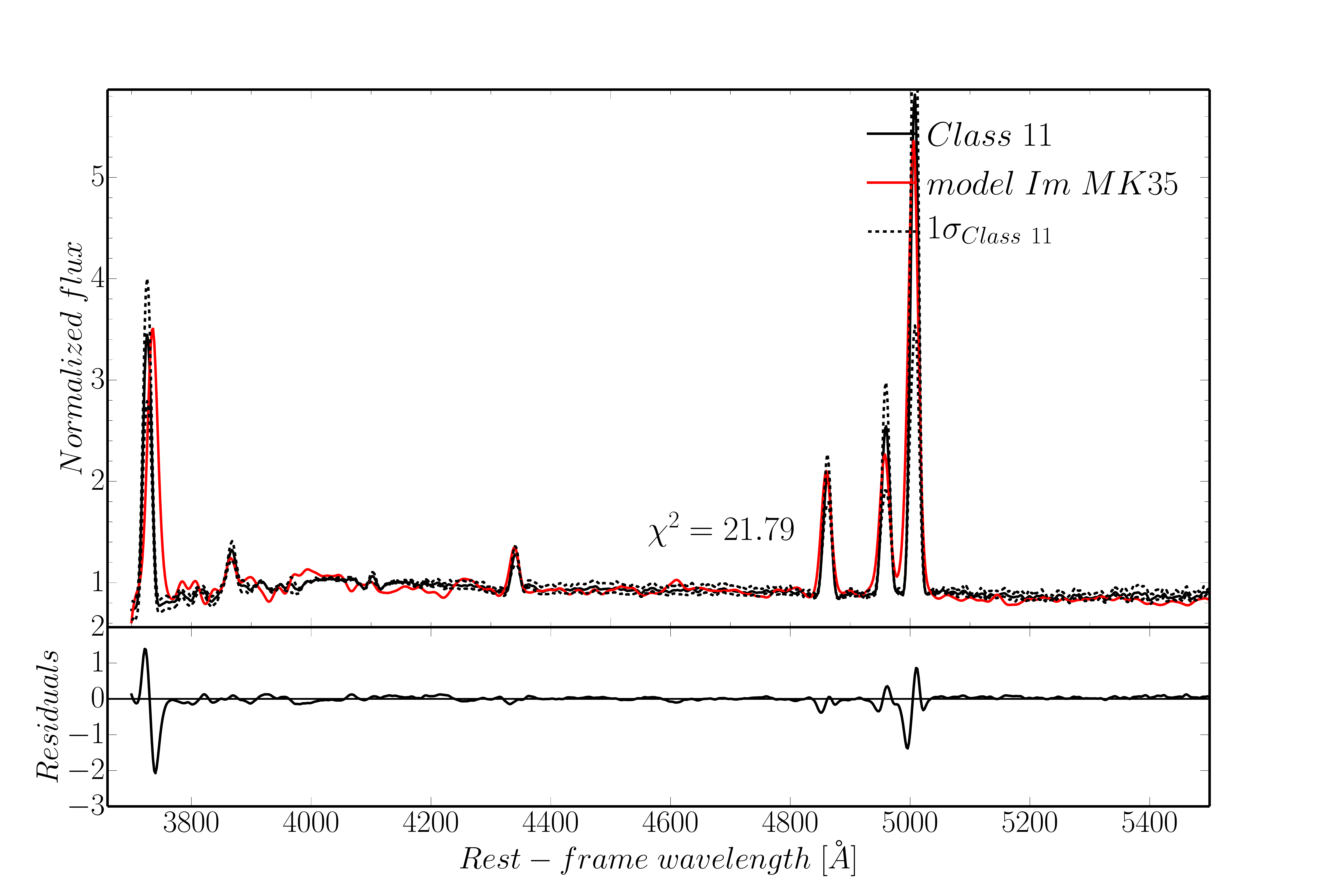}   
                \includegraphics[width=0.49\textwidth]{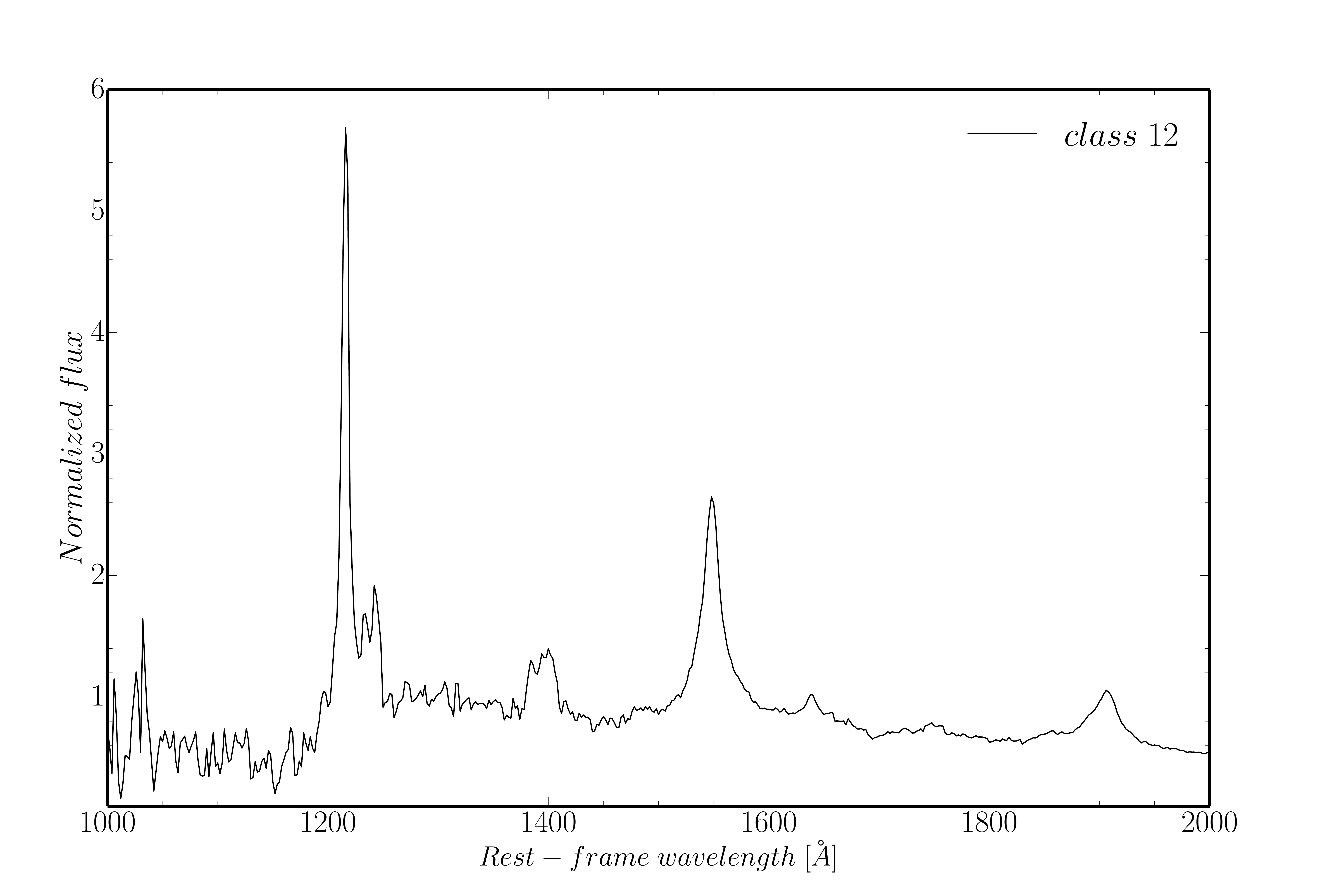}                                                                  
                \caption{Continuation of Fig.~\ref{fig:atlas}. The
                  composite spectrum of galaxies within the
                    12th class is built using only objects  observed
                  at $z\sim2$. The wavelength range of spectra in the spectral atlas of~\cite{kennicutt92} does not cover the observed wavelength range. }
                \label{fig:atlas2}
        \end{figure*}
        
        Figures~\ref{fig:atlas} and~\ref{fig:atlas2} show the 11 FEM representative stacked spectra of VIPERS galaxies observed in the redshift range of $0.4 < z < 1.3$ (marked with the black solid line). 
        The $1\sigma$ stacked spectrum is marked with a dashed line, and the best template is over-plotted in red.   
        Red passive galaxies (classes 1--3) show no difference with respect to their spectral type, since for each representative spectrum the best spectral template corresponds to the Sab type of $NGC$ $3368$ galaxy from the Kennicutt Atlas (see the first three panels in Fig.~\ref{fig:atlas}).  
        The models fit the observed spectra remarkably well,
        especially the main features that can be attributed to old stellar populations, including $D4000$, $H\delta$, and $G-band$ showing no difference with respect to the template spectrum.  
        The family of Sab galaxies is dominated by evolved giant stars, but there can also be found a contribution from younger stellar populations~\citep{kennicutt92}.
        When it comes to their visual appearance, the Sa group contains non-barred galaxies, whereas Sb is assigned to barred galaxies, and Sab refers to intermediate sources, showing weak signs of a bar~\cite[an oval one in the case of $NGC3368$ according to][]{kennicutt92}.
        However, $NGC3368$ is also classified as unbarred Sa  according to other authors~\citep{sandage, kormendy04}.  
        It should be noted that the VIPERS stacked spectra of red
        classes (1--3) are very well fitted by the templates of E galaxies (especially with E1 $NGC3379$).
        Among the best-fitting templates for stacked spectra of classes 1--3, types E/S0/Sa dominate. 
        The template spectra of elliptical galaxies do not show strong differences with respect to spectra of   Sab/Sa/S0 galaxies, as all of them are dominated by strong absorption lines typical for red galaxies with old stellar populations.
        Those templates present strong $4000\AA$ breaks and strong absorption lines (G-band, $H\delta$ line) typical for old stellar populations. 
        The best template was assigned to Sab mainly because of the
        best fit to the $[OII]\lambda3727$ line, which is not seen in absorption as in elliptical galaxies according to spectroscopic Atlas of~\cite{kennicutt92}.   
        However, the spectrum of this intermediate family appears to have features typical for old stellar populations (i.e. strong $4000\AA$ break and strong absorption line $[H\delta]\lambda4102$). 
        The stacked spectra of green galaxies assigned to class 4 and 5 are best fitted with  the  template of the spiral galaxy Sb $NGC3327$.
        The classification of $NGC3227$ galaxy is tentative, as it is classified as a peculiar object within a family of Sab sources according to~\cite{Vaucouleurs91}. 
        \cite{kennicutt92} has assigned this template as the spectrum of strongly interacting/merging galaxy with a Seyfert 2 nucleus. 
        Although the BPT diagram (see Fig.~\ref{fig:bpt}) does not confirm that these galaxies are Seyfert galaxies, we do not exclude the possibility of AGNs belonging to those groups. 
        Although the spectra do not present any broad Balmer lines indicating clearly the presence of active nuclei, 
        the high-excitation emission-lines combined with red continuum can be interpreted as a signature of AGN~\citep{kennicutt92}. 
       
        The irregular galaxy $NGC3077$ template shows the best fit to the representative stacked spectrum of galaxies in classes~6 and~7. 
        Although some of the templates of Sb and Sc families show
        comparable values of $\chi^2$, neither of them are able to fit
        well the $H\beta$, $[OIII]$ lines and the $4000\AA$ break at the same time. 
        Irregular galaxies do not fall into a regular classification
        showing some unusual features, mainly involving peculiar
        asymmetries or shapes, however the spectrum of $NGC3077$ is not distinguishable from a spectrum of normal Sc galaxy based only on the spectrum~\citep{kennicutt92}.  
        Interestingly, according to~\cite{buta15}, this galaxy appears as an early-type galaxy in $3.6\mu m$ images with a very scattered dust distribution. 
        The spectrum is very similar to the spectra of the blue "E+A" galaxies, except of the presence of the $[NII]$ and [OII] emission lines. 

        Stacked spectra of star-forming galaxies in classes~8 and~9
        are  well fitted with the spectrum of the Sc galaxy $NGC4775$. 
        According to the morphological properties, the $NGC4775$ galaxy is classified as the late-type spiral galaxy with flocculent spiral arms~\citep{buta15}.
        Spectra of the Sc family qualitatively differ from those of earlier Hubble types with respect to their continuum shape, absorption and emission features.  
        They present strong principal emission lines, like $H\beta$, and $[OIII]$. 
        
        The stacked spectrum of galaxies assigned to class~10 is best-fit with a template spectrum of the peculiar I galaxy $UGC6697$.
        This family is built from star-forming galaxies with much
        stronger emission lines than the average strengths observed in Sb or Sc families~\citep[3-10 times higher EW($H\alpha,NII$)][]{kennicutt92}. 
        Therefore, class 10 represents galaxies which are undergoing global bursts of star formation. 
        
        The Im galaxy $MK35$ fits best the representative stacked spectrum of class~11. 
        This template characterises an extreme emission-line galaxy, and might represent a new-born galaxy. 
        The galaxy is filled with gas and is in the phase of the global burst of star formation. 
        
        The stacked spectrum of broad-line AGNs (class 12) is shown in the last panel in Fig.~\ref{fig:atlas}, but is not compared to the Atlas of~\cite{kennicutt92} due to a lack of broad-line AGN templates.
        
        In conclusion, the 11 FEM classes follow the Hubble sequence
        according to spectroscopic types given by the Atlas of~\cite{kennicutt92}.
        Classes 1--3 present spectra typical for early-type galaxies, while classes 7--11 demonstrate spectra of actively star-forming galaxies with extreme emission-line galaxies in the highest class, with green galaxies showing spectra of intermediate types.

\end{document}